\documentclass[toc,a4paper]{PoS}

\title{Review of hard scattering and jet analysis}

\ShortTitle{Review of hard scattering and jet analysis}

\author{\speaker{Michael J. Tannenbaum}%
         \thanks{Supported by the U.S. Department of Energy, Contract No. DE-AC02-98CH1-886.}\\
        Physics Dept., 510c, Brookhaven National Laboratory, Upton, NY 11973-5000, USA\\
        E-mail: \email{mjt@bnl.gov}}

\abstract{A review of hard-scattering and jet analysis in p-p and heavy ion collisions at RHIC is presented in the context of earlier work at the CERN ISR in the 1970's which utilized inclusive single or pairs of hadrons to establish that high 
transverse momentum particles in p-p collisions are produced from states with two roughly back-to-back jets which are 
the result of scattering of constituents of 
the nucleons as described by Quantum Chromodynamics (QCD), which was developed during the course of those measurements. These techniques have been used extensively and further developed at RHIC since they are the only practical method to study hard-scattering and jet phenomena in Au+Au central collisions at RHIC energies. }

\FullConference{Correlations and Fluctuations in Relativistic Nuclear Collisions\\
		 July 7-9 2006\\
		 Florence, Italy}
\def\lsim{\raise0.3ex\hbox{$<$\kern-0.75em\raise-1.1ex\hbox{$\sim$}}}
\def\gsim{\raise0.3ex\hbox{$>$\kern-0.75em\raise-1.1ex\hbox{$\sim$}}}

\def\mean#1{\left<#1\right>}
\def\Journal#1#2#3#4{ {\it{#1}} {\bf #2}, #3 (#4)}
\def\IJMPA{{Int. J. Mod. Phys.}~{\rm A}}
\def\EPJC{{Eur. Phys. J.}~{\rm C}}
\def\JPG{{J. Phys}~{\rm G}}
\def\JPCS{{J. Phys: Conf. Series\ }}

\def\NPA{{Nucl. Phys.}~{\rm A}}
\def\NPB{{Nucl. Phys.}~{\rm B}}
\def\PLB{{Phys. Lett.}~{\rm B}}
\def\PLC{Phys. Repts.\ }
\def\PRL{Phys. Rev. Lett.\ }
\def\PRD{{Phys. Rev.}~{\rm D}}
\def\PRC{{Phys. Rev.}~{\rm C}}

\def\ZPC{{Z. Phys.}~{\rm C}}
\def\ARNPS{{Ann. Rev. Nucl. Part. Sci.\ }} 
\def\RMP{Rev. Mod. Phys.\ }
\begin{document}
\section{Introduction}

  In 1998, at the QCD workshop in Paris, Rolf Baier asked me whether jets could be measured in Au+Au collisions because he had a prediction of a QCD medium-effect (energy loss via  soft gluon radiation induced by multiple scattering~\cite{BDPS}) on color-charged partons traversing a hot-dense-medium composed of screened color-charges~\cite{BaierQCD98}. I told him~\cite{MJTQCD98} that there was a general consensus~\cite{Strasbourg} that for Au+Au central collisions at $\sqrt{s_{NN}}=200$ GeV, leading particles are the only way to find jets, because in one unit of the nominal jet-finding cone,  $\Delta r=\sqrt{(\Delta\eta)^2 + (\Delta\phi)^2}$, there is an estimated $\pi\Delta r^2\times{1\over {2\pi}} {dE_T\over{d\eta}}\sim 375$ GeV of energy !(!) The good news was that hard-scattering in p-p collisions was originally observed by the method of leading particles and that these techniques could be used to study hard-scattering and jets in Au+Au collisions. In fact, in several recent talks~\cite{MJTEPS04,MJTHP04,MJTCF05} and in talks from earlier years~\cite{MJTQCD98, MJTRHIC97}, some as long ago as 1979~\cite{MJTDPF79}, I have been on record describing ``How everything you want to know about jets can be found using 2-particle correlations''. This past year, I had to soften the statement to {\em almost everything} because we found by explicit calculation in PHENIX~\cite{ppg029} that the two-particle opposite-side correlation is actually quite insensitive to the fragmentation function---overturning a belief dating from the seminal paper of Feynman, Field and Fox in 1977~\cite{FFF}. However, we also found that the opposite-side correlation function {\em is}  sensitive to the ratio of the transverse momentum of the away-side jet ($\hat{p}_{T_a}$)  to that of the trigger-side jet ($\hat{p}_{T_t}$) and thus provides a way to measure the relative energy loss of the two jets from a hard-scattering which
escape from the medium in an A+A collision.   
  
     \section{Status of theory and experiment, circa 1982}
 
Hard-scattering was visible both at the ISR and at FNAL fixed-target-energies 
via inclusive single particle production at large $p_T\geq$ 2-3 
GeV/c~\cite{egsee1}. Scaling and dimensional arguments~\cite{Bj,BBK,CIM,CGKS} for plotting 
data revealed the systematics and underlying physics. The theorists had the 
basic underlying physics correct; but many (inconvenient) details remained to 
be worked out, several by experiment. The transverse momentum 
imbalance of outgoing parton-pairs, the ``$k_T$-effect", was 
discovered by experiment~\cite{CCHK,MJT79}, and clarified by Feynman and collaborators~\cite{FFF}. The first modern QCD calculation and 
prediction for high $p_T$ single particle inclusive cross sections, including 
non-scaling and initial state radiation was done in 1978, by Jeff 
Owens and collaborators~\cite{Owens78} under the assumption that high $p_T$ particles  
are produced from states with two roughly back-to-back jets
which are the result of scattering of constituents of the nucleons (partons).

	The overall p-p hard-scattering cross section in ``leading logarithm'' pQCD~\cite{Owens}    
is the sum over parton reactions $a+b\rightarrow c +d$ 
(e.g. $g+q\rightarrow g+q$) at parton-parton center-of-mass (c.m.) energy $\sqrt{\hat{s}}$:   
\begin{equation}
\frac{d^3\sigma}{dx_1 dx_2 d\cos\theta^*}=
\frac{s d^3\sigma}{d\hat{s} d\hat{y} d\cos\theta^*}=
\frac{1}{s}\sum_{ab} f_a(x_1) f_b(x_2) 
\frac{\pi\alpha_s^2(Q^2)}{2x_1 x_2} \Sigma^{ab}(\cos\theta^*)
\label{eq:QCDabscat}
\end{equation} 
where $f_a(x_1)$, $f_b(x_2)$, are parton distribution functions, 
the differential probabilities for partons
$a$ and $b$ to carry momentum fractions $x_1$ and $x_2$ of their respective 
protons (e.g. $u(x_2)$), and where $\theta^*$ is the scattering angle in the parton-parton c.m. system. 
The parton-parton c.m. energy squared is $\hat{s}=x_1 x_2 s$,
where $\sqrt{s}$ is the c.m. energy of the p-p collision. The parton-parton 
c.m. system moves with rapidity $\hat{y}=1/2 \ln (x_1/x_2)$ in the p-p c.m. system. The quantities $f_a(x_1)$ and $f_b(x_2)$, the ``number'' 
distributions of the constituents, are related 
(for the electrically charged quarks) to the structure functions measured in 
Deeply Inelastic lepton-hadron Scattering (DIS), e.g. 
\begin{equation}
F_{1}(x,Q^2)={1\over 2} \sum_a e_a^2\; f_a(x,Q^{2}) \;\;\;\mbox{and}\;\;\;
F_{2}(x,Q^2)=x\sum_a e_a^2\; f_a(x,Q^{2})
\label{eq:F2}
\end{equation}
where $e_a$ is the electric charge on a quark. 

The Mandelstam invariants $\hat{s}$, $\hat{t}$ and 
$\hat{u}$ of the constituent scattering have a clear definition in terms of the
scattering angle $\theta^*$ in the constituent c.m. system:  
\begin{equation}
\hat{t} = -\hat{s}\; \frac{(1-\cos\theta^*)}{2} \qquad\mbox{ and }\qquad
\hat{u} = -\hat{s}\; \frac{(1+\cos\theta^*)}{2} 
\quad .
\label{eq:hatt}
\end{equation} 
The transverse momentum of a scattered constituent is:
\begin{equation}
p_T = p_T^* = { \sqrt{\hat{s}} \over 2 } \; \sin\theta^* \: ,
\label{eq:cpT}
\end{equation}
and the scattered constituents $c$ and $d$ in the outgoing parton-pair have equal and opposite momenta in the parton-parton (constituent) c.m. system. 
A naive experimentalist would think of $Q^2 = -\hat{t}$ for a scattering 
subprocess and $Q^2 = \hat{s}$ for a Compton or annihilation subprocess.

\begin{figure}[ht]
\begin{center}
\begin{tabular}{cc}
\hspace*{-0.1in}\includegraphics[width=0.75\linewidth]{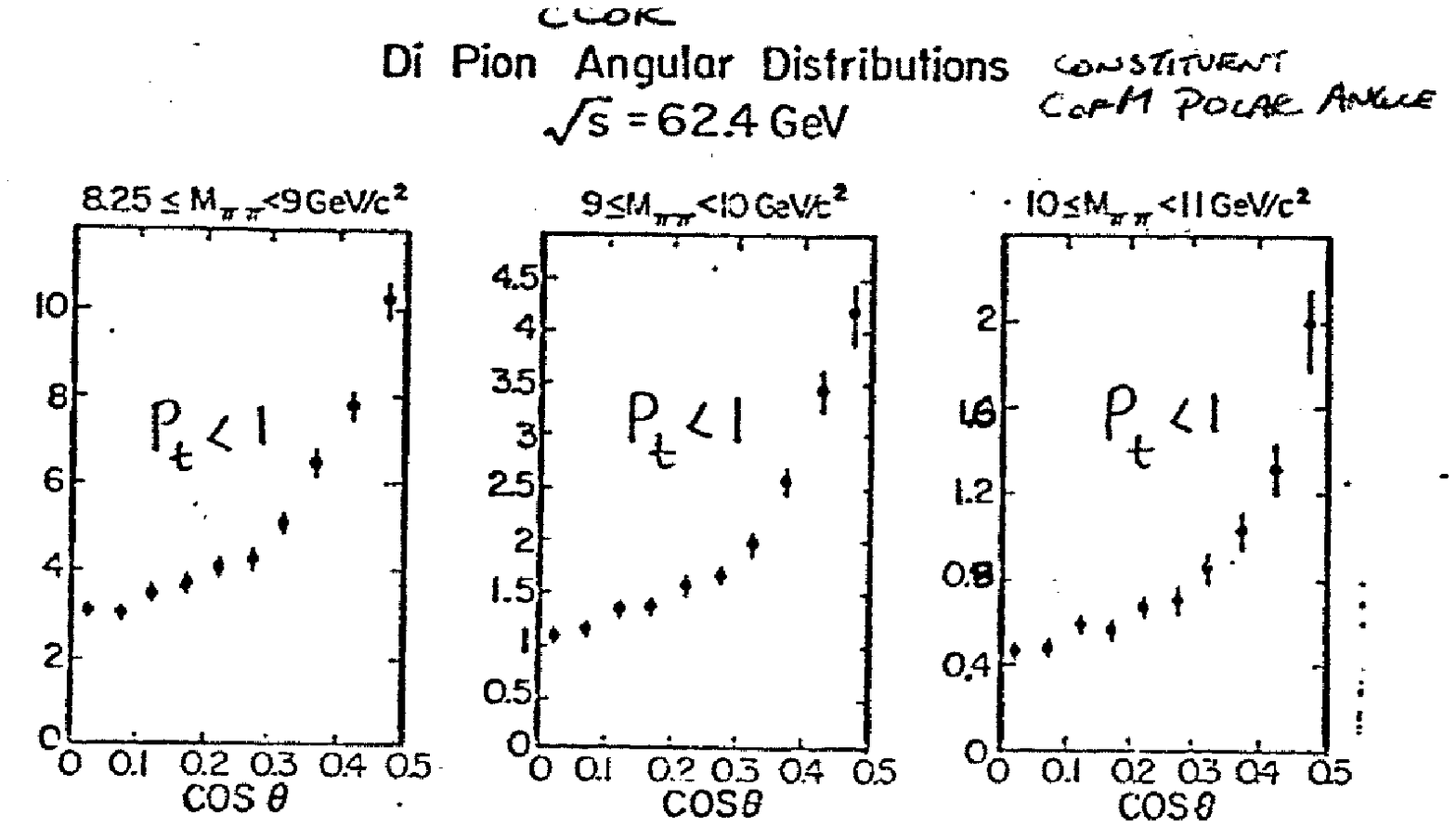} &
\hspace*{-0.35in}\includegraphics[width=0.288\linewidth]{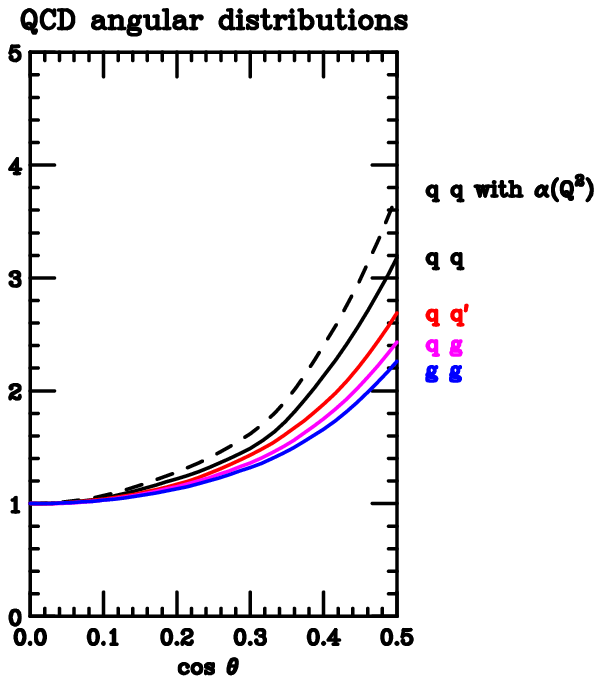}
\end{tabular}
\end{center}
\caption[]
{a) (left 3 panels) CCOR measurement~\cite{Paris82,CCOR82NPB} of polar angular distributions of $\pi^0$ pairs with net $p_T < 1$ GeV/c at mid-rapidity in p-p collisions with $\sqrt{s}=62.4$ GeV for 3 different values of $\pi\pi$ invariant mass $M_{\pi \pi}$. b) (rightmost panel) QCD predictions for $\Sigma^{ab}(\cos\theta^*)$ for the elastic scattering of $gg$, $qg$, $qq'$, $qq$, and $qq$ with $\alpha_s(Q^2)$ evolution.    
\label{fig:mjt-ccorqq} }
\end{figure}

Equation~\ref{eq:QCDabscat} gives the $p_T$ spectrum of outgoing parton $c$, which then
fragments into a jet of hadrons, including e.g. $\pi^0$.  The fragmentation function
$D^{\pi^0}_{c}(z)$ is the probability for a $\pi^0$ to carry a fraction
$z=p^{\pi^0}/p^{c}$ of the momentum of outgoing parton $c$. Equation~\ref{eq:QCDabscat}
must be summed over all subprocesses leading to a $\pi^0$ in the final state weighted by their respective fragmentation functions. In this formulation, $f_a(x_1)$, $f_b(x_2)$ and $D^{\pi^0}_c (z)$ 
represent the ``long-distance phenomena" to be determined by experiment;
while the characteristic subprocess angular distributions,
{\bf $\Sigma^{ab}(\cos\theta^*)$} (see Fig.~\ref{fig:mjt-ccorqq}) 
and the coupling constant,
$\alpha_s(Q^2)=\frac{12\pi}{25\, \ln(Q^2/\Lambda^2)}$,
are fundamental predictions of QCD~\cite{CutlerSivers,Combridge:1977dm} 
for the short-distance, large-$Q^2$, phenomena. 
When higher order effects are taken into account, it is necessary to specify  factorization scales $\mu$ for the distribution and fragmentation functions in addition to renormalization scale $\Lambda$ which governs the running of 
$\alpha_s(Q^2)$. 
As noted above, the momentum scale $Q^2\sim p_T^2$ for the scattering
subprocess, while $Q^2\sim\hat{s}$ for a Compton or annihilation subprocess,
but the exact meaning of $Q^2$ and $\mu^2$ tend  to be treated as parameters  rather than as dynamical
quantities.

\begin{figure}[ht]
\begin{center}
\includegraphics[width=0.80\linewidth]{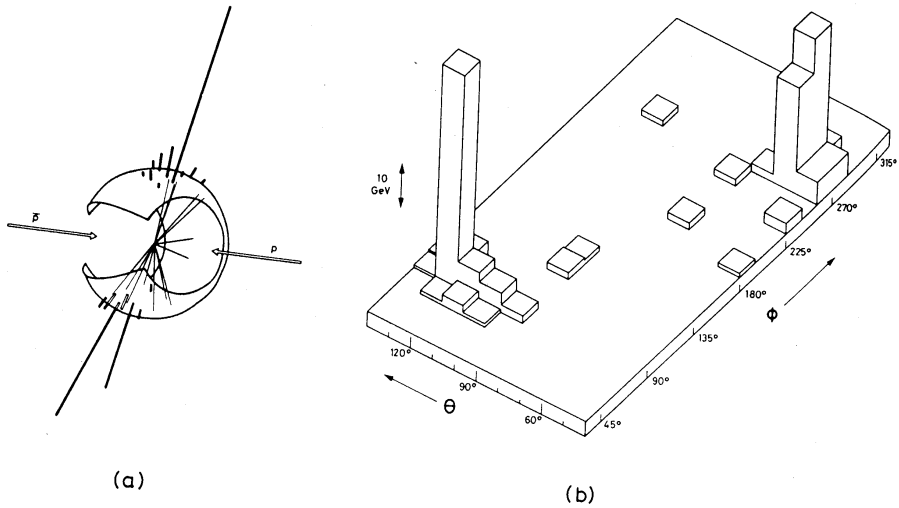} 
\end{center}
\caption[]
{ UA2 jet event from 1982 ICHEP~\cite{Paris82}. a) event shown in geometry of detector. b) ``Lego" plot of energy in calorimeter cell as a function of angular position of cell in polar ($\theta$) and azimuthal ($\Phi$) angle space.
\label{fig:UA2jet} }
\end{figure}

 Due to the fact (which was unknown in the 1970's) that jets in $4\pi$ calorimeters at ISR 
energies or lower are invisible below $\sqrt{\hat{s}}\sim E_T \leq 25$ 
GeV~\cite{Gordon}, there were many false claims of jet observation in the period 1977-1982. This led to skepticism 
about jets in hadron collisions, particularly in the USA~\cite{MJTIJMPA}. 
A `phase change' in belief-in-jets was produced by one UA2 event 
at the 1982 ICHEP in Paris~\cite{Paris82} (Fig.~\ref{fig:UA2jet}), 
which, together with the first direct measurement of the QCD constituent-scattering angular distribution, $\Sigma^{ab}(\cos\theta^*)$ (Eq.~\ref{eq:QCDabscat}), using two-particle correlations~\cite{CCOR82NPB}, presented at the same meeting (Fig.~\ref{fig:mjt-ccorqq}), gave universal credibility to the pQCD description of high $p_T$ hadron physics~\cite{Owens,Darriulat,DiLella}.    

\section{Mid-rapidity $p_T$ spectra from p-p collisions---$x_T$-scaling}
Equation \ref{eq:QCDabscat} leads to a general `$x_T$-scaling' form for the invariant cross
section of high-$p_T$ particle production: 
\begin{equation}
E \frac{d^3\sigma}{d^3p} = \frac{1}{p_T^{n(x_T,\sqrt{s})}} F({x_T}) = 
 \frac{1}{\sqrt{s}^{\, n(x_T,\sqrt{s})}} G({x_T}) \: ,
 \label{eq:bbg}
 \end{equation} 
where $x_T = 2p_T/\sqrt{s}$. 
The cross section has two factors, a function $F({x_T})$ ($G({x_T})$) which `scales',
i.e. depends only on the ratio of momenta, and a dimensioned factor,
${1/p_T^{n(x_T,\sqrt{s})}}$ ($1/\sqrt{s}^{\, n(x_T,\sqrt{s})}$),   
where $n(x_T,\sqrt{s})$ equals 4  in lowest-order (LO) calculations, analogous to the $1/q^4$ form
of Rutherford Scattering in QED. The structure and fragmentation
functions   
are all in the
$F(x_T)$ ($G(x_T)$) term. Due to higher-order effects such as the running of
the coupling constant, $\alpha_s(Q^2)$, the evolution of the
structure and fragmentation functions, and the initial-state
transverse momentum
$k_T$, $n(x_T,\sqrt{s})$ is not a constant but is a function of $x_T$, $\sqrt{s}$.
Measured values of ${\,n(x_T,\sqrt{s})}$ for $\pi^0$ in p-p 
collisions are between 5 and 8~\cite{MJTEPS04}.

\begin{figure}[!thb]
\begin{center}
\begin{tabular}{cc}

\includegraphics[width=0.50\linewidth]{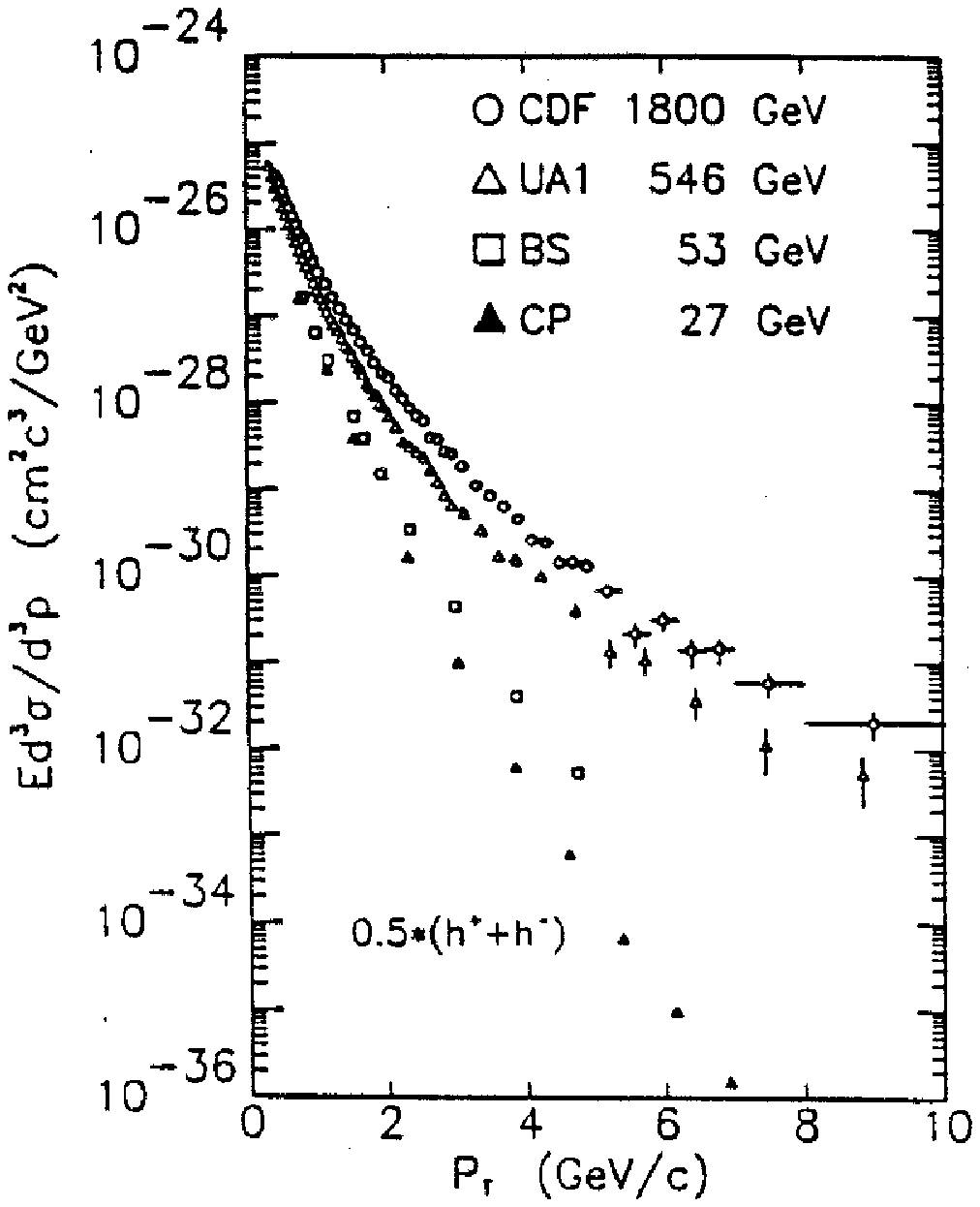}&
\includegraphics[width=0.44\linewidth]{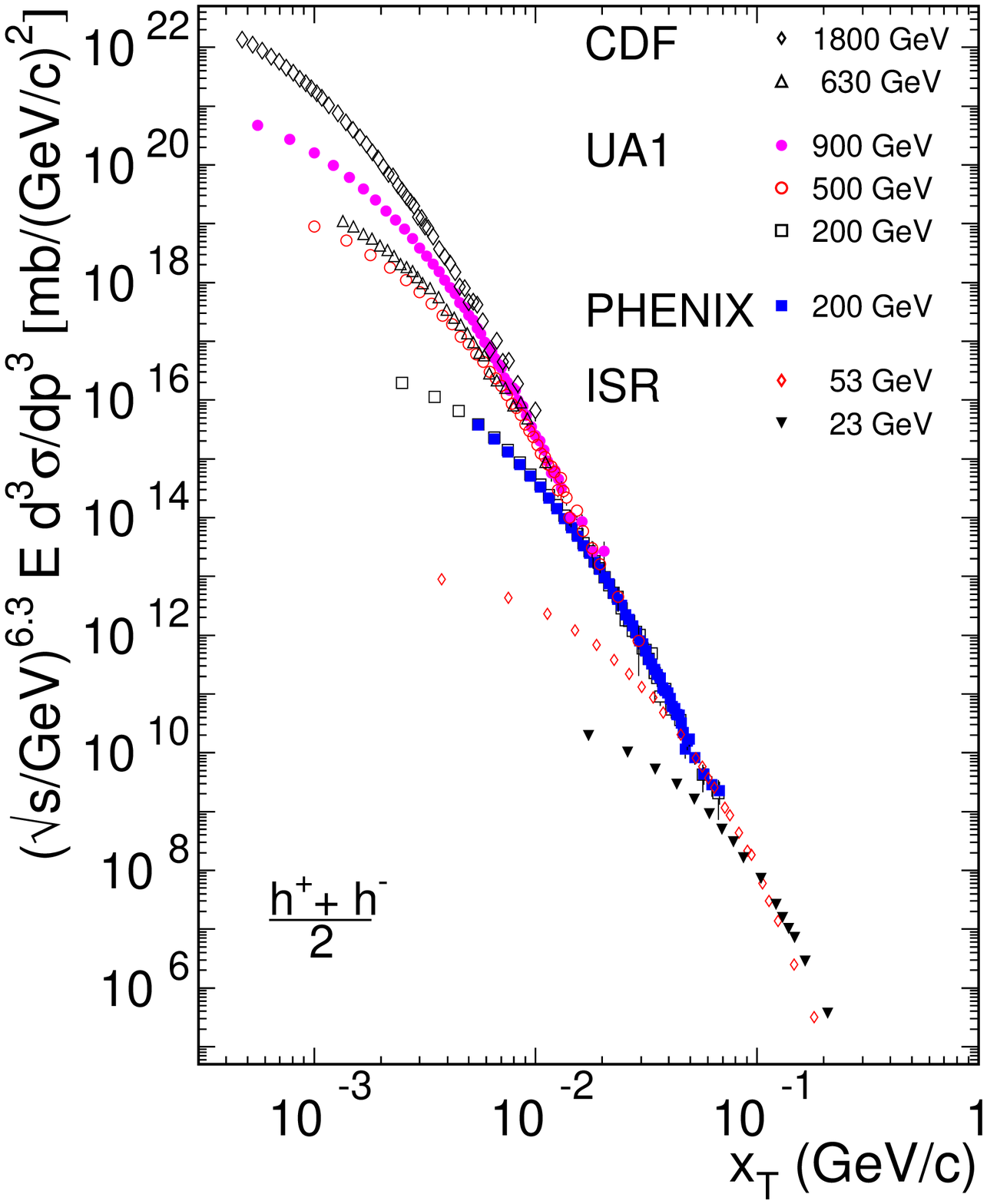}

\end{tabular}
\end{center}\vspace*{-0.25in}
\caption[]{a) (left) $E {d^3\sigma}(p_T)/{d^3p}$ at mid-rapidity as a function of $\sqrt{s}$ in p-p collisions~\cite{CDF}. b) (right) log-log plot of $\sqrt{s}({\rm GeV})^{6.3}\times Ed^3\sigma/d^3p$ vs $x_T=2{p_T}/\sqrt{s}$~\cite{Adler:2003pb}. }
\label{fig:hpTxT}

\end{figure}

\begin{figure}[ht]
\begin{center}
\begin{tabular}{cc}

\includegraphics[width=0.48\linewidth]{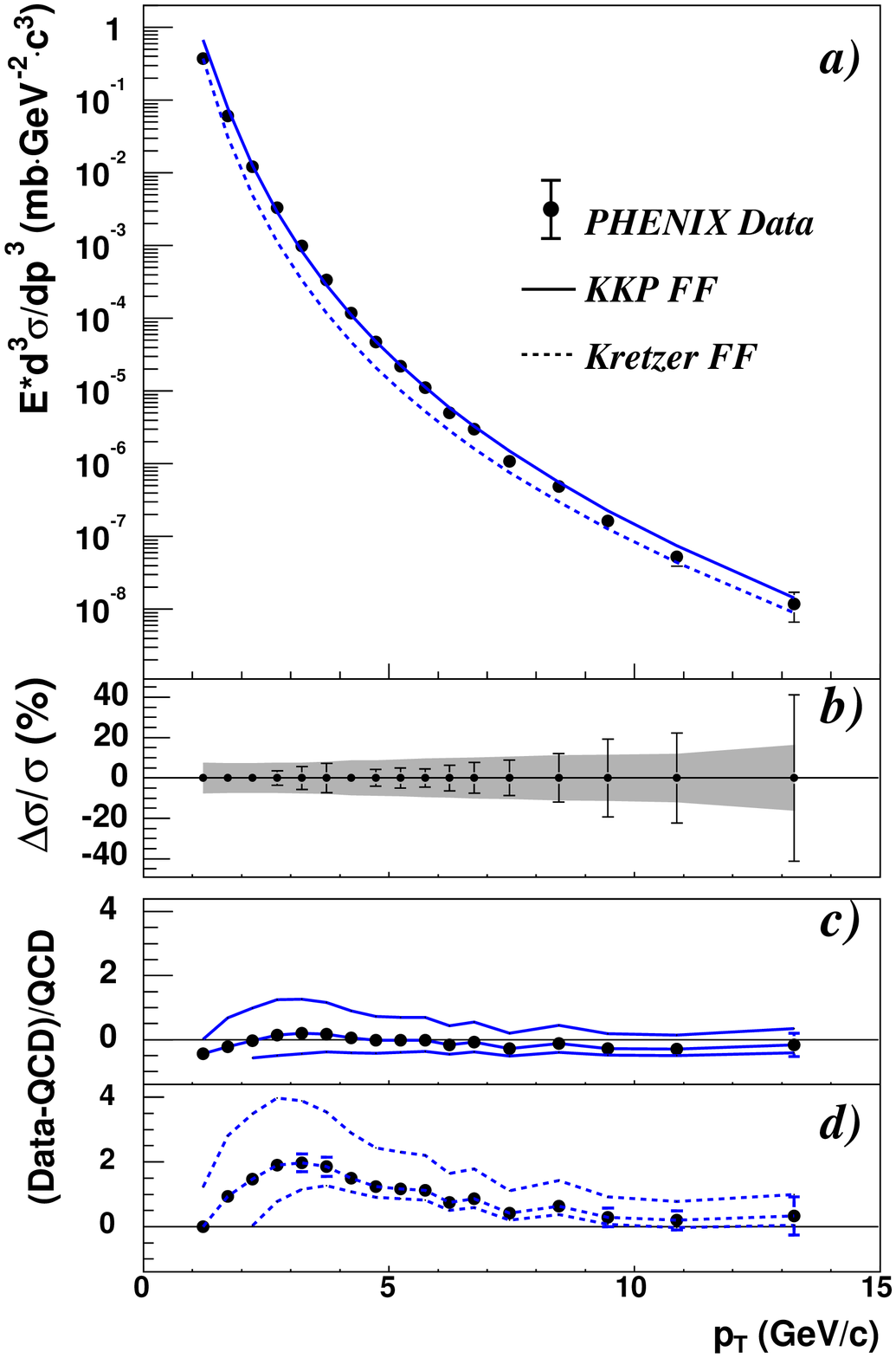} \hspace*{-0.2in}& 
e)\hspace*{-0.1in}\vspace*{-0.2in}\includegraphics[width=0.50\linewidth]{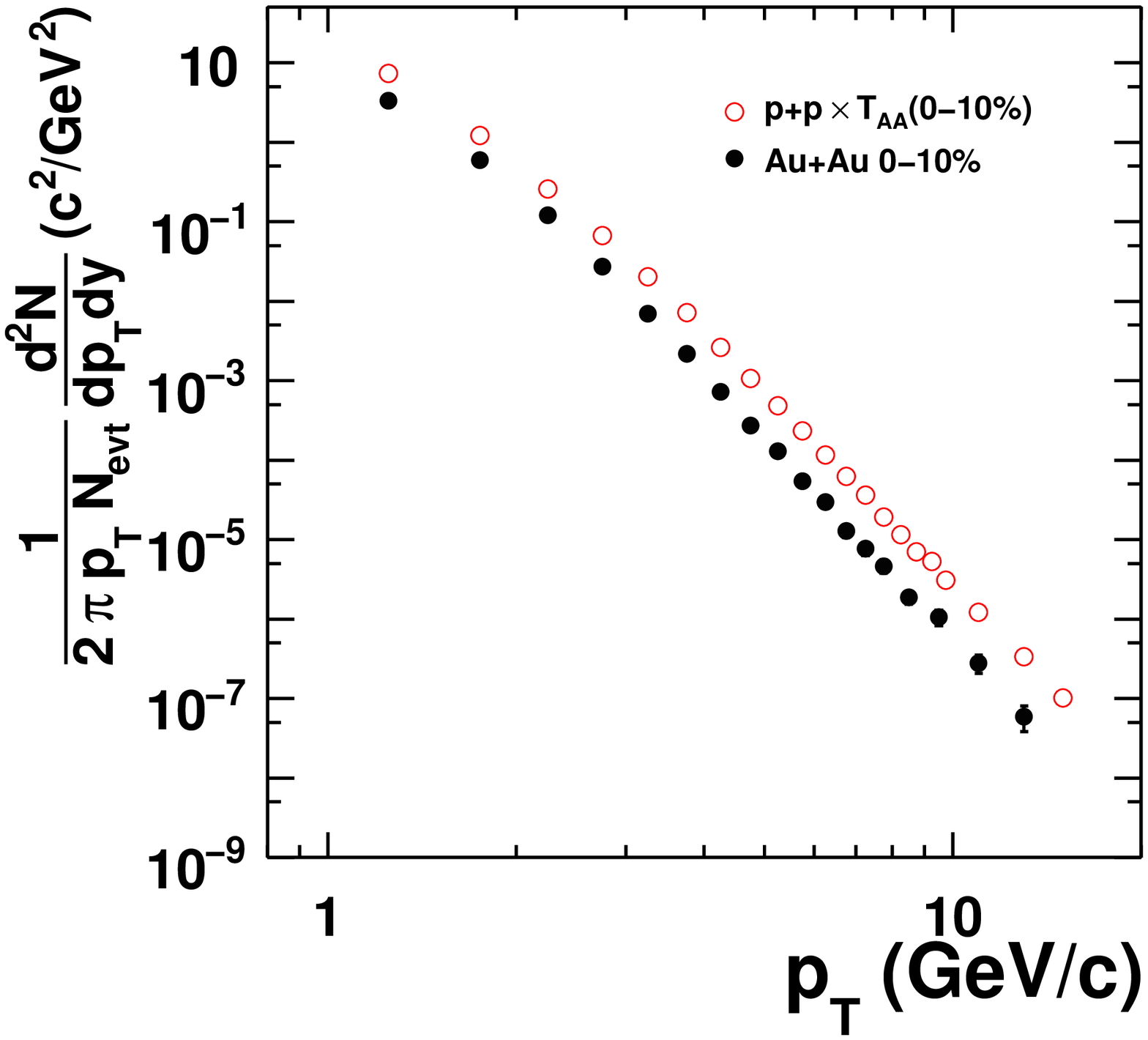}  
\end{tabular}

\end{center}
\caption[]{(left) PHENIX~\cite{Adler:2003pb}  $\pi^0$ invariant cross section at mid-rapidity from p-p collisions
at $\sqrt{s} = 200$ GeV, together with NLO pQCD predictions from Vogelsang \cite{Aversa:1988vb,Jager:2002xm}. 
a) The invariant differential cross section for inclusive $\pi^{\circ}$ 
production (points) and the results from 
NLO pQCD calculations with equal renormalization and factorization scales of $p_T$ 
using the ``Kniehl-Kramer-P\"{o}tter'' (solid line)  and ``Kretzer'' (dashed line) 
sets of fragmentation functions.  
b) The relative statistical (points) and point-to-point systematic (band) errors.  
c,d)  The relative difference between
the data and the theory using KKP (c) and Kretzer (d) fragmentation functions with
scales of $p_T$/2 (lower curve), $p_T$, and 2$p_T$
(upper curve).  In all figures, 
the normalization error of 9.6\% is not shown. (right) e) p-p data from a) multiplied by the nuclear thickness function, $T_{AA}$, for Au+Au central (0-10\%) collisions plotted on a log-log scale (open circles) together with the measured semi-inclusive $\pi^0$ invariant yield in Au+Au central collisions at $\sqrt{s_{NN}}=200$ GeV~\cite{ppg054} } \label{fig:pizpp200} 
\end{figure}

The scaling and power-law behavior of hard scattering are evident from the $\sqrt{s}$ dependence of the $p_T$ dependence of the p-p invariant cross sections.  This is
shown for nonidentified charged hadrons,
$(h^+ + h^-)/2$, in Fig. \ref{fig:hpTxT}a. 
At low $p_T\leq 1$ GeV/$c$ the cross sections exhibit a ``thermal'' 
$\exp {(-6 p_T)}$ dependence, which is largely independent of $\sqrt{s}$, while at high $p_T$
there is a power-law tail, due to hard scattering, which depends strongly on $\sqrt{s}$. 
The characteristic variation with $\sqrt{s}$ at high $p_T$ is produced by the fundamental
power-law and scaling dependence of Eqs. \ref{eq:QCDabscat}, \ref{eq:bbg}. This is best
illustrated by a plot of 
\begin{equation}
\sqrt{s}^{{\,n(x_T,\sqrt{s})}} \times E \frac{d^3\sigma}{d^3p} = G(x_T) \: ,
\label{eq:xTscaling}
\end{equation}
as a function of $x_T$, with ${{\,n(x_T,\sqrt{s})}} = 6.3$, which is valid for the $x_T$
range of the present RHIC measurements (Fig. \ref{fig:hpTxT}b).  The data show an
asymptotic power law with increasing $x_T$. Data at a given $\sqrt{s}$ fall
below the asymptote at successively lower values of $x_T$ with
increasing $\sqrt{s}$, corresponding to the transition region from
hard to soft physics in the $p_T$ region of about 2 GeV/$c$. 
Although $x_T$-scaling provides a rather general test of the validity QCD without reference to details, the agreement of the PHENIX measurement of the invariant cross section for $\pi^0$ production in p-p 
collisions at $\sqrt{s} = 200$ GeV~\cite{Adler:2003pb} with NLO pQCD predictions
over the range $2.0\leq p_T\leq 15$ GeV/$c$ (Fig. \ref{fig:pizpp200}) is, nevertheless, impressive.   

\subsection{The importance of the power law}
\label{sec:power}
   A log-log plot of the $\pi^0$ spectrum from Fig.~\ref{fig:pizpp200}a in p-p collisions, shown in Fig.~\ref{fig:pizpp200}e along with corresponding data from Au+Au collisions~\cite{ppg054}, illustrates that the inclusive single particle hard-scattering cross section is a pure power law for $p_T\geq 3$ GeV/c. 
The invariant cross section for $\pi^0$ production can be fit to the form 
\begin{equation}
E d^3 \sigma/dp^3\propto p_T^{-n}
\label{eq:simpower}
\end{equation}
 with $n=8.10\pm 0.05$~\cite{ppg054}. It is important to emphasize that this $n$ is different from the $n(x_T,\sqrt{s})$ in Eqs.~\ref{eq:bbg}, \ref{eq:xTscaling}. The power $n$ in Eq.~\ref{eq:simpower} measures the pure power law shape of the cross section at a fixed $\sqrt{s}$ represented by the function $G(x_T)$ in Eqs.~\ref{eq:bbg}, \ref{eq:xTscaling}, while $n(x_T,\sqrt{s})$ represents the $p_T$ dependence at fixed $x_T$ when $\sqrt{s}$ is varied. Clearly, from Eq.~\ref{eq:bbg} and Fig. \ref{fig:hpTxT}b, the simple power $n$ (Eq.~\ref{eq:simpower}) is greater than or equal to $n(x_T, \sqrt{s})$ (Eq.~\ref{eq:xTscaling}). 
 
       The steeply-falling power-law spectrum at a given $\sqrt{s}$ has many important consequences for single particle inclusive measurements of hard-scattering. The most famous properties in this regard are the ``Bjorken parent-child relationship''~\cite{BjPRD8} and the ``leading-particle effect'', which also goes by the unfortunate name ``trigger bias''~\cite{JacobLandshoff,BjPRD8}. These will be discussed below in the section on correlations. The power-law also makes the calculation of the inclusive photon spectrum from the decay $\pi^0\rightarrow \gamma +\gamma$ very easy, but nevertheless very precise~\cite{prec-note,CCRSNPB113}, when expressed as the ratio of photons from $\pi^0$ to $\pi^0$ at the same $p_T$:
       \begin{equation} 
\left . {\gamma\over \pi^{0}}\right |_{\pi^0}\!\! (p_T)=2/(n-1) \qquad.         \label{eq:gampi}
       \end{equation}
Similarly, the inclusive electron spectrum from internal or external conversion of these photons has a simple formula when expressed as the ratio to $\pi^0$ at the same $p_T$:
\begin{equation} 
 \left . {e^{-}\over \pi^{0}}\right |_{\pi^0}\!\! (p_T)= 
\left . {(e^{-} + e^{+})\over 2\pi^{0}}\right |_{\pi^0}\!\! (p_T) = 
\left ({ \delta_{2} \over 2} + {t \over { {9\over 7} X_0} } \right )  \times {2\over {(n-1)^2}} \qquad, 
\label{eq:eoverpi}
\end{equation}
where $\delta_{2}/2=$ Dalitz (internal conversion) branching ratio per photon 
and $t/X_{0}$ is the thickness of the external converter in radiation lengths ($X_{0}$)~\cite{CCRSPLB53,MJTqcd2003}. 

\section{Measurement of the medium effect in A+A collisions with hard-scattering by comparison to baseline measurements in p-p and d+A collisions}

    Since hard scattering is point-like, with distance scale $1/p_T< 0.1$ fm, the cross section in p+A (B+A) collisions, compared to p-p, should be simply proportional to the relative number of possible point-like encounters~\cite{MMay}, a factor of $A$ ($BA$) for p+A (B+A) minimum bias collisions. For semi-inclusive reactions in centrality class $f$ at impact parameter $b$, the scaling is proportional to $T_{AB}(b)$, the overlap integral of the nuclear thickness functions~\cite{Vogt99}, where $\mean{T_{AB}}_{f}$ averaged over the centrality class is:
    \begin{equation}
\langle T_{AB}\rangle_{f}=\frac{\displaystyle\int_{f} T_{AB}(b)\,d^2b}{\displaystyle\int_{f} (1- e^ {-\sigma_{NN}\,T_{AB}(b)})\, d^2 b}=\frac{\langle N_{coll}\rangle_f}{\sigma_{NN}} \quad, 
\label{eq:TABf}
\end{equation}
and where $\langle N_{coll}\rangle_f$ is the average number of binary nucleon-nucleon
inelastic collisions, with cross section $\sigma_{NN}$, in the centrality class $f$.
This leads to the description of the scaling for point-like processes as binary-collision
(or $N_{coll}$)  scaling. This description is convenient, but confusing, because the scaling has nothing to do with the inelastic hadronic collision probability, it is proportional only to the geometrical factor $\mean{T_{AB}}_{f}$ (Eq.~\ref{eq:TABf}). 

    Effects of the nuclear medium, either in the initial or final state, may modify the point-like scaling. This is shown rather dramatically in Fig.~\ref{fig:pizpp200}e where the Au+Au data are suppressed relative to the scaled p-p data by a factor of $\sim 4-5$ for $p_T\geq 3$ GeV/c. A quantitative evaluation of the  suppression is made using the ``nuclear modification factor'', $R_{AB}$, the ratio of the measured semi-inclusive yield to the point-like scaled p-p cross section: 
\begin{equation}
R_{AB} = \frac{dN_{AB}^P}{\langle T_{AB} \rangle_{f} \times d\sigma_{NN}^P}
       = \frac{dN_{AB}^P}{\langle N_{coll} \rangle_{f} \times dN_{NN}^P} \label{eq:RAB}
\end{equation}
where $dN_{AB}^P$ is the differential yield of a point-like process $P$
in an $A+B$ collision and $d\sigma_{NN}^P$ is the cross section of $P$ in an $NN$ (usually p-p)  collision. For point-like scaling, $R_{AB}=1$. 

\begin{figure}[ht]
	\begin{center}
	\includegraphics[width=0.7\linewidth]{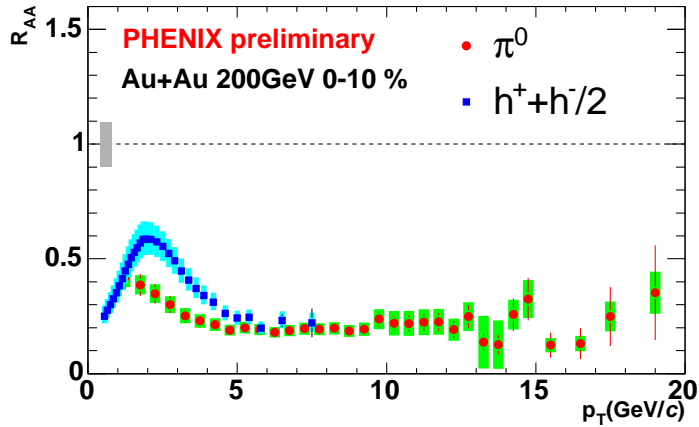}
	\end{center}
	\vspace*{-0.24in}

	\caption[] {PHENIX measurement of nuclear modification factor $R_{AA}$ for identified $\pi^0$ and-non identified charged hadrons $(h^+ + h^-)/2$ for central (0-10\%) Au+Au collisions at $\sqrt{s_{NN}}=200$ GeV~\cite{MayaQM05}.}
	\label{fig:RAA-pi-h}
	\end{figure}

	While the suppression of $\pi^0$ at a given $p_T$ in Au+Au compared to the scaled p-p spectrum may be imagined as a loss of these particles due to, for instance, the stopping or absorption of a certain fraction of the parent partons in an opaque medium, it is evident from Fig.~\ref{fig:pizpp200}e that an equally valid quantitative representation can be given by a downshift of the scaled p-p spectrum due to, for instance, the energy loss of the parent partons in the medium---a particle with $p{'}_T$ in the scaled p-p spectrum is shifted in energy by an amount $S(p_T)$ to a measured value $p_T=  p{'}_{T} -S(p_T)$ in the Au+Au spectrum~\cite{explain1}. 
	The fact that the Au+Au and reference p-p spectra are parallel on Fig.~\ref{fig:pizpp200}e provides graphical evidence that the fractional $p_T$ shift in the spectrum, $S(p_T)/p_T$ is a constant for $p_T > 3$ GeV/c, which, due to the power law, results in a constant ratio of the $\pi^0$ spectra $R_{AA}(p_T)$ as shown in Fig.~\ref{fig:RAA-pi-h}. 
  
    The nuclear modification factors are clearly different for $\pi^0$ and $(h^+ + h^-)/2$ for $p_T < 6$ GeV/c in Fig.~\ref{fig:RAA-pi-h}. Is it possible to tell whether one or both of these reactions obey QCD? 
      \subsection{$x_T$ scaling in A+A collisions as a test of QCD}  
\begin{figure}[tbhp]
   \begin{center}
\includegraphics[width=0.8\linewidth]{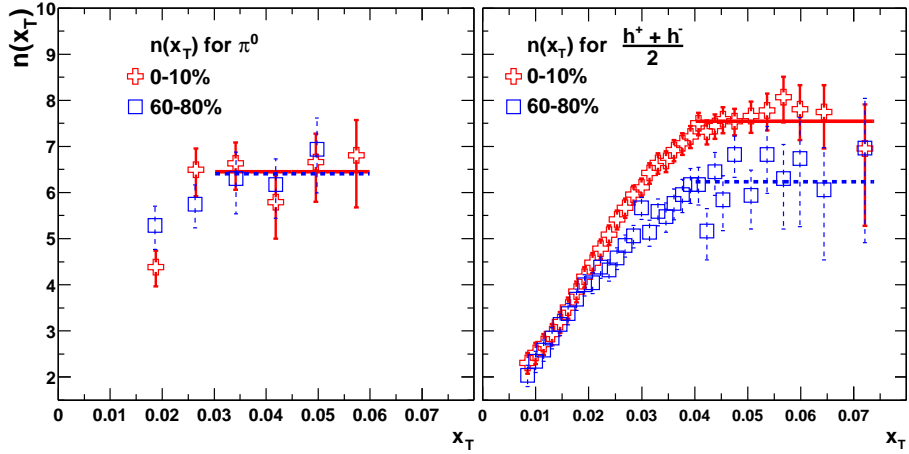}
\end{center}
	\vspace*{-0.24in}
\caption[]{Power-law exponent $n(x_T)$ for $\pi^0$ and $h$ spectra in central
and peripheral Au+Au collisions at $\sqrt{s_{NN}} = 130$ and 200 GeV \cite{Adler:2003au}. } 
\label{fig:nxTAA}
\end{figure}

If the production of high-$p_T$ particles in Au+Au collisions is the
result of hard scattering according to pQCD, then $x_T$ scaling should
work just as well in Au+Au collisions as in p-p collisions and should
yield the same value of the exponent $n(x_T,\sqrt{s})$.  The only assumption
required is that the structure and fragmentation functions in Au+Au
collisions should scale, in which case Eq. \ref{eq:xTscaling} still
applies, albeit with a $G(x_T)$ appropriate for Au+Au. In
Fig. \ref{fig:nxTAA}, $n(x_T,\sqrt{s_{NN}})$ in Au+Au is derived from
Eq. \ref{eq:xTscaling}, for peripheral and central collisions, by
taking the ratio of $E d^3\sigma/dp^3$ at a given $x_T$ for
$\sqrt{s_{NN}} = 130$ and 200 GeV, in each case. 
The $\pi^0$'s exhibit
$x_T$ scaling, with the same value of $n = 6.3$ as in p-p collisions,
for both Au+Au peripheral and central collisions, while the
non-identified charged hadrons $x_T$-scale with $n = 6.3$ for peripheral
collisions only. Notably, the $(h^+ +h^-)/2$ in Au+Au central collisions
exhibit a significantly larger value of $n(x_T,\sqrt{s})$, indicating different
physics, which will be discussed below.  The $x_T$ scaling establishes
that high-$p_T$ $\pi^0$ production in peripheral and central Au+Au
collisions and $(h^+ +h^-)/2$ production in peripheral Au+Au collisions
follow pQCD as in p-p collisions, with parton distributions and 
fragmentation functions that scale with $x_T$, at least within
the experimental sensitivity of the data. The fact that the fragmentation functions scale for $\pi^0$ in Au+Au central collisions indicates that the effective energy loss must scale, i.e. $S(p_T)/p_T=$ is constant, which is  consistent with the parallel spectra on Fig.~\ref{fig:pizpp200}e and the constant value of $R_{AA}$ as noted in the discussion above. 

\begin{figure}[ht]
\vspace*{0.1in}
\begin{center}
\begin{tabular}{cc}
\includegraphics[width=0.5\linewidth]{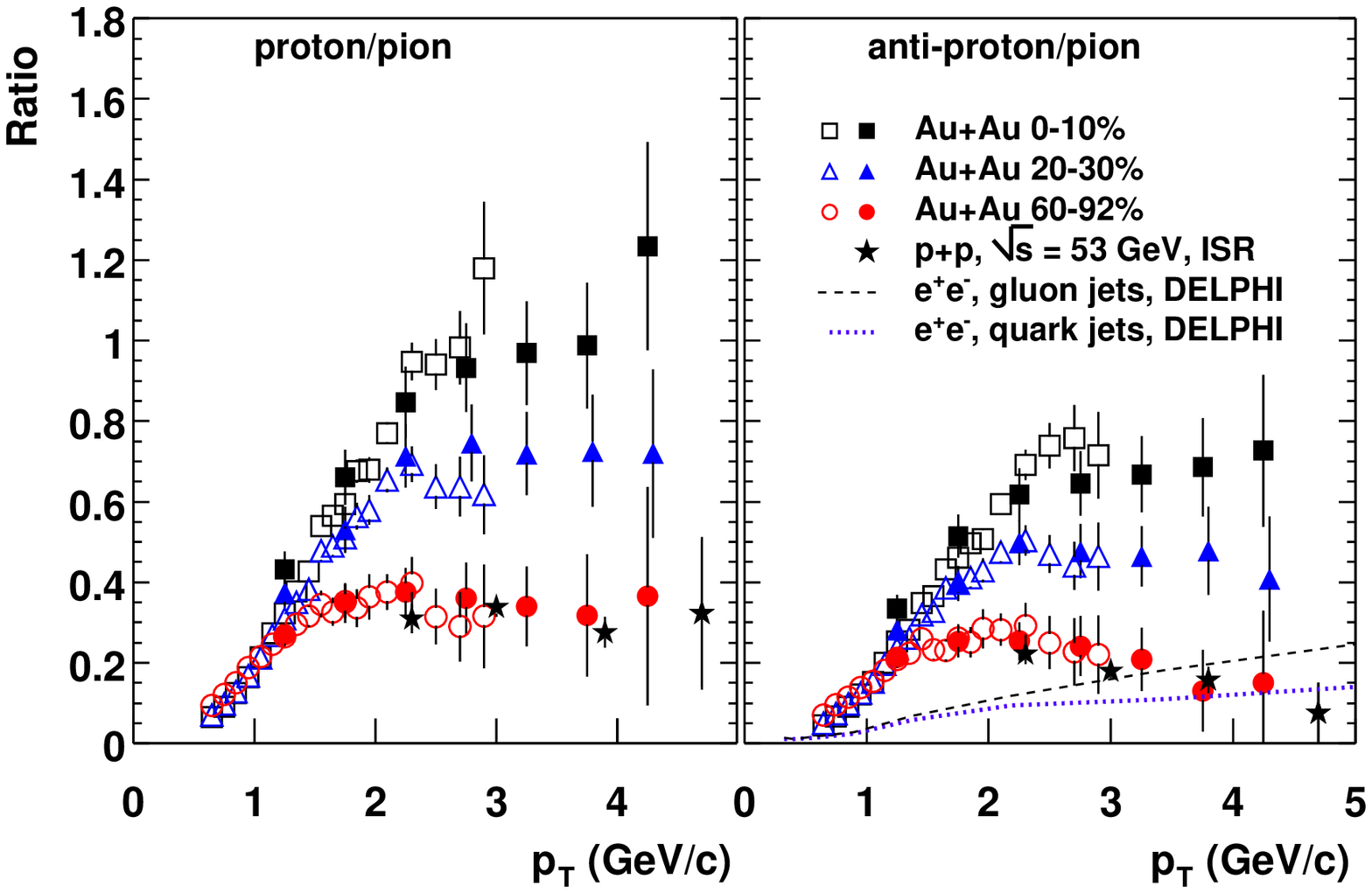} &
\includegraphics[width=0.45\linewidth]{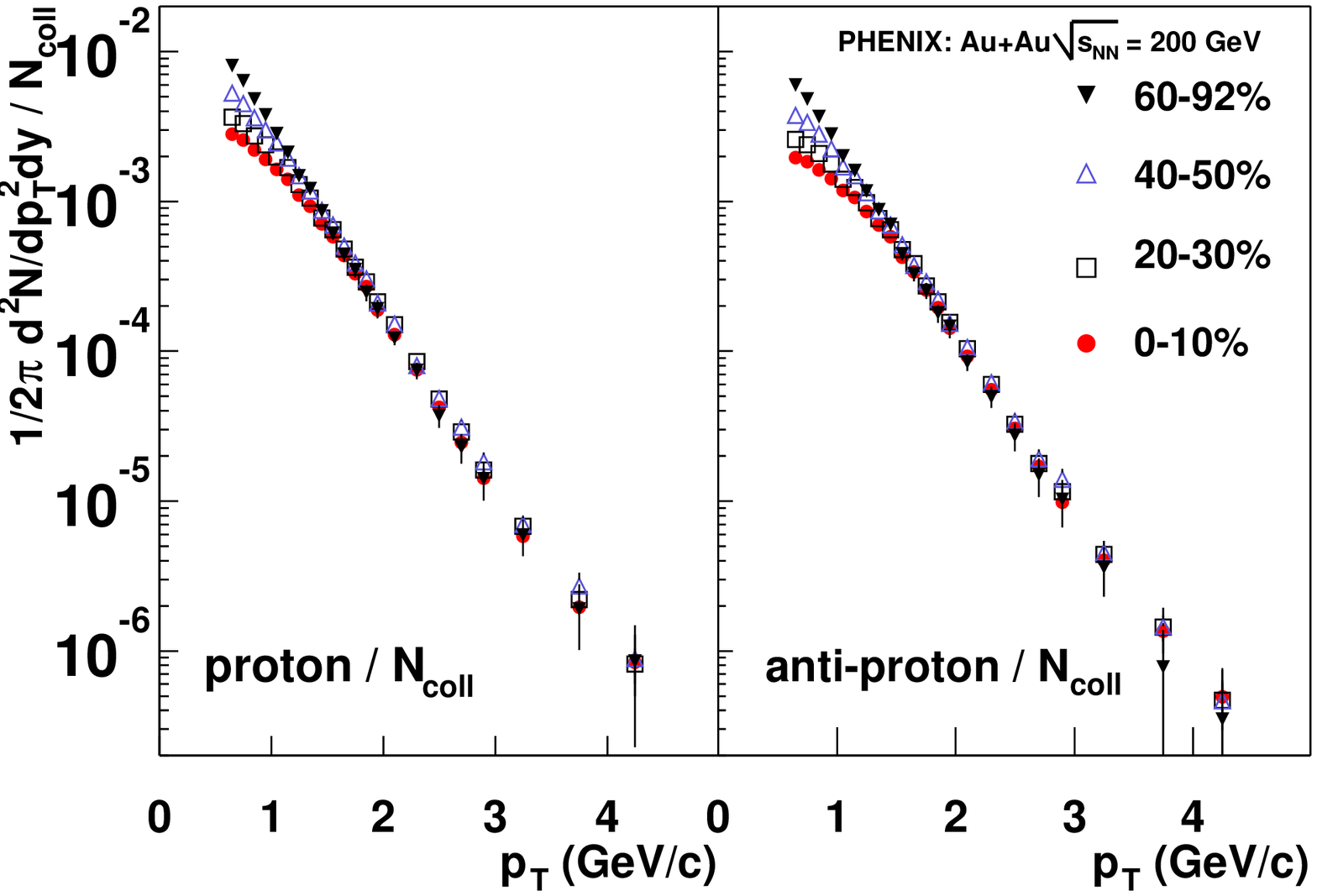} 
\end{tabular}
\end{center}
\caption[]{(left) $p/\pi$ and $\bar{p}/\pi$ ratio as a function of $p_T$ and centrality from Au+Au collisions at $\sqrt{s_{NN}}=200$ GeV~\cite{PXscalingPRL91}. Open (filled) points are for $\pi^{\pm}$ ($\pi^0$), respectively. (right) Invariant yield of $p$ and $\bar{p}$, from the same data,  as a function of centrality scaled by the number of binary-collisions ($N_{coll}$) }
\label{fig:banomaly}
\end{figure}

    The deviation of $(h^+ + h^-)/2$ from $x_T$ scaling in central Au+Au collisions is indicative of and consistent with the strong non-scaling modification of particle composition of identified charged-hadrons observed in Au+Au collisions compared to that of p-p collisions in the range $2.0\leq p_T\leq 4.5$ GeV/c, where particle production is the result of jet-fragmentation. As shown in Fig.~\ref{fig:banomaly}-(left) the $p/\pi^{+}$ and $\bar{p}/\pi^{-}$ ratios as a function of $p_T$ increase dramatically to values $\sim$1 as a function of centrality in Au+Au collisions at RHIC~\cite{PXscalingPRL91} which was totally unexpected and is still not fully understood.  
Interestingly, the $p$ and $\bar{p}$ in this $p_T$ range appear to follow the $N_{coll}$ scaling expected for point-like processes (Fig~\ref{fig:banomaly}-(right)), while the $\pi^0$ are suppressed, yet this effect is called the `baryon anomaly', possibly because of the non-$x_T$ scaling. An elegant explanation of this effect as due to coalescence of quarks from a thermal distribution~\cite{Greco,Fries,Hwa}, which would be prima facie evidence of a Quark Gluon Plasma, is not in agreement with the jet correlations observed in both same and away-side particles associated with both meson and baryon triggers~\cite{PXPRC71} (see discussion of Fig.~\ref{fig:Sickles} below). 

\subsection{Direct photon production}
   Direct photon production is one of the best reactions to study QCD in hadron collisions, since there is direct and unbiased access to one of the interacting constituents, the photon, which can be measured to high precision, and production is predominantly via a single subprocess ~\cite{QCDcompton}: 
\begin{equation}
g + q\rightarrow \gamma + q \;\;\;\;\; ,
\label{eq:QCDcompton}
\end{equation}
with $q + \bar q\rightarrow \gamma + g$ contributing on the order of 10\%. 
   However, the measurement is difficult experimentally due to the huge  background of photons from $\pi^0\rightarrow \gamma+\gamma$ and $\eta\rightarrow \gamma+\gamma$ decays. This background can be calculated using Eq.~\ref{eq:gampi} and can be further reduced by `tagging'---eliminating direct-photon candidates which reconstruct to the invariant mass of a $\pi^0$ when combined with other photons in the detector, and/or by an isolation cut---e.g. requirement of less than 10\% additional energy within a cone of radius $\Delta r=\sqrt{(\Delta\eta)^2 + (\Delta\phi)^2}=0.5$ around the candidate photon direction---since the direct photons emerge from the constituent reaction with no associated fragments. 
   
\begin{figure}[th]
 \vspace*{-0.12in}
\begin{center}
\begin{tabular}{cc}
\hspace*{-0.1in}\includegraphics[width=0.5\linewidth]{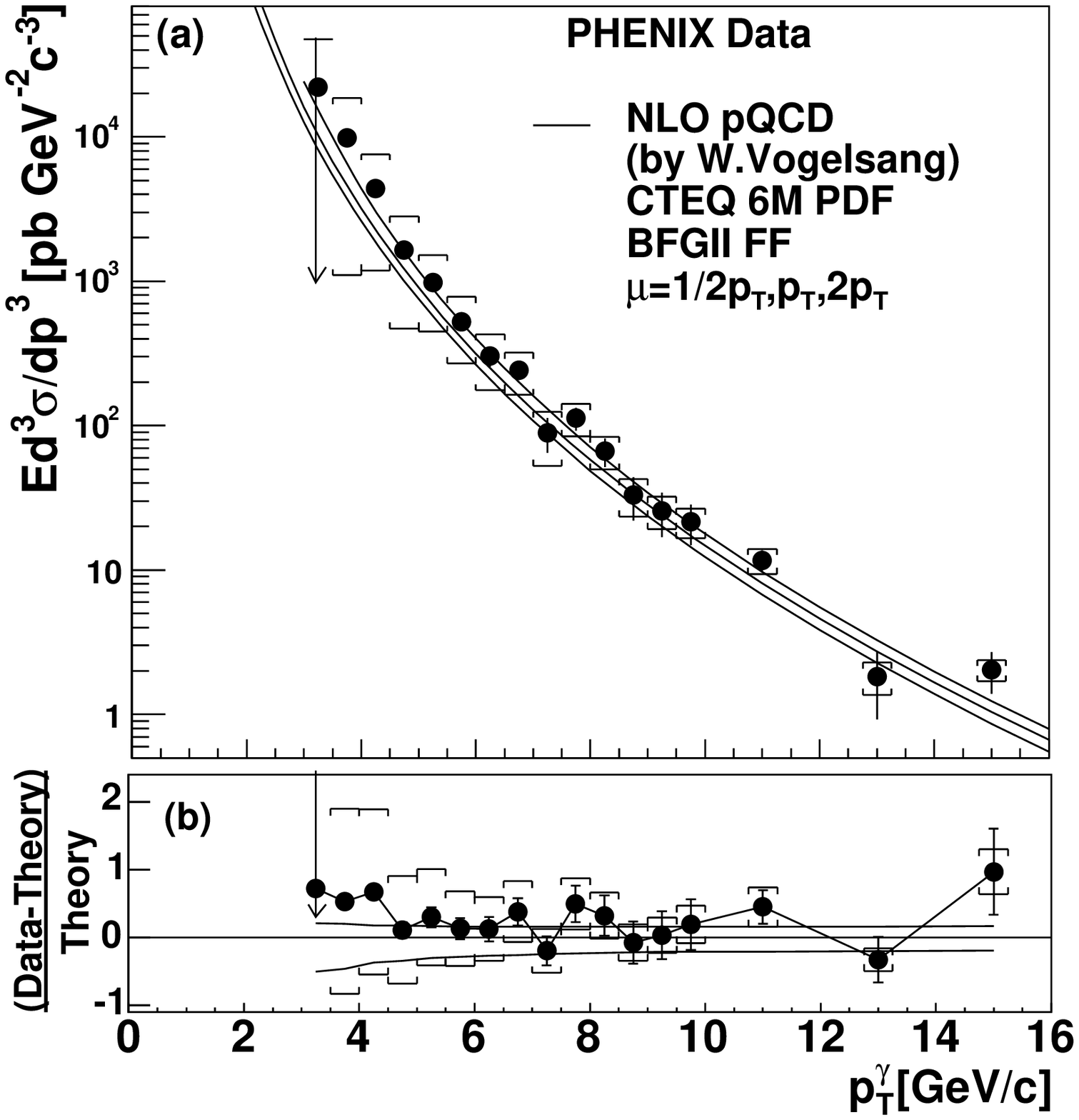} &
\hspace*{-0.2in}\includegraphics[width=0.5\linewidth]{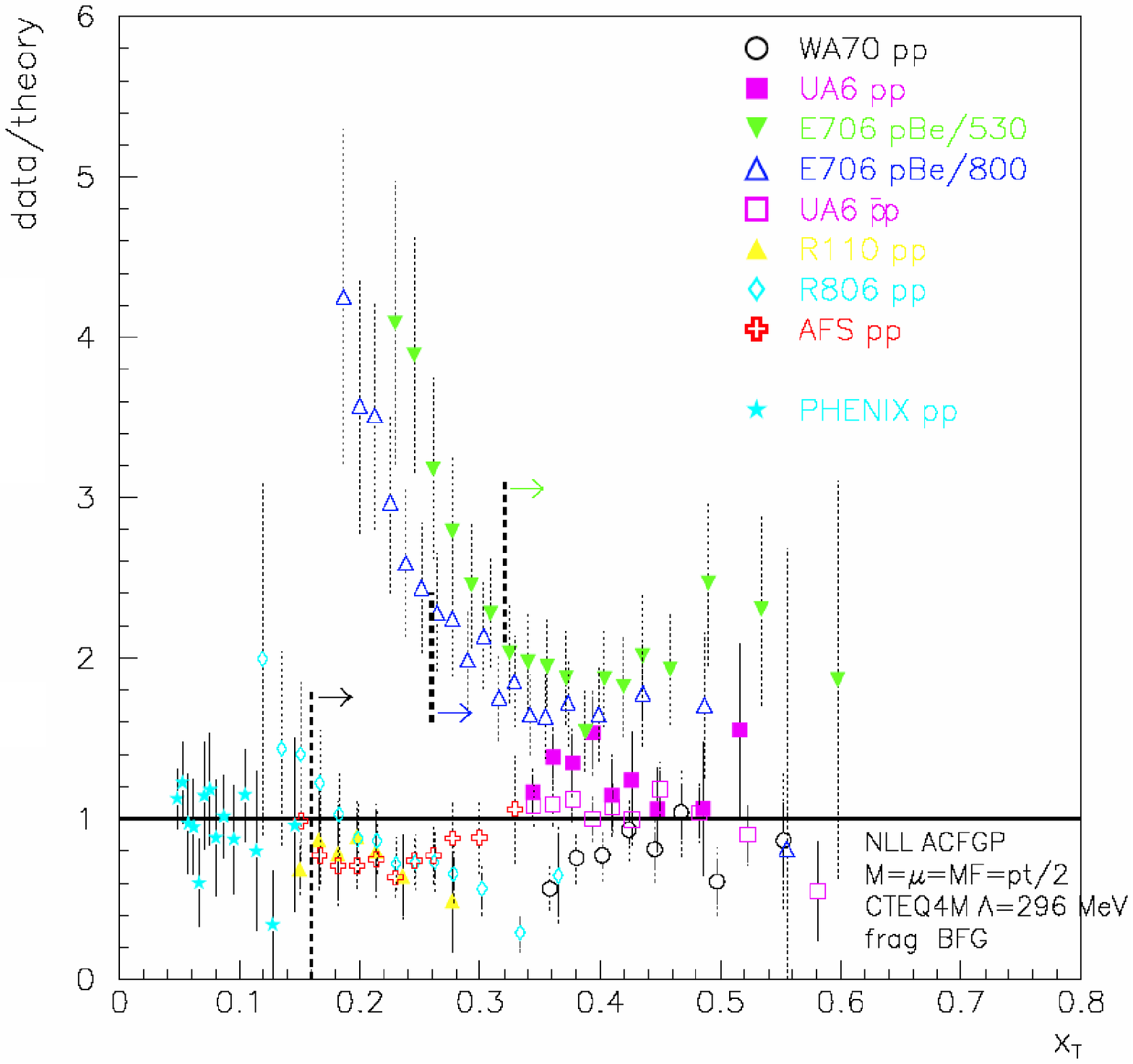} 
\end{tabular}
\end{center}
\vspace*{-0.12in}
\caption[]{(left) (a) Direct photon spectra~\cite{ppg060} with NLO pQCD calculations
for three theory scales, $\mu$. Brackets around data
points show systematic errors. (b) Comparison to the NLO
pQCD calculation for $\mu=p_T$ , with upper and lower curves
for $\mu=p_T/2$ and $2p_T$. (right) Ratio of direct photon measurements to theoretical calculation~\cite{Werlen,Aurenche99}}
\label{fig:dirphot}
\end{figure}

    The exquisite segmentation of the PHENIX Electromagnetic calorimeter ($\Delta\eta \times \Delta\phi\sim 0.01\times 0.01$) required in order to operate in the high multiplicity environment of RHI collisions also provides excellent $\gamma$ and $\pi^0$ separation out to $p_T\sim 25$ GeV/c. This will be useful in making spin-asymmetry measurements of direct photons in polarized p-p collisions for determination of the gluon spin structure function~\cite{PSU},  but, in the meanwhile, has provided a new direct photon measurement in p-p collisions which clarifies a longstanding puzzle between theory and experiment in this difficult measurement. 
In Fig.~\ref{fig:dirphot}-(left) the new measurement of the direct photon cross section in p-p collisions at $\sqrt{s}=200$ GeV from PHENIX~\cite{ppg060} is shown compared to a NLO pQCD calculation, with excellent agreement for $p_T > 3$ GeV/c. This data has resolved a longstanding discrepancy in extracting the gluon structure function from previous direct photon data~\cite{Werlen,Aurenche99} (see Fig.~\ref{fig:dirphot}-(right)) by its agreement with ISR data and the theory at low $x_T$.  
 
 \subsection{$x_T$-scaling in direct photon, jet and identified proton production in p-p collisions}
\begin{figure}[ht]
\vspace*{-0.12in}
\begin{center}
\begin{tabular}{cc}
\includegraphics[width=0.5\linewidth]{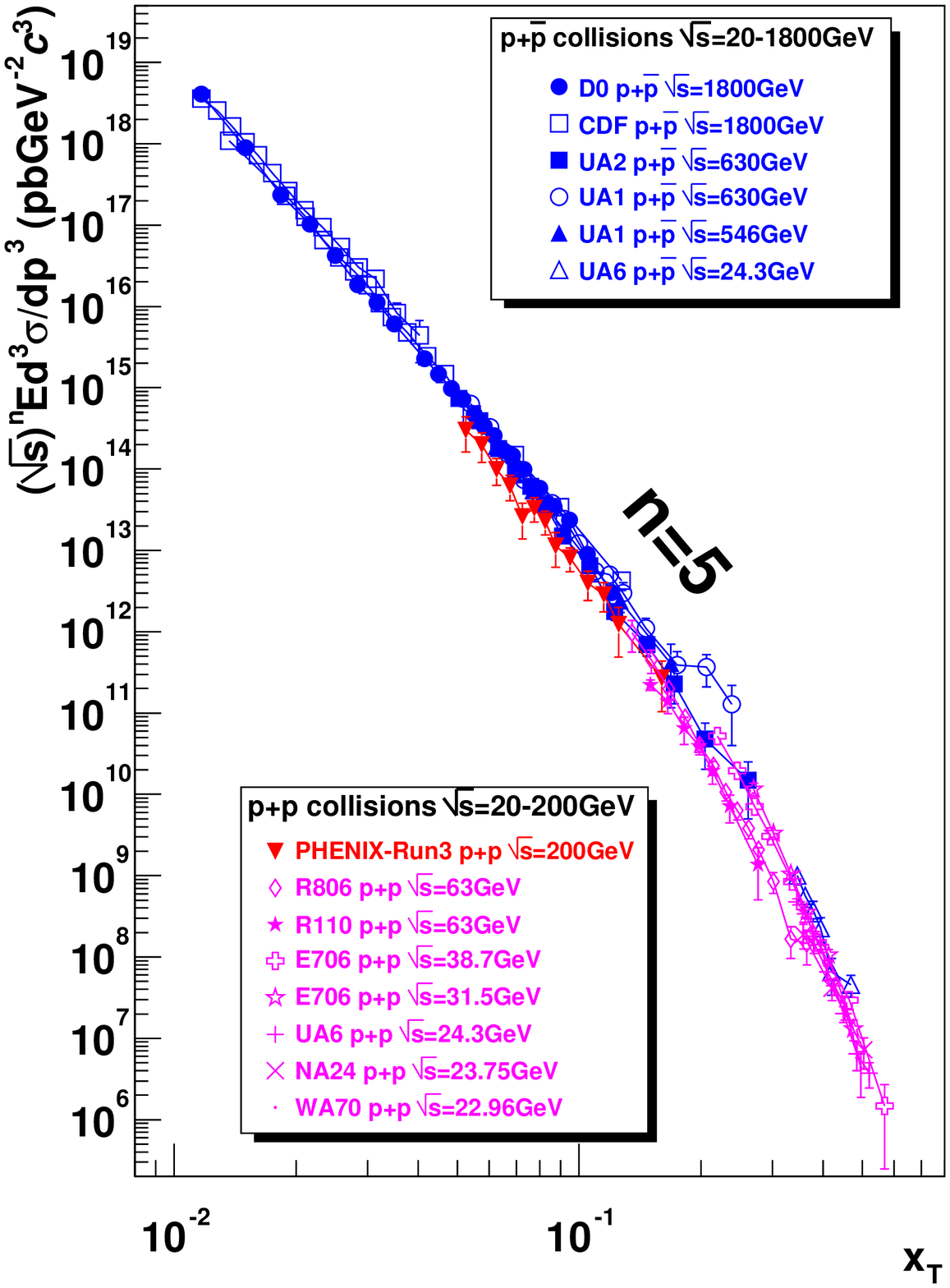} &
\includegraphics[width=0.5\linewidth]{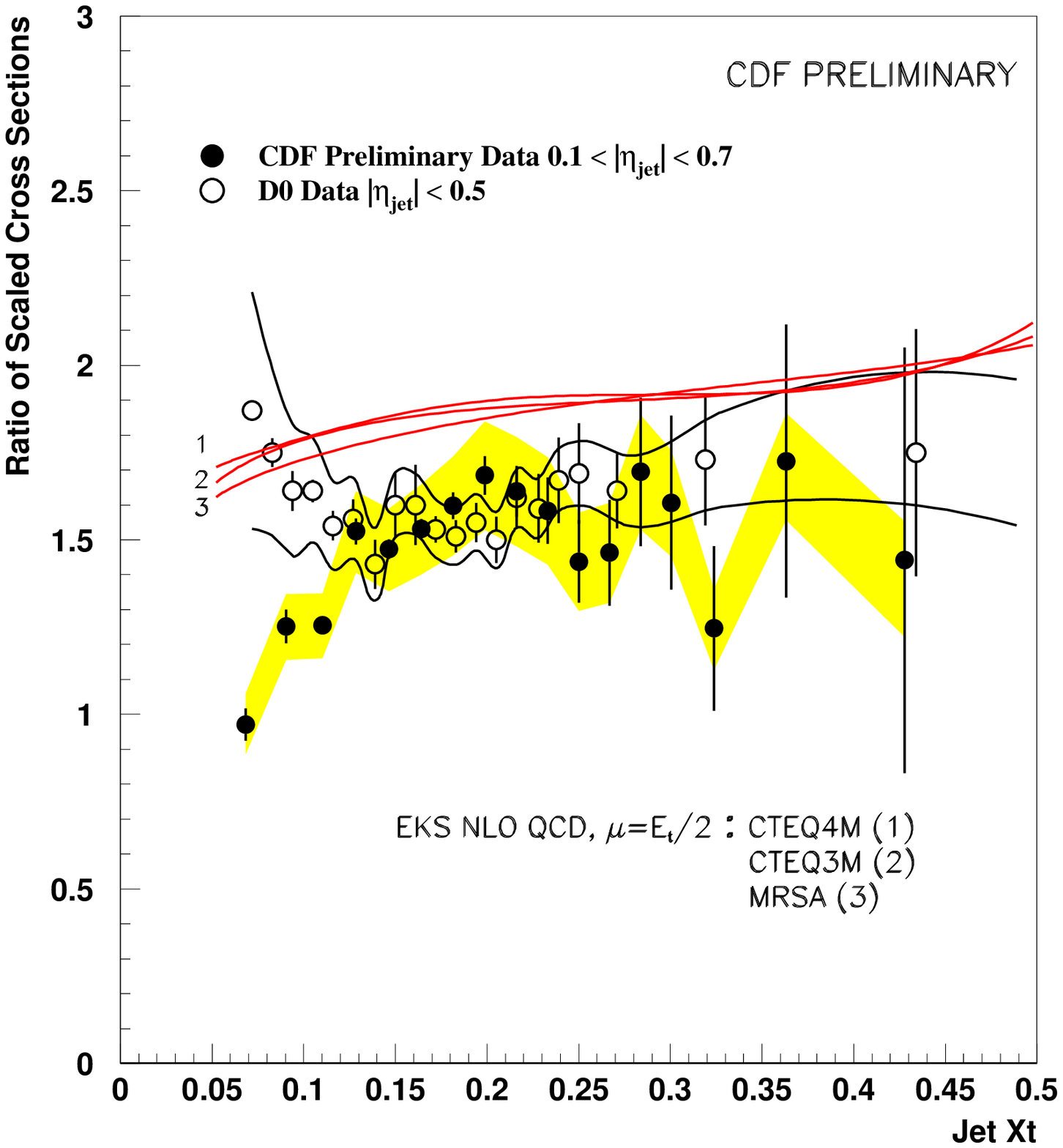} 
\end{tabular}
\end{center}
\vspace*{-0.12in}
\caption[]{(left) $x_T$ scaling~\cite{ppg060} of direct photon data in p-p and p-$\bar{\rm p}$ collisions. The quantity plotted is $(\sqrt{s})^n \times Ed^3\sigma/dp^3 (x_T)$ with $n=5.0$. (right) $x_T$ scaling of jet cross sections measured in p-$\bar{\rm p}$ collisions by CDF and D0~\cite{Blazey}. The quantity plotted is the ratio of $p_T^4$ times the invariant cross section as a function of $x_T$ for $\sqrt{s}=630$ and 1800 GeV. Note that the theory curves are plotted in the same way in order to avoid as much as possible uncertainties from the various parton distribution functions used. }
\label{fig:xTgamjet}
\end{figure}

    The new direct photon measurement also shows nice $x_T$ scaling with previous measurements (Fig.~\ref{fig:xTgamjet}-(left)) with a value $n(x_T, \sqrt{s})=5.0$. This is closer to the asymptotic value of $n(x_T, \sqrt{s})=4$ than the $\pi^0$ measurements in this range of $p_T, \sqrt{s}$ but is still not as close as the $n(x_T, \sqrt{s})=4.5$ from jet measurements~\cite{Blazey} at the Tevatron (Fig.~\ref{fig:xTgamjet}-(left)).  

\begin{figure}[ht]
\begin{center}
\includegraphics[width=0.95\linewidth]{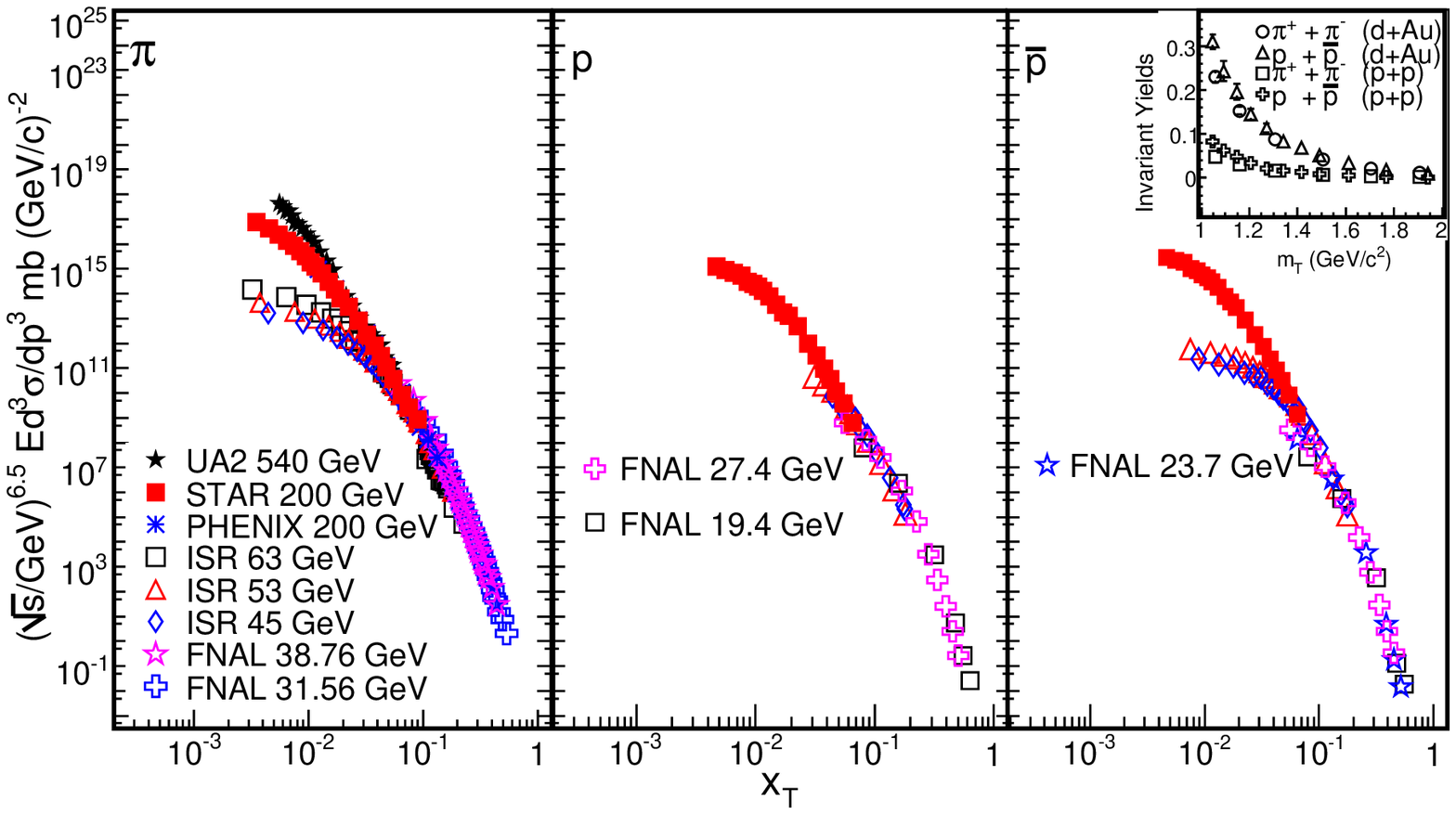} 
\end{center}
\caption[]{$x_T$-scaling of pions, protons and anti-protons as presented by STAR~\cite{STARPLB637} in comparison to previous measurements. The inset shows the $m_T$-scaling of the invariant yields of charged pions and protons+ anti-protons in p-p and d+Au collisions.}
\label{fig:STARxT}
\end{figure}

	A new measurement by STAR~\cite{STARPLB637} of $x_T$ scaling of identified $\pi^{\pm}$, $p$ and $\bar{p}$ in p-p collisions at $\sqrt{s}=200$ GeV in comparison to previous measurements (Fig.~\ref{fig:STARxT}) 
	gives $n(x_T,\sqrt{s})=6.8\pm 0.5$ for pions in agreement with the PHENIX measurement~\cite{Adler:2003au} and provides the first measurement of $x_T$ scaling of $p$ and $\bar{p}$ in p-p collisions, with $n(x_T,\sqrt{s})=6.5\pm 1.0$ in agreement with the value for pions. This result shows that $p$ and $\bar{p}$ are produced by fragmentation of hard-scattered partons in p-p collisions for $p_T \geq 2$ GeV/c, which contradicts a recent proposal~\cite{BPR06} to explain the `baryon anomaly' in A+A collisions as due to the possibility that protons and pions in the range $2.0\leq p_T\leq 4.5$ GeV/c in p-p collisions are produced by different mechanisms. As shown on the inset in Fig.~\ref{fig:STARxT}, the pion and proton spectra
follow transverse mass scaling for $m_T < 2$ GeV/c$^2$ in both p-p and d+Au collisions, suggesting the transition region from soft to hard process domination occurs at $p_T \sim 2$ GeV/c in these collision systems.
\subsection{The state of jet-suppression measurements in Au+Au and d+Au collisions at RHIC}
	  \begin{figure}[h]
\begin{center}
\includegraphics[width=0.9\linewidth]{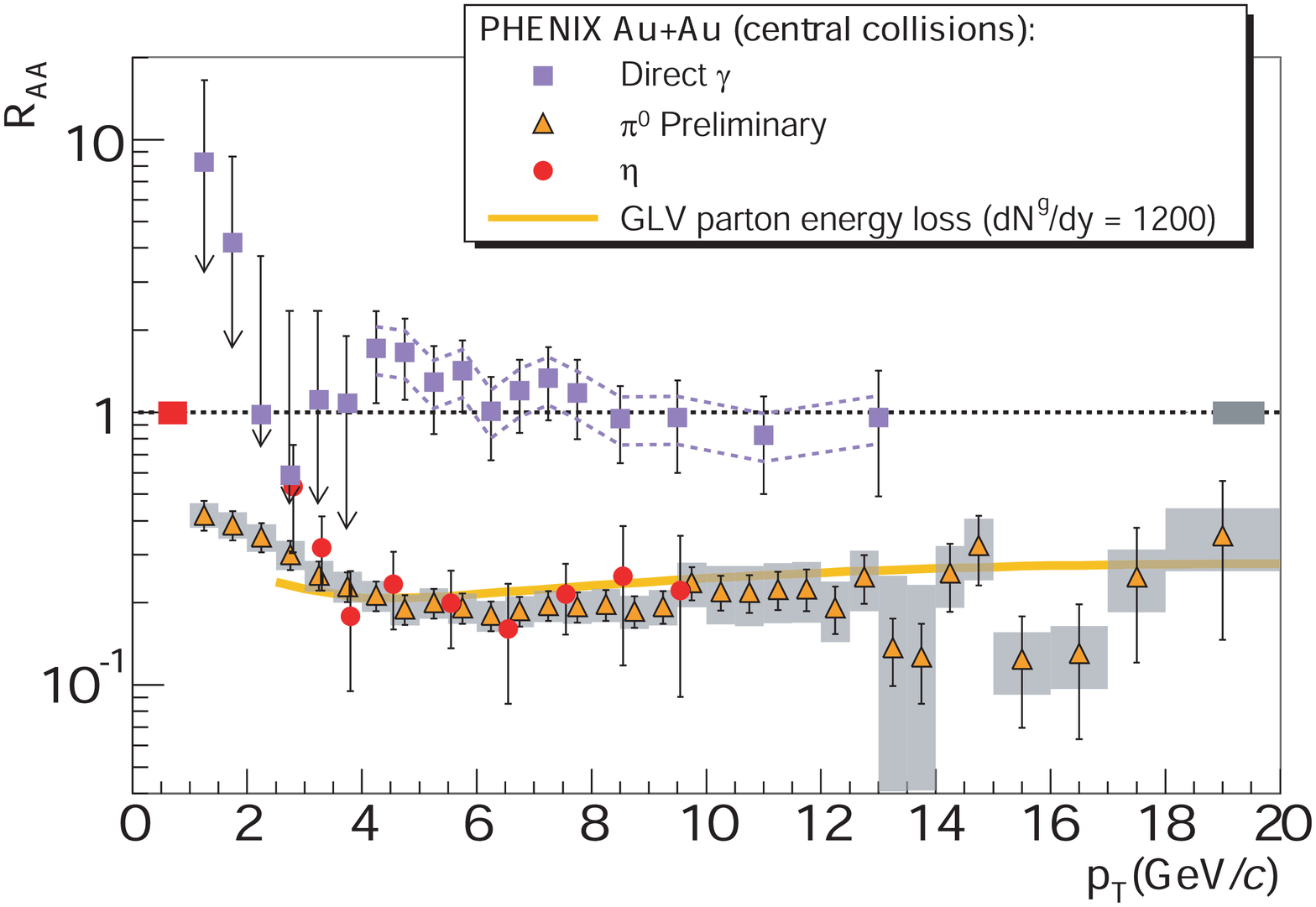} 
\end{center}
\caption[]{Nuclear modification factor, $R_{AA}$ for direct photons, $\pi^0$ and $\eta$ in Au+Au central collisions at $\sqrt{s_{NN}}=200$ GeV~\cite{AkibaQM05}}
\label{fig:3raa}
\end{figure}
   The state of $R_{AA}$ measurements at RHIC is beautifully summarized in Fig.~\ref{fig:3raa} where the nuclear modification factor is the same for $\pi^0$ and $\eta$ in Au+Au central collisions at $\sqrt{s_{NN}}=200$ GeV, both are suppressed relative to point-like scaled p-p data by a factor of $\sim 5$ which appears to be constant for $p_T\geq 4$ GeV/c, while the direct photons are not suppressed at all. Since the direct photons do not interact (strongly) with the medium, while the $\pi^0$ and $\eta$ are fragments of outgoing hard-scattered partons which do interact with the medium, this plot proves that the suppression is a medium effect. The curve on the plot shows a theoretical prediction~\cite{VG} from a model of parton energy loss. The model assumes an inital parton density $dN/dy = 1200$, which corresponds to an energy density of approximately 15 GeV/fm$^3$. The theory curve appears to show a reduction in suppression with increasing $p_T$, while, as noted above, the data appear to be flat to within the errors, which clearly could still be improved.  
   
\begin{figure}[ht]
\begin{center}
\includegraphics[width=0.99\linewidth]{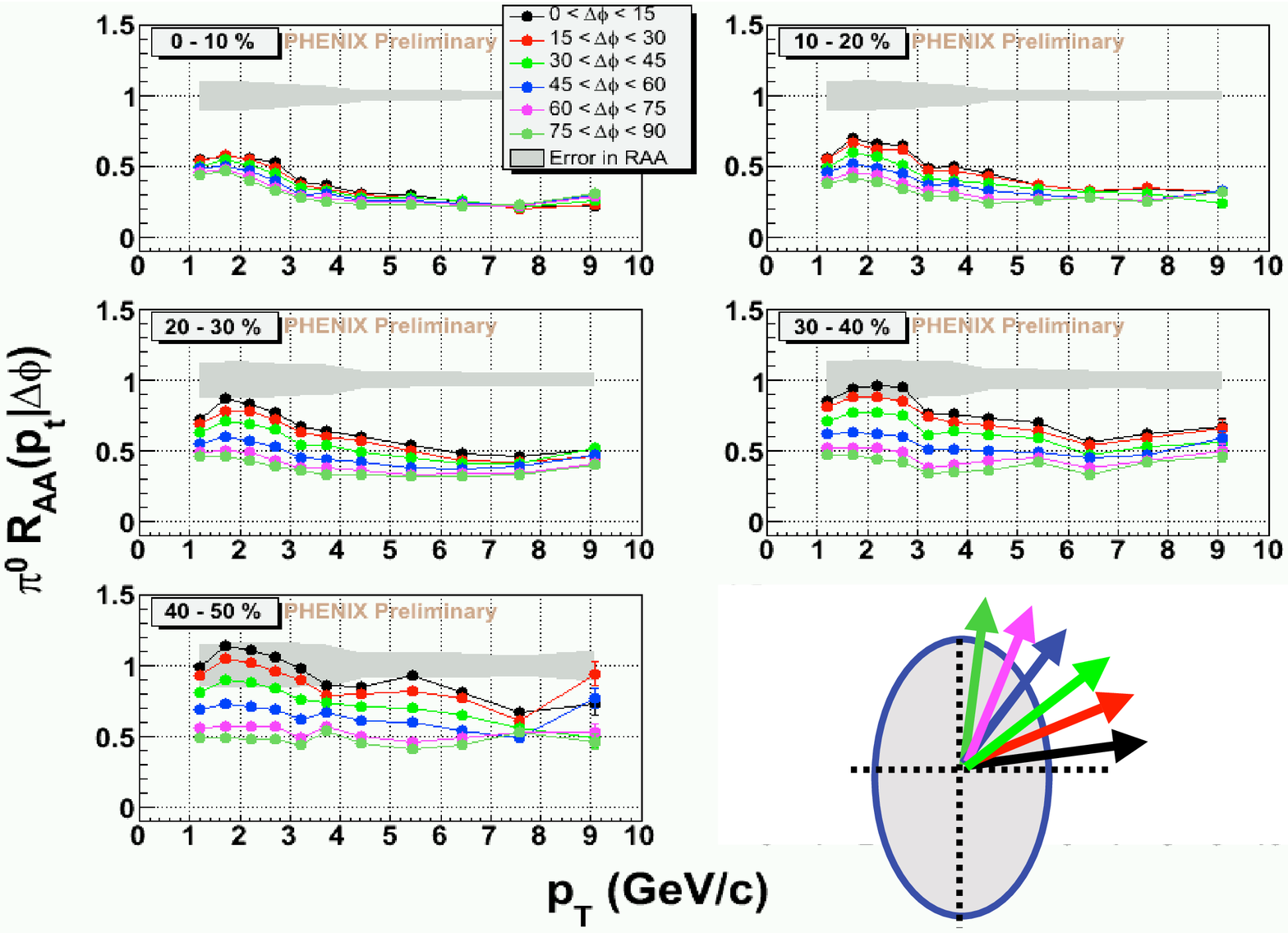} 
\end{center}
\caption[]{Nuclear modification factor, $R_{AA}(p_T)|_{\Delta\phi}$, of $\pi^0$'s as a function of $p_T$ for several values of centrality and angle to the reaction plane ($\Delta\phi$) in Au+Au collisions at $\sqrt{s_{NN}}=200$ GeV~\cite{ppg054,coleQM05}. The shaded region around $R_{AA}=1$ indicates the systematic error due to the reaction plane resolution correction. The arrows sketched on  the ellipse indicate the different path-lengths traversed for centrality $\sim 30-40$\% . }
\label{fig:cole}
\end{figure}

   It is unreasonable to believe that the properties of the medium have been determined by a theorist's line through the data which constrains a few parameters of a model. The model and the properties of the medium must be able to be verified by more detailed and differential measurements. All models of medium induced energy loss~\cite{egVG} predict a characteristic dependence of the average energy loss on the length of the medium traversed. This is folded into the theoretical calculations with added complications that the medium expands during the time of the collision, etc~\cite{egseeHP06}.  In an attempt to separate the effects of the density of the medium and the path length traversed, PHENIX~\cite{ppg054,coleQM05} has studied the dependence of the $\pi^0$ yield as a function of the angle ($\Delta\phi$) to the reaction plane in Au+Au collisions (see Fig.~\ref{fig:cole}).  For a given centrality, variation of $\Delta\phi$ gives a variation of the path-length traversed for fixed initial conditions, while varying the centrality allows the initial conditions to vary.   
Clearly these data reveal much more activity than the reaction-plane-integrated $R_{AA}$ (Fig.~\ref{fig:3raa}) and merit further study by both experimentalists and theorists. 

\begin{figure}[ht]
\begin{center}
\begin{tabular}{cc}
\includegraphics[width=0.5\linewidth]{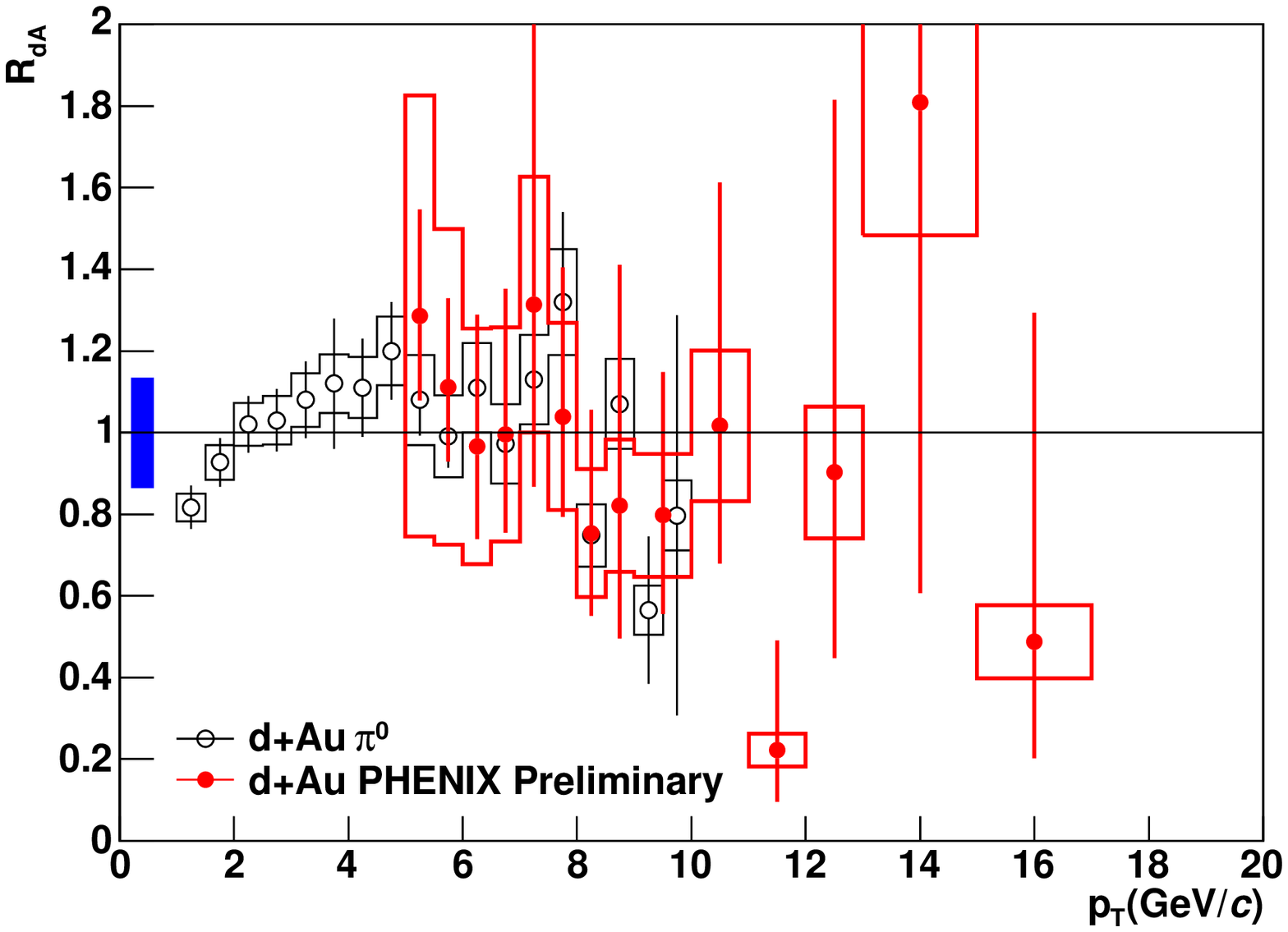} &
\hspace*{-0.25in}\includegraphics[width=0.5\linewidth]{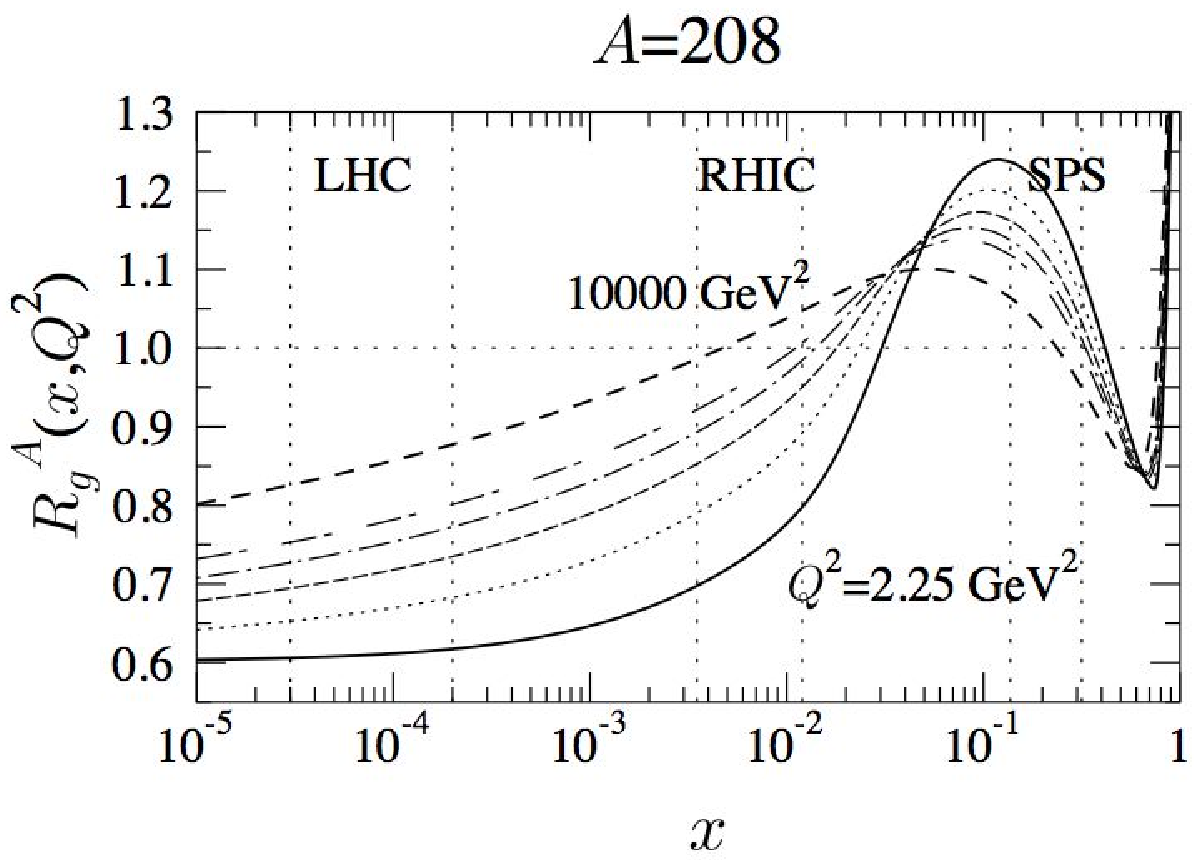} 
\end{tabular}
\end{center}
\caption[]{(left)-Nuclear modification factor $R_{dA}$ of direct photons (red solid circles) and neutral pions (open circles) measured in minimum bias d+Au collisions at $\sqrt{s_{NN}}=200$ GeV~\cite{PXdA}. (right) Theoretical prediction for gluon structure function ratio $R^A_g(x)$ in Pb~\cite{EKS,EKV}. Note that $x\approx x_T=p_T({\rm GeV/c})/100$ for comparing the left and right figures.} 
\label{fig:dirgdA}
\end{figure}

    The point-like scaling of direct photon production in Au+Au collisions indicated by the absence of suppression in Fig.~\ref{fig:3raa} implies that there also should be no suppression of direct photons in d+A collisions. The further implication is that the gluon structure function in a nucleus ($g^A(x)$) scales like $A$, i.e. $R^A_g(x)=g^A(x)/Ag^N(x)=1$, where $g^N(x)$ is the parton distribution function of gluons in a nucleon, since it is known from measurements of deeply inelastic scattering of muons in nuclei that there is only a slight effect in the quark-structure functions, $R^A_{F_2}(x)=F^A_{2}(x)/AF^N_{2}(x)\lsim 1$~\cite{MMay,EMC,NMC} (see Eq.~\ref{eq:F2}).   However, until now there has been no direct measurement of the gluon structure function in nuclei. A first attempt in this direction has been presented by the PHENIX collaboration as a measurement of $R_{dA}(p_T)$ of direct photons in d+Au collisions at $\sqrt{s_{NN}}=200$ GeV~\cite{PXdA} (Fig.~\ref{fig:dirgdA}-(left)). 

At mid-rapidity there is a simple relationship between $R_{dA}$ for direct photon production and the structure function ratios~\cite{QCDcompton}
\begin{equation}
R_{dA}={{d\sigma_{\gamma}^{dA}(p_T)/dp_T} \over {(2 \times A)\times d\sigma_{\gamma}^{pp}(p_T)}/dp_T}={1\over 2}\left({ {F^A_{2}(x_T)} \over {A F^N_{2}(x_T)}}+{ {g^{A}(x_T)} \over {A g^{N}(x_T)}}\right)={1\over 2}\left(R^A_{F_2}(x_T)+R^A_g(x_T)\right) \qquad .
\end{equation}
The measurement is consistent with $R^A_g(x)=1$, but clearly the statistical errors must be improved by an order of magnitude before the data can be compared in detail to the theoretical prediction~\cite{EKS,EKV} (Fig.~\ref{fig:dirgdA}-(right)) which is used in all calculations for RHI collisions. 
   \section{Correlations}  
          As noted above (section \ref{sec:power}), the steeply-falling power-law spectrum at a given $\sqrt{s}$ has many important and helpful consequences for single particle inclusive and two-particle correlation measurements of hard-scattering. The most famous properties in this regard are the ``Bjorken parent-child relationship''~\cite{BjPRD8} and the ``leading-particle effect'', which also goes by the unfortunate name ``trigger bias''~\cite{JacobLandshoff,BjPRD8}. 
   \subsection{Why single particle inclusive measurements accurately measure hard-scattering---the leading-particle effect, also known as ``trigger bias.''}
   \label{sec:leading}     
       Due to the steeply falling power-law transverse momentum ($\hat{p}_{T_t}$) spectrum of the scattered parton, the inclusive single particle (e.g. $\pi$) $p_{T_t}$ spectrum from jet fragmentation is dominated by fragments with large $z_t$, where $z_t=p_{T_t}/\hat{p}_{T_t}$ is the fragmentation variable. 
       The joint probability for a fragment pion, with $p_{T_t}=z_t\hat{p}_{T_t}$, originating from a parton with $\hat{p}_{T_t}=p_{T{\rm jet}}$ is:
   \begin{eqnarray}
    {{d^2\sigma_{\pi} (\hat{p}_{T_t},z_t) }\over {\hat{p}_{T_t} d\hat{p}_{T_t} dz_t }}&=&{{d\sigma_q}\over {\hat{p}_{T_t} d\hat{p}_{T_t}}}\times D^{\pi}_q (z_t) \nonumber \\[0.12in]
    &=& f_q(\hat{p}_{T_t}) \times D^{\pi}_q(z_t) 
 \qquad ,  \label{eq:mjt-zgivenq} 
   \end{eqnarray}
where $f_q(\hat{p}_{T_t})$ represents the final-state scattered-parton invariant spectrum ${{d\sigma_{q} }/{\hat{p}_{T_t} d\hat{p}_{T_t}}}$  and $D^{\pi}_q (z_t)$  
represents the fragmentation function. The first term in Eq.~\ref{eq:mjt-zgivenq} is the probability of finding a parton with transverse momentum $\hat{p}_{T_t}$ and the second term corresponds to the conditional probability that the parton fragments into a particle of momentum $p_T=z_t \hat{p}_{T_t}$. A simple change of variables, $\hat{p}_{T_t}={{p}_{T_t}/z_t}$,  ${d\hat{p}_{T_t}}/{d{p}_{T_t}}|_{z_t}= {1/z_t}$,  
then gives the joint probability of a pion with  transverse momentum $p_{T_{t}}$ which is a fragment with momentum fraction $z_t$ from a parton with $\hat{p}_{T_t}=p_{T_t}/z_t$: 
\begin{equation}
{{d^2\sigma_{\pi} (p_{T_{t}},z_t)} \over {p_{T_t} dp_{T_{t}} dz_t}} 
= f_q({{p}_{T_t} \over {z_t}}) \times D^{\pi}_q(z_t)\times {1 \over z_t^2} \qquad . 
\label{eq:mjt-zgivenpi}
\end{equation}
The $p_{T_t}$  and $z_t$ dependences do not factorize. However, the $p_{T_t}$ spectrum may be 
found by integrating over all values of $\hat{p}_{T_t}$ from $p_{T_t}\leq\hat{p}_{T_t}\leq\sqrt{s}/2$, 
which corresponds to values of $z_t$ from $x_T=2 p_T/\sqrt{s}$ to 1:  
\begin{equation}
\label{eq:dsigma_integral}
{1\over p_{T_t}}{d\sigma_{\pi}\over dp_{T_t}}  =  
 \int_{x_T}^{1} f_{q}({p_{T_t}\over z_t})\, D^{\pi}_q(z_t)\; {dz_t\over {z_t}^{2}}\qquad . 
\end{equation} 
Also, for any fixed value of $p_{T_t}$  one can 
evaluate the $\mean{z_t(p_{T_t})}$, integrated over the parton spectrum: 
\begin{equation} \label{meanz_inc}
\mean{z_t(p_{T_t})}= {{\int_{x_T}^{1}z_t\;D^{\pi}_q(z_t)\; f_{q}(p_{T_t}/z_t) {dz_t\over z_t^{2}}} \over
{\int_{x_T}^{1} D^{\pi}_q(z_t)\; f_{q}(p_{T_t}/z_t) {dz_t\over z_t^{2}}}}\qquad .
\end{equation}

     Since the observed $\pi^0$ spectrum is a power-law for $p_{T_t}\geq 3$ GeV/c, one can deduce from Eq.~\ref{eq:dsigma_integral} that the partonic $\hat{p}_{T_t}$ spectrum is also a power-law with the same power---this is the `Bjorken parent-child relationship''~\cite{BjPRD8}. If we take:
     \begin{equation}
      {{d\sigma_{q} }\over{\hat{p}_{T_t} d\hat{p}_{T_t}}}=f_{q}(\hat{p}_{T_t})=A \hat{p}_{T_t}^{-n} \qquad, 
      \label{eq:parton_power}
      \end{equation}
            then  
     \begin{eqnarray}
\label{eq:power_law}
{1\over p_{T_t}}{d\sigma_{\pi}\over dp_{T_t}} & = &
\int_{x_T}^{1} A\, D^{\pi}_q(z_t)\, ({p_{T_t}\over z_t})^{-n} {dz_t\over z_t^{2}} \nonumber\\
& = & {1\over {p_{T_t}^{n}}} \int_{x_T}^{1} A\, D^{\pi}_q(z_t)\, {z_t}^{n-2} dz_t \qquad ,
\end{eqnarray}
where the last integral depends only weakly on $p_{T_t}$ due to the small value of $x_T$. Eq.~\ref{eq:power_law} also indicates that the effective fragmentation function for a detected inclusive single particle (with $p_{T_t}$) is weighted upward in $z_t$ by a factor $z_t^{n-2}$, where $n$ is the simple power fall-off of the jet invariant cross section (i.e. not the $n(x_T, \sqrt{s})$ of Eq.~\ref{eq:bbg}). This is the so called ``trigger bias'' although it doesn't actually involve a hardware trigger. Any particle selected from an inclusive $p_{T_t}$ spectrum will most likely carry a large fraction of its parent parton transverse momentum; and it was commonly accepted that this would define the hard scattering kinematics ($\hat{s}$ in Eq.~\ref{eq:cpT} or $\hat{p}_{T_t}$ in Eq.~\ref{eq:mjt-zgivenq}) so that the jet from the other outgoing parton in the hard-scattered parton-pair would be unbiased~\cite{DarriulatNPB107,FFF}, so that its properties such as the fragmentation function and the fragmentation transverse momentum could be measured.  

 \subsubsection{Fragmentation Formalism---Single Inclusive}
 \label{sec:frag-single}
     For an exponential fragmentation function, 
       \begin{equation}
      D(z)=B e^{-bz} \qquad, 
   \label{eq:ff1}
   \end{equation}
calculation of the ``trigger bias'' and the ``parent-child'' factor is straightforward~\cite{JacobLandshoff}. 
The mean multiplicity of fragments in the jet is:
\begin{equation}
\mean{m}=\int_0^1 D(z) dz= {B\over b} (1-e^{-b}) 
\label{eq:ff1-mm}
\end{equation}
and these fragments carry the total momentum of the jet: 
\begin{equation}
\int_0^1 z D(z) dz={B \over b^2} (1-e^{-b}(1+b))\equiv 1 \qquad , 
\label{eq:ff1-total}
\end{equation}
where the $\mean{z}$ per fragment is:
\begin{equation}
\mean{z}={{\int_0^1 z D(z) dz}\over {\int_0^1 D(z) dz}} ={1\over \mean{m}} \qquad . 
\label{eq:ff1-mz}
\end{equation}
The results are:
\begin{equation}
B={{b^2} \over {1-e^{-b}(1+b)}}\approx b^2
\label{eq:ff-B}
\end{equation}
\begin{equation}
\mean{m}={{b (1-e^{-b})}\over {1-e^{-b}(1+b)}}\approx b   \qquad ,
\label{eq:ff-mm}
\end{equation}
\begin{equation}
\mean{z}={ {1-e^{-b}(1+b)} \over {b (1-e^{-b})}}\approx {1\over b} \qquad.
\label{eq:ff-mz}
\end{equation}    
The mean multiplicity of charged particles in the jet is $\mean{m}\approx b$, which is 8--10 at RHIC (see below). 

    Substitution of Eq.~\ref{eq:ff1} into Eq.~\ref{eq:power_law} for the $p_{T_t}$ spectrum of the $\pi$ gives:  
    \begin{equation}\label{eq:int_mjt_sig_inclus}
{1\over p_{T_t}}{d\sigma_\pi \over d p_{T_t}} =  {AB\over p_{T_t}^{n}}\int_{x_{T_t}}^{1} dz_t z_t^{n-2} \exp -bz_t \qquad ,
\end{equation}
which can be written as:
\begin{equation}
{1\over p_{T_t}}{d\sigma_\pi \over dp_{T_t}} =  {AB\over p_{T_t}^{n}} 
{1 \over {b^{n-1}}}  \left[ \Gamma({n-1},b x_{T_t}) - \Gamma (n-1, b) \right] \qquad, 
\label{eq:ans1_int_mjt_sig_inclus}
\end{equation}
where 
\begin{equation}
\Gamma(a,x)\equiv \int_{x}^\infty t^{a-1} \, e^{-t} \, dt   \qquad 
\label{eq:inc_gamma}
\end{equation}
is the Complementary or upper Incomplete Gamma function, and   
$\Gamma(a,0)=\Gamma(a)$ is the Gamma function, where $\Gamma(a)=(a-1)!$ for $a$ an integer. 

	A reasonable approximation for small $x_T$ values is obtained by taking the lower limit of Eq.~\ref{eq:int_mjt_sig_inclus} to zero and the upper limit to infinity, with the result that:
 \begin{equation}
{1\over p_{T_t}}{d\sigma_\pi \over dp_{T_t}} \approx {\Gamma(n-1) \over b^{n-1}}  {AB\over p_{T_t}^{n}} \qquad.
 \label{eq:result2_int_mjt_sig_inclus}
\end{equation}
The parent-child ratio, the ratio of the number of $\pi$ at a given $p_{T_t}$ to the number of partons at the same $p_{T_t}$ is just given by the ratio of Eq.~\ref{eq:result2_int_mjt_sig_inclus} to Eq.~\ref{eq:parton_power} at $\hat{p}_{T_t}=p_{T_t}$:
\begin{equation}
\left . {\pi^0 \over q }\right|_{\pi^0} (p_{T_t})={B\,\Gamma(n-1) \over b^{n-1}}  \approx {\mean{m} \Gamma(n-1) \over b^{n-2}} \qquad . 
\label{eq:pioverjet}
\end{equation}
Similarly, the same substitutions in Eq.~\ref{meanz_inc} for $\mean{z_t(p_{T_t})}$ give:
      \begin{equation}
\mean{z_t(p_{T_t})}= {{\int_{x_{T_t}}^{1} dz_t z_t^{n-1} \exp -bz_t}  \over {\int_{x_{T_t}}^{1} dz_t z_t^{n-2} \exp -bz_t }}= {1\over b} {  { \left[ \Gamma({n},b x_{T_t}) - \Gamma (n, b) \right]}\over { \left[ \Gamma({n-1},b x_{T_t}) - \Gamma (n-1, b) \right]}} \approx {{n-1} \over b}\qquad .
\label{eq:ans1_meanz_inc}
\end{equation}

	This shows the ``trigger-bias'' quantitatively. The $\mean{z_t(p_{T_t})}$ of an inclusive  single particle (e.g $\pi^0$) with transverse momentum $p_{T_t}$,  which is a fragment with momentum fraction $z_t$ from a parent parton with $\hat{p}_{T_t}=p_{T_t}/z_t$ (Eq~\ref{eq:ans1_meanz_inc}), is $n-1$ times larger than the unconditional $\mean{z}$ of fragmentation (Eq.~\ref{eq:ff-mz})~\cite{noteonn}.   The prevailing opinion from the early 1970's until early this year was that although the inclusive single particle (e.g. pizero) spectrum from jet fragmentation is dominated by trigger fragments with large $\mean{z_t}\sim 0.7-0.8$ the away-jets should be unbiased and would measure the fragmentation function, once the correction is made for $\mean{z_t}$ and the fact that the jets don't exactly balance $p_T$ due to the $k_T$ smearing effect~\cite{egsee1}.    

\subsection{Almost everything you want to know about jets can be found using 2-particle correlations.} \label{sec:almost}

\begin{figure}[ht]
\begin{center}
\includegraphics[width=0.50\linewidth]{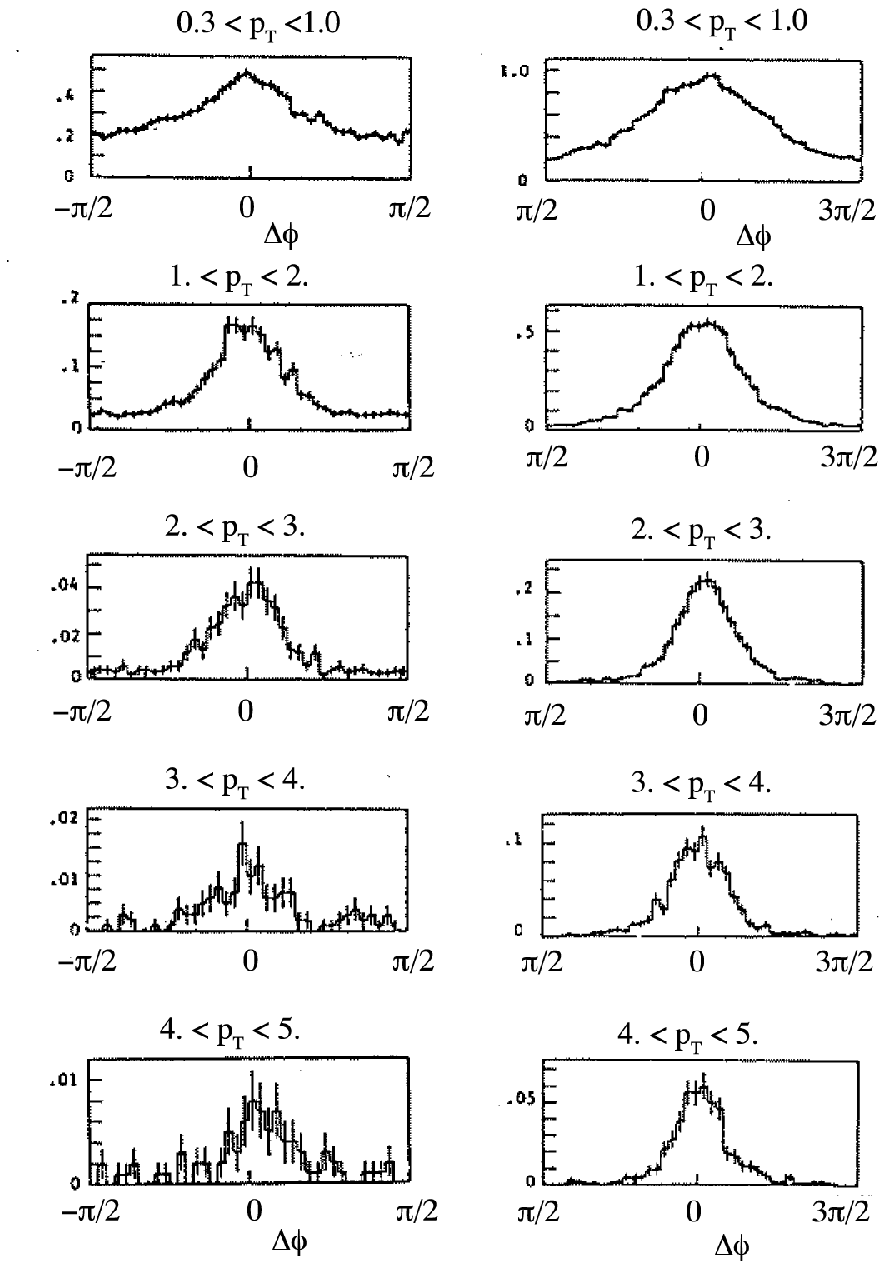} 
\includegraphics[width=0.48\linewidth]{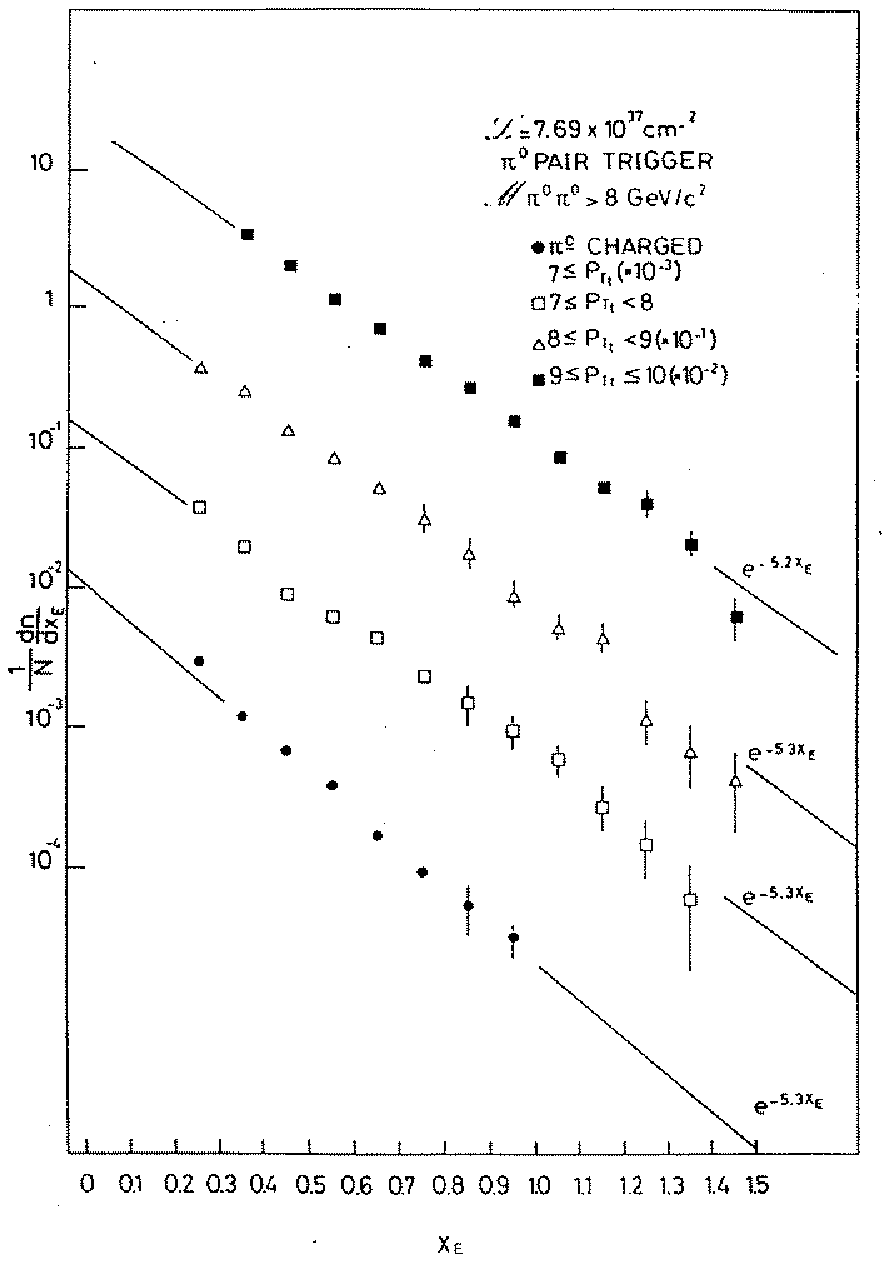}
\end{center}
\vspace*{-0.12in}
\caption[]
{a,b) Azimuthal distributions of charged particles of transverse momentum $p_T$, with respect to a trigger $\pi^0$ with $p_{Tt}\geq 7$ GeV/c, for 5 intervals of $p_T$: a) (left-most panel) for $\Delta\phi=\pm \pi/2$ rad about the trigger particle, and b) (middle panel) for $\Delta\phi=\pm \pi/2$ about $\pi$ radians (i.e. directly opposite in azimuth) to the trigger. The trigger particle is restricted to $|\eta|<0.4$, while the associated charged particles are in the range $|\eta|\leq 0.7$. c) (right panel) $x_E$ distributions (see text) corresponding to the data of the center panel.   
\label{fig:mjt-ccorazi} }
\end{figure}

\begin{figure}[ht]
\begin{center}
\begin{tabular}{cc}
\includegraphics[width=0.47\linewidth]{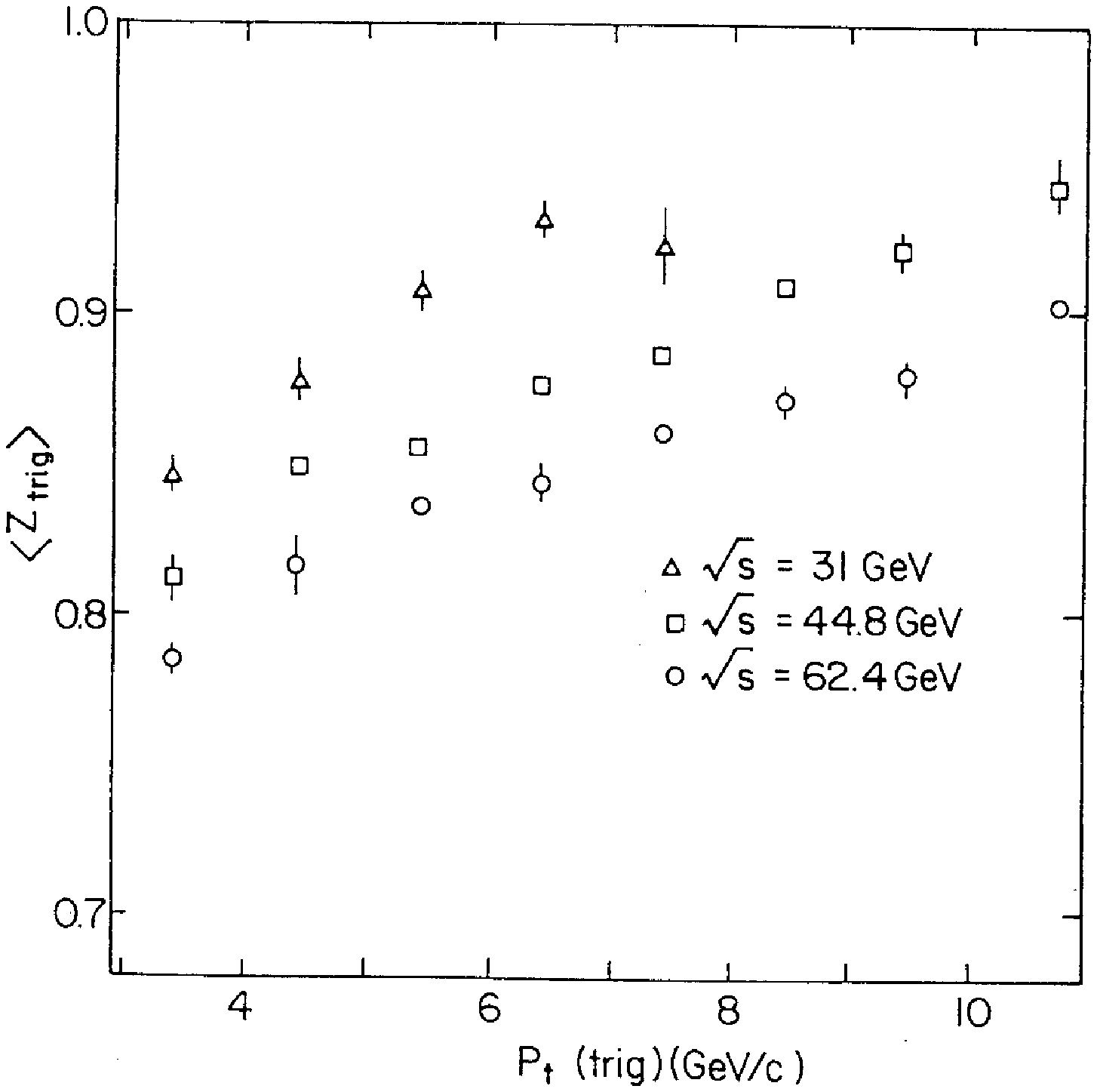} &
\includegraphics[width=0.45\linewidth]{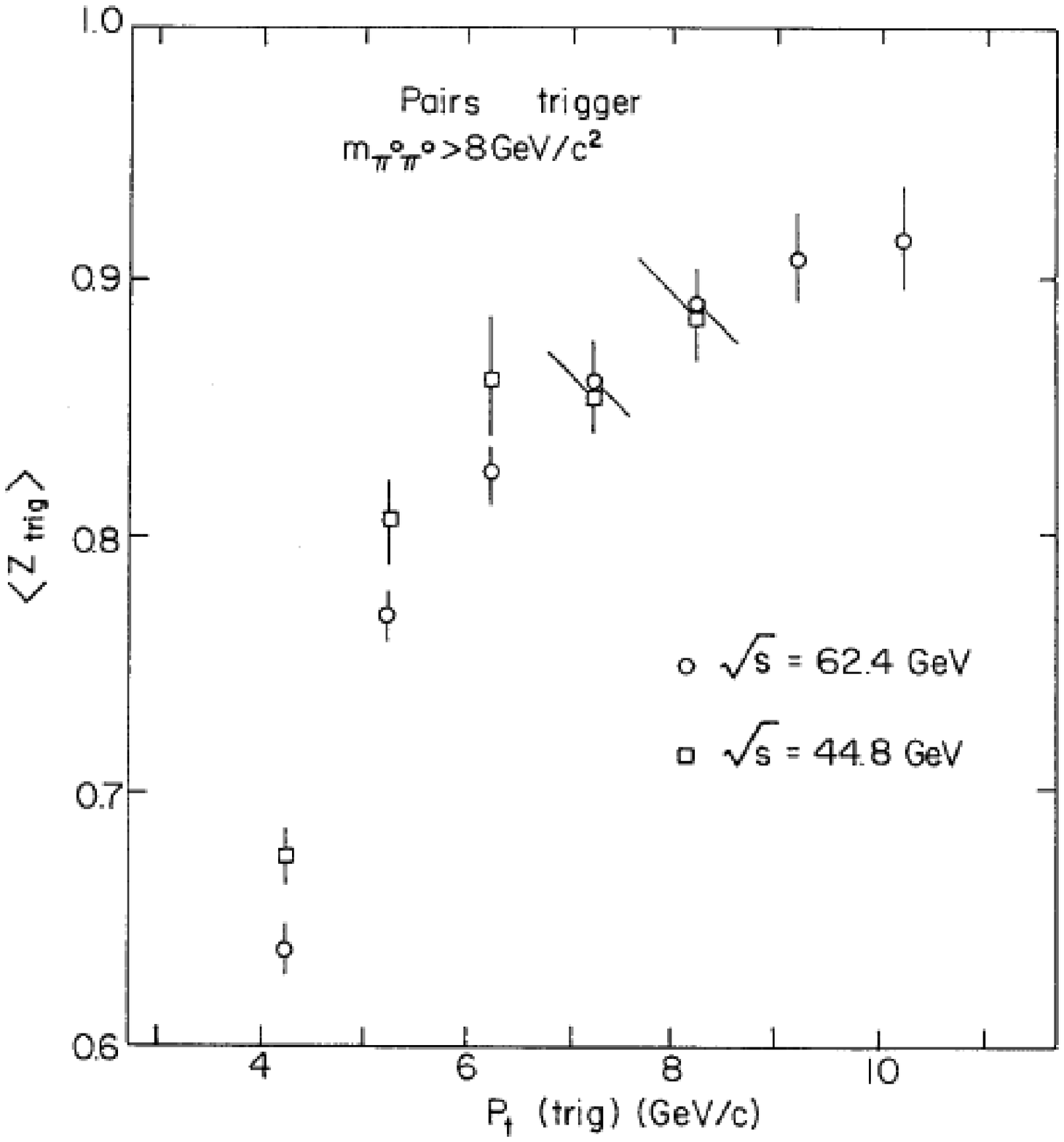}
\end{tabular}
\end{center}
\vspace*{-7mm}
\caption[]
{CCOR~\cite{CCOR82NPB} measurement of $\langle z_{\rm trig}\rangle$ as a function of $p_{T_t}$: (left) for inclusive $\pi^0$'s and (right) for a $\pi^0$-pair trigger with $m_{\pi^0 \pi^0} >8$ GeV/c$^{\rm 2}$  
\label{fig:mjt-ccormeanz} }
\end{figure}

   The outgoing jet-pair of hard-scattering obeys the kinematics of elastic-scattering (of partons) in a parton-parton c.m. frame which is longitudinally moving with rapidity $y=1/2 \ln(x_1/x_2)$ in the p-p c.m. frame. Hence, the jet-pair formed from the scattered partons should be co-planar with the beam axis, with two jets of equal and opposite transverse momentum. Thus, the outgoing jet-pair should be back-to-back in azimuthal projection. It is not necessary to fully reconstruct the jets in order to measure their properties. In many cases two-particle correlations are sufficient to measure the desired properties, and in some cases, such as the measurement of the net transverse momentum of a jet-pair, may be superior, since the issue of the systematic error caused by missing some of the particles in the jet is not-relevant.   Many ISR experiments provided excellent 2-particle correlation measurements~\cite{Moriond79}. However, the CCOR experiment~\cite{Angelis79} was the first to provide charged particle measurement with full and uniform acceptance over the entire azimuth, with pseudorapidity coverage $-0.7\leq \eta\leq 0.7$, so that the jet structure of high $p_T$ scattering could be easily seen and measured. In  Fig.~\ref{fig:mjt-ccorazi}a,b, the azimuthal distributions of associated charged particles 
relative to a $\pi^0$ trigger with transverse momentum $p_{T_t} > 7$ GeV/c are shown for five intervals of associated particle transverse momentum $p_T$. In all cases, strong correlation peaks on flat backgrounds are clearly visible, indicating the di-jet structure which is contained in an interval $\Delta\phi=\pm 60^\circ$ about a direction towards and opposite the to trigger for all values of associated $p_T\, (>0.3$ GeV/c) shown. The width of the peaks about the trigger direction (Fig.~\ref{fig:mjt-ccorazi}a), or opposite to the trigger (Fig.~\ref{fig:mjt-ccorazi}b) indicates out-of-plane activity from the individual fragments of jets.
The trigger bias was directly measured from these data by reconstructing the trigger jet from associated charged particles with $p_{T}\geq 0.3$ Gev/c, within $\Delta\phi=\pm 60^\circ$ from the trigger particle, using the algorithm $p_{T{\rm jet}}=p_{T_t}+1.5\sum p_{T}\cos(\Delta\phi)$, where the factor 1.5 corrects the measured charged particles for missing neutrals. The measurements of $\langle z_{\rm trig}\rangle=\langle p_{T_t}/p_{T{\rm jet}}\rangle$ as a function of $p_{T_t}$ for 3 values of $\sqrt{s}$ (Fig.~\ref{fig:mjt-ccormeanz}-(left)) show a variation which is consistent with scaling as a function of $x_T$, which was not expected~\cite{JacobEPS79,egsee1}.  Another observation~\cite{CCOR82NPB}, not much emphasized at the ISR but relevant to recent observations at RHIC, is that the measured $\langle z_{\rm trig}\rangle$ is different for single particle inclusive triggers and pair triggers (Fig.~\ref{fig:mjt-ccormeanz}-(right)). 
A recent measurement by STAR at RHIC~\cite{STARPRL95} gives $\mean{z_t}\approx 0.78\pm 0.04$ for inclusive $p_{T_t}=7.0$ GeV/c at $\sqrt{s}=200$ GeV.

  	Following the analysis of previous CERN-ISR experiments~\cite{DarriulatNPB107,CCHK}, the away jet azimuthal angular distributions  of Fig.~\ref{fig:mjt-ccorazi}b, which were thought to be unbiased, were analyzed in terms of the two variables: $p_{\rm out}=p_T \sin(\Delta\phi)$, the out-of-plane transverse momentum of a track;  
 and $x_E$, where:\\ 
\begin{minipage}[c]{0.5\linewidth}
\vspace*{-0.30in}
\begin{equation}	
x_E=\frac{-\vec{p}_T\cdot \vec{p}_{Tt}}{|p_{Tt}|^2}=\frac{-p_T \cos(\Delta\phi)}{p_{Tt}}\simeq \frac {z}{z_{\rm trig}}  
\label{eq:mjt-xE}
\end{equation}
\vspace*{0.06in}
\end{minipage}
\hspace*{0.01\linewidth}
\begin{minipage}[b]{0.50\linewidth}
\vspace*{0.06in}
\includegraphics[scale=0.6]{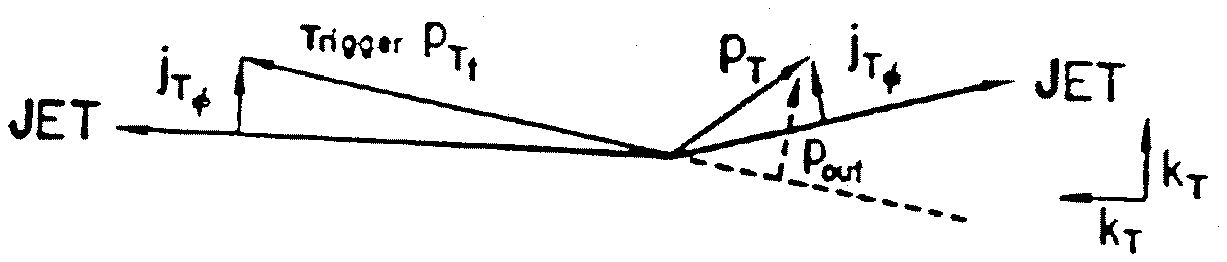}
\vspace*{-0.12in}
\label{fig:mjt-poutxe}
\end{minipage}\vspace*{-0.12in}
$z_{\rm trig}\simeq p_{Tt}/p_{T{\rm jet}}$ is the fragmentation variable of the trigger jet, and $z$ is the fragmentation variable of the away jet. Note that $x_E$ would equal the fragmenation fraction $z$ of the away jet, for $z_{\rm trig}\rightarrow 1$, if the trigger and away jets balanced transverse momentum. 
The $x_E$ distributions~\cite{Angelis79,JacobEPS79b} for the data of Fig.~\ref{fig:mjt-ccorazi}b are shown in Fig.~\ref{fig:mjt-ccorazi}c and show the fragmentation behavior expected at the time, $e^{-6z}\sim e^{-6 x_E \langle z_{\rm trig}\rangle}$. If the width of the away distributions (Fig.~\ref{fig:mjt-ccorazi}b) corresponding to the out of plane activity were due entirely to jet fragmentation, then  
$\langle |\sin(\Delta\phi)|\rangle=\langle |j_{T_{\phi}}|/p_T \rangle$ would decrease in direct proportion to $1/p_T$, where $j_{T_{\phi}}$ is the component of $\vec{j}_T$ in the azimuthal plane, since the jet fragmentation transverse momentum, $\vec{j}_T$, should be independent of $p_T$.  This is clearly not the case, as originally shown by the CCHK collaboration~\cite{CCHK}, which inspired Feynman, Field and Fox (FFF)~\cite{FFF} to introduce, $\vec{k}_T$, the transverse momentum of a parton in a nucleon. In this formulation, the net transverse momentum of an outgoing parton pair is $\sqrt{2} k_T$, which is composed of two orthogonal components, $\sqrt{2} k_{T_{\phi}}$, out of the scattering plane, which makes the jets acoplanar, i.e. not back-to-back in azimuth, and $\sqrt{2} k_{T_x}$, along the axis of the trigger jet, which makes the jets unequal in energy. Originally, ${k}_T$ was thought of as having an `intrinsic' part from confinement, which would be constant as a function of $x$ and $Q^2$, and a part from NLO hard-gluon emission, which would vary with $x$ and $Q^2$, however now it is explained as `resummation' to all orders of QCD~\cite{Sterman}. 

	FFF~\cite{FFF,Levin} gave the approximate formula to derive $k_T$ from the measurement of $p_{\rm out}$ as a function of $x_E$:
\begin{equation}
\langle |p_{\rm out}|\rangle^2=x_E^2 [2\langle |k_{T_{\phi}}|\rangle^2 +  \langle |j_{T_{\phi}}|\rangle^2 ] + \langle |j_{T_{\phi}}|\rangle^2 \qquad .
\label{eq:mjt-FFFpoutkT}
\end{equation}
CCOR~\cite{CCOR80} used this formula to derive $\langle |k_{T_{\phi}}|\rangle$ and $\langle |j_{T_{\phi}}|\rangle$ as a function of $p_{Tt}$ and $\sqrt{s}$ from the data of Fig.~\ref{fig:mjt-ccorazi}b.  
This important result showed that $\langle |j_{T_{\phi}}|\rangle$ is constant, independent of $p_{T_t}$ and $\sqrt{s}$, as expected for fragmentation, but that $\langle |k_{T_{\phi}}|\rangle$ varies with both $p_{T_t}$ and $\sqrt{s}$, suggestive of a radiative, rather than an intrinsic origin for $k_T$. The analysis was repeated, this year, by PHENIX for p-p collisions at $\sqrt{s}=200$ GeV~\cite{ppg029}. 
\subsection{Why `everything' became `almost everything' due to a new understanding of $x_E$ distributions}
\begin{figure}[htb]
\begin{center}
\begin{tabular}{cc}
\includegraphics[width=0.48\linewidth]{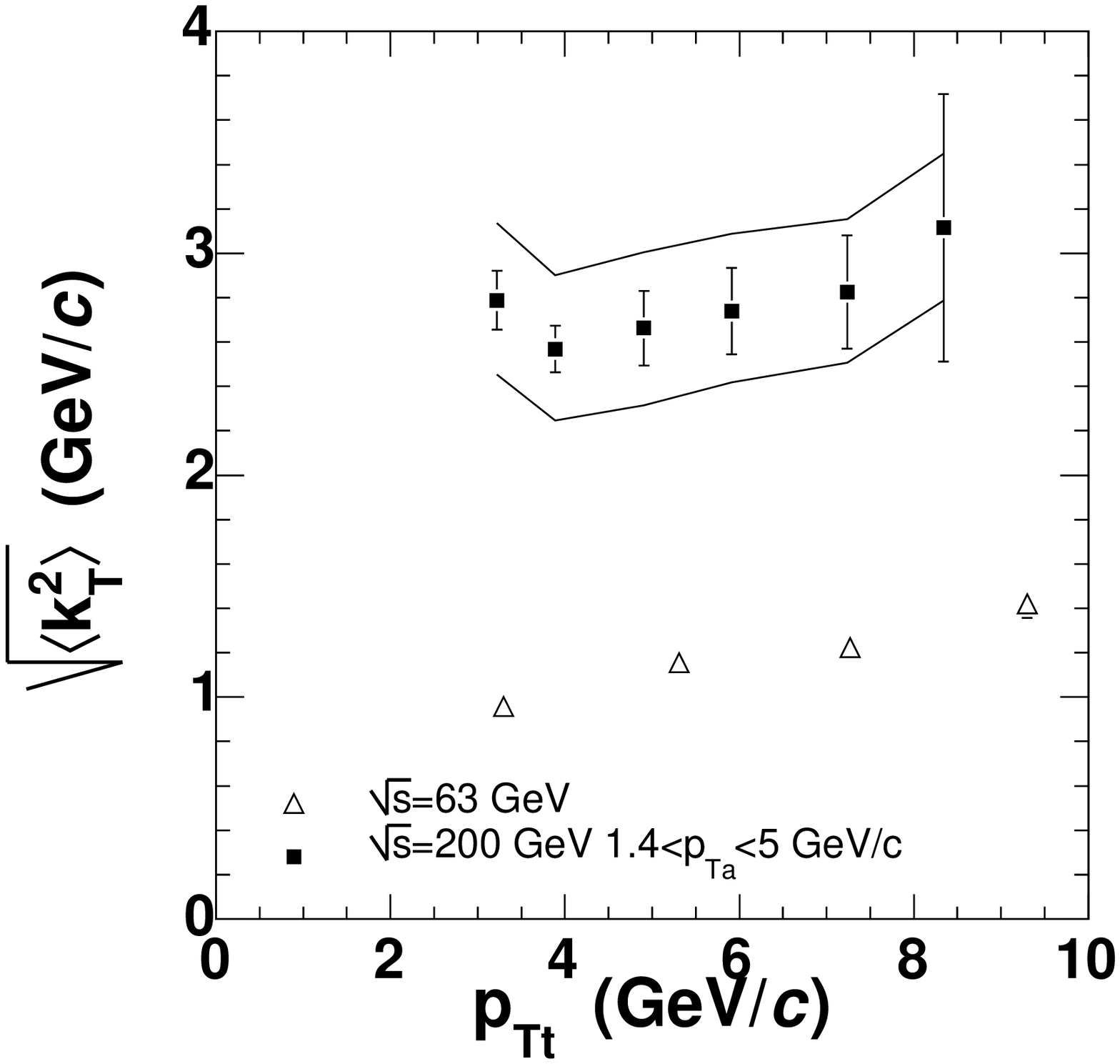} &
\includegraphics[width=0.48\linewidth]{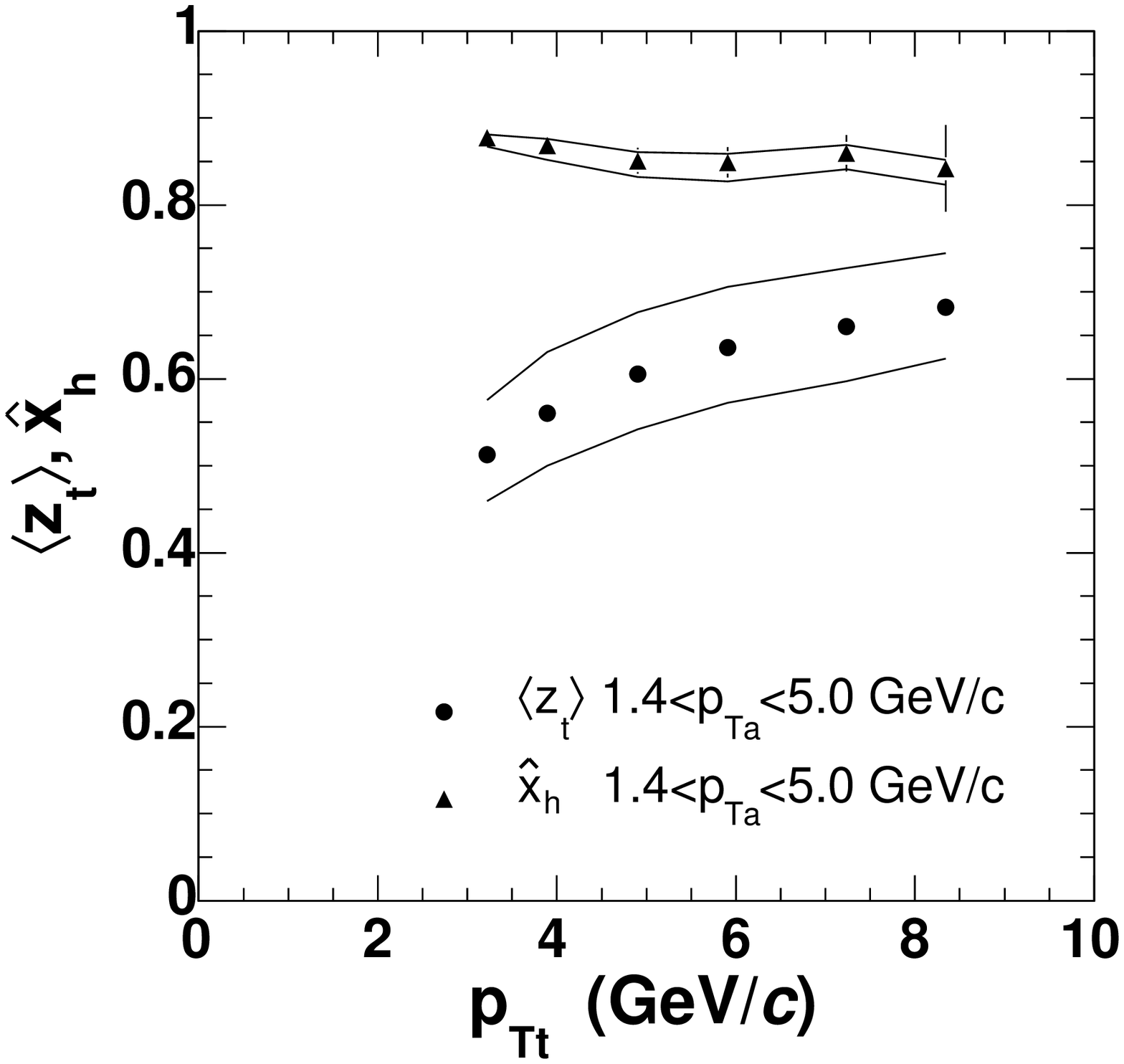}
\end{tabular}
\end{center}
\vspace*{-7mm}
\caption[]
{(left) $\sqrt{\mean{k^2_T}}$ values in p-p collisions for associated charged particles with $1.4<p_{T_a}<5$ GeV/c as a function of $p_{T_t}$ of a $\pi^0$ trigger (solid symbols) with statistical and systematic errors, measured by PHENIX~\cite{ppg029}. The CCOR masurement at $\sqrt{s}=62.4$ GeV~\cite{CCOR80} (open triangles) is also shown. (right) $\mean{z_t}$ and $\mean{\hat{x}_h}$ as a function of $p_{T_t}$ used in the PHENIX measurement, shown with statistical and systematic errors.      \label{fig:ppg029kT}}
\end{figure}

   The new measurement of $\mean{j_T}$, $\mean{k_T}$ and the $x_E$ distribution, this year~\cite{ppg029}, led to several complications and surprises, most notably that the shape of the $x_E$ distribution is not sensitive to the fragmentation function. The complications concern the fact that while the effect of $\mean{z_t}$ could be neglected at the ISR, where $\mean{z_t}\sim 1$ due to the larger value of $n$, it had to be taken into account at RHIC, with the result that the already complicated formula (Eq.~\ref{eq:mjt-FFFpoutkT}) for deriving $k_T$ became even more complicated: 
   \begin{equation}
   \label{new-FFFpoutkT}
   {{\mean{z_t(k_T,x_h)}\sqrt{\mean{k_T^2}}}\over {\mean{\hat{x}_h(k_T,x_h)}}}=
   {1\over x_h}\sqrt{\mean{p^2_{\rm out}}-\mean{j^2_{T_{\phi}}}(1+x_h^2)} 
   \qquad ,
   \end{equation}
where $x_h$ ($\hat{x}_h$) is the ratio of the associated particle (parton) transverse momentum to the trigger particle (parton) transverse momentum:
\begin{equation}
x_h\equiv {p_{T_a} \over p_{T_t}} \qquad \hat{x}_h=\hat{x}_h({k_T},x_h) 
\equiv { \hat{p}_{T_a} \over \hat{p}_{T_t}} \qquad .
\label{eq:defxh-hat}
\end{equation}     
Note that the hadronic variable $x_h$ is measured on every event and that the partonic variable $\hat{x}_h$ is a function of both $k_T$ and $x_h$ (Fig.~\ref{fig:ppg029kT}-(right)), as is the ``trigger bias'' $\mean{z_t}$~\cite{ppg029}.  Thus, the solution of Eq.~\ref{new-FFFpoutkT} for $\sqrt{\mean{k^2_T}}$ is an iterative process. The results~\cite{ppg029} are shown in Fig.~\ref{fig:ppg029kT}.

    In order to evaluate $\mean{z_t}$ the fragmentation function must be known. Based on the longstanding belief that the away jet was unbiased, PHENIX attempted to derive the fragmentation function from the measured $x_E$ distribution. 
\subsubsection{Fragmentation Formalism---two-particle correlations from a jet-pair}    
\begin{figure}[ht]
\begin{center}
\begin{tabular}{cc}
\includegraphics[width=0.50\linewidth]{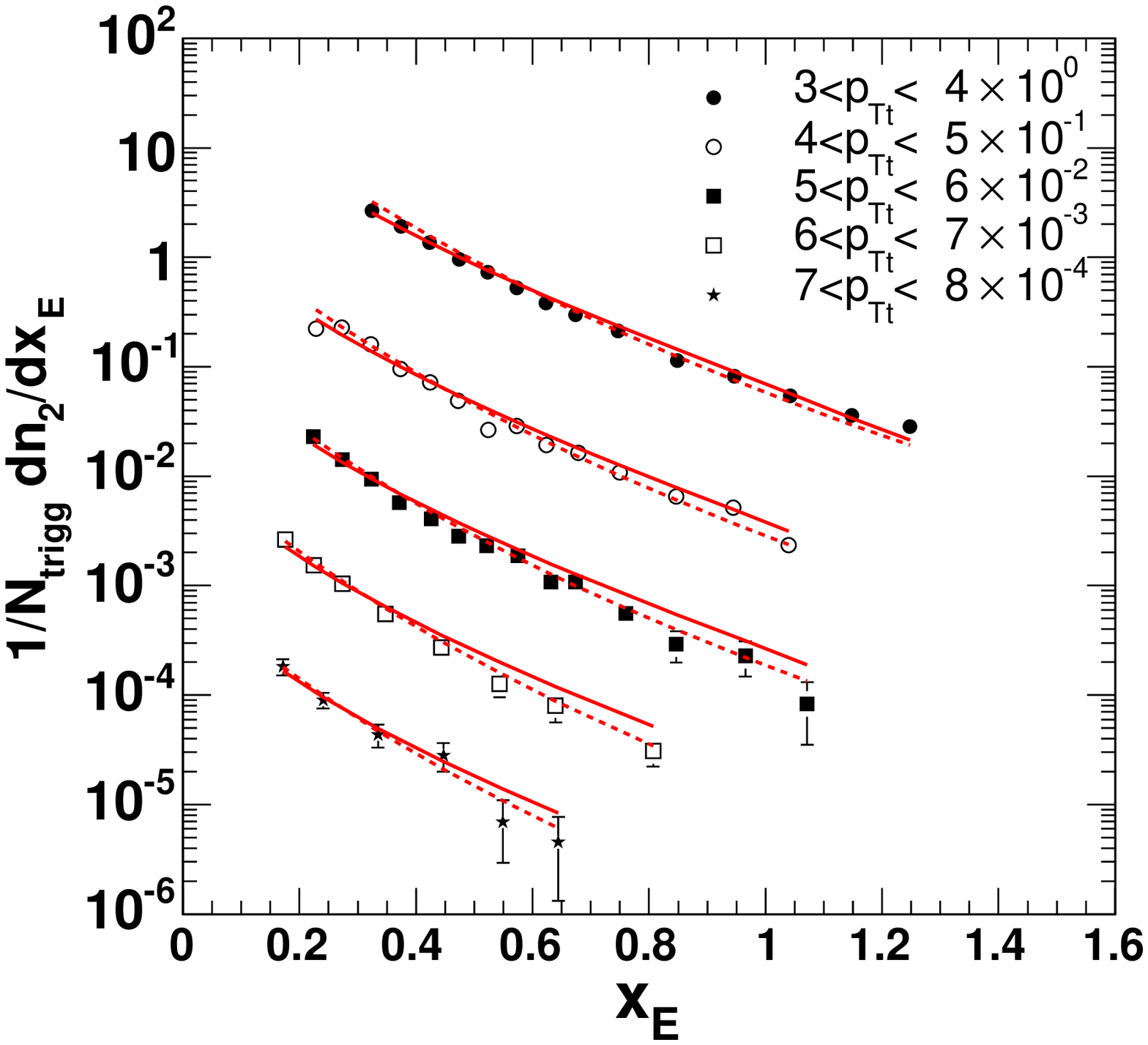} &
\includegraphics[width=0.48\linewidth]{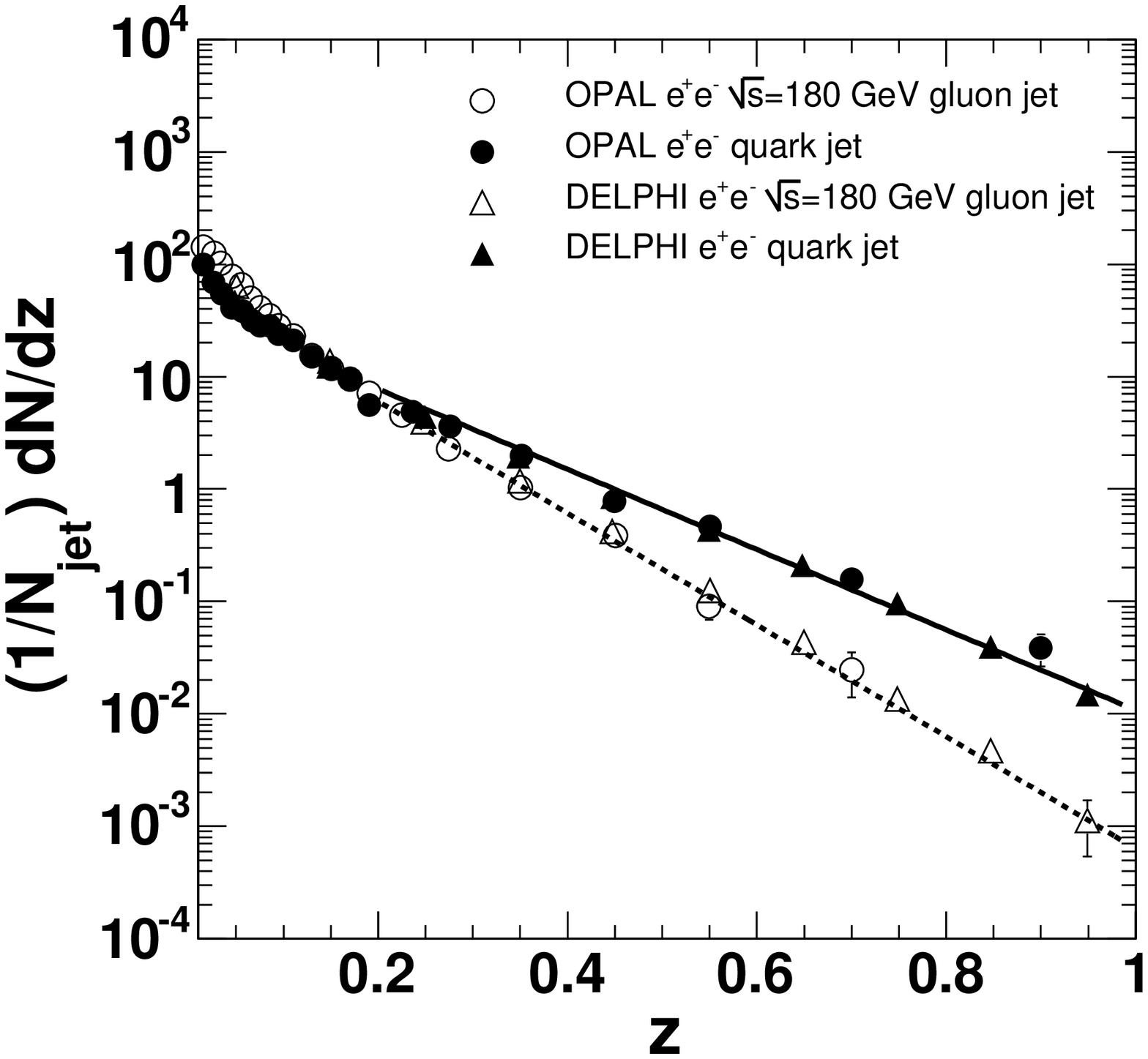}
\end{tabular}
\end{center}
\vspace*{-7mm}
\caption[]
{(left) $x_E$ distributions from PHENIX~\cite{ppg029} in p-p collisions at $\sqrt{s}=200$ GeV for several values of $p_{T_t}$. The solid and dashed lines represent calculations of the distribution from the integral of Eq.~\ref{eq:mikes_dsig} for quark (solid lines) and gluon (dashed lines) fragmentation functions based on exponential fits to the LEP measurements~\cite{OPAL,DELPHI} shown on the right panel.
\label{fig:wow} }
\end{figure}

    First recall the joint probability for a fragment pion, with $p_{T_t}=z_t\hat{p}_{T_t}$, originating from a parton with $\hat{p}_{T_t}$ (Eq.~\ref{eq:mjt-zgivenq}):
   \begin{eqnarray}
    {{d^2\sigma_{\pi} (\hat{p}_{T_t},z_t) }\over {\hat{p}_{T_t} d\hat{p}_{T_t} dz_t }}&=&{{d\sigma_q}\over {\hat{p}_{T_t} d\hat{p}_{T_t}}}\times D^{\pi}_q (z_t) \nonumber \\[0.12in]
    &=& f_q(\hat{p}_{T_t}) \times D^{\pi}_q(z_t) 
 \qquad .  \label{eq:mjt-zgivenq-rpt} 
   \end{eqnarray}
Here we make explicit that $f_q(\hat{p}_{T_t})$ represents the $k_T$-smeared final-state scattered-parton invariant spectrum ${{d\sigma_{q} }/{\hat{p}_{T_t} d\hat{p}_{T_t}}}$ and $D^{\pi}_q (z_t)$  
represents the fragmentation function. Due to the $k_T$ smearing, the transverse momentum $\hat{p}_{T_a}$ of the away parton in the hard-scattered parton-pair is less than the transverse momentum of the trigger parton $\hat{p}_{T_t}$~\cite{ppg029}. The probability that the parton with $\hat{p}_{T_a}$ fragments to a particle with $p_{T_a}=z_a \hat{p}_{T_a}$ in interval $dz_a$ is given by $D^{\pi}_q (z_a)$.  Thus, the joint probability for a fragment pion with $p_{T_t}=z_t\hat{p}_{T_t}$, originating from a parton with $\hat{p}_{T_t}$, and a fragment pion with $p_{T_a}=z_a\hat{p}_{T_a}$, originating from the other parton in the hard-scattered pair with $\hat{p}_{T_a}$ is:  
\begin{equation}
 { {d^3\sigma_{\pi} (\hat{p}_{T_t},z_t,z_a) }\over {\hat{p}_{T_t}d\hat{p}_{T_t}dz_t dz_a}}=
 {{d\sigma_q}\over {\hat{p}_{T_t}d\hat{p}_{T_t}}}\times D^{\pi}_q(z_t) \times D^{\pi}_q(z_a) \qquad , 
\label{eq:mjt-zaztgivenq}
\end{equation}
where 
\[ z_a={p_{T_a}\over \hat{p}_{T_a}}={p_{T_a}\over {\hat{x}_h\hat{p}_{T_t}}}=
{{z_t p_{T_a}}\over {\hat{x}_h {p}_{T_t}}} \]
and $\hat{x}_h=\hat{p}_{T_a}/\hat{p}_{T_t}$ (Eq.~\ref{eq:defxh-hat}). 
 Changing variables from $\hat{p}_{T_t}$, $z_t$ to $p_{T_t}$, $z_t$ as above and similarly from $z_a$ to $p_{T_a}$ yields:
\begin{equation} 
{ {d^3\sigma_{\pi} }\over {d{p}_{T_t}dz_t dp_{T_a}}}= {1\over {\hat{x}_{\rm h}\, p_{T_t}}}
 {{d\sigma_q}\over {d({p}_{T_t}/z_t})}\,D^{\pi}_q(z_t) D^{\pi}_q({{z_t p_{T_a}} \over {\hat{x}_{\rm h} p_{T_t}}}) 
 \label{eq:mikes_dsig}
 \end{equation}
where for integrating over $z_t$ or finding $\mean{z_t}$ for fixed
$p_{T_t}$, $p_{T_a}$, the minimum value of $z_t$ is $z_t^{\rm min}=2p_{T_t}/\sqrt{s}=x_{T_t}$
and the maximum value is:
\[
z_t^{\rm max}=\hat{x}_h {p_{T_t}\over p_{T_a}}={\hat{x}_h\over x_h} \qquad ,
\]
where $\hat{x}_h(p_{T_t},p_{T_a})$ is also a function of $k_T$ (Eq.~\ref{eq:defxh-hat}). Integrating over $dz_t$ in Eq.~\ref{eq:mikes_dsig} gives the $x_E$ distribution in the collinear limit, where $p_{T_a}=x_E p_{T_t}$, and it was thought~\cite{ppg029} that a simply parameterized fragmentation function could be extracted  from a joint fit to the measured $x_E$ and inclusive $p_{T_t}$ distributions (Eq.~\ref{eq:power_law}). However, there were serious difficulties with convergence which took a while to sort out. Eventually, the $x_E$ distributions were calculated from Eq.~\ref{eq:mikes_dsig} using LEP measurements for quark and gluon fragmentation functions, with shocking results (see Fig.~\ref{fig:wow})---the $x_E$ distributions calculated with quark $D^{\pi}_q \approx exp (-8.2\cdot z)$ or gluon $D^{\pi}_g\approx exp (-11.4\cdot z)$  fragmentation functions do not differ significantly!  Clearly, the $x_E$ distributions are rather insensitive to the fragmentation function of the away jet in contradiction to the conventional wisdom dating from the early 1970's.    

	The evidence of this explicit counter example led to an attempt to perform the integral of Eq.~\ref{eq:mikes_dsig} analytically which straightforwardly confirmed that the shape of the $x_E$ distribution is not sensitive to the shape of the fragmentation function. However, it was found that $x_E$ distribution is sensitive to $\hat{x}_h$, the ratio of the transverse momentum of the away-side jet ($\hat{p}_{T_a}$)  to that of the trigger-side jet ($\hat{p}_{T_t}$). This can be put to use in A+A collisions to measure the relative energy loss of the two jets from a hard-scattering which escape from the medium.  
\subsubsection{Analytical formula for the $x_E$ distribution} 
   With a substitution of a power-law parton $\hat{p}_{T_t}$ spectrum (Eq.~\ref{eq:parton_power}) and an exponential fragmentation function (Eq.~\ref{eq:ff1}), as in section~\ref{sec:frag-single}, the integral of  Eq.~\ref{eq:mikes_dsig} over $z_t$ becomes:
  \begin{equation}\label{eq:int_mikes_dsig}
 {d^2\sigma_\pi \over dp_{T_t} dp_{T_a}} = 
{B^2\over \hat{x}_h} {A\over p_{T_t}^{n}}
\int_{x_{T_t}}^{\hat{x}_h {p_{T_t}\over p_{T_a}}} d z_t z_t^{n-1}
exp [-b z_t(1+ {p_{T_a} \over {\hat{x}_h p_{T_t}}})]\qquad .
\end{equation}
This is again an incomplete gamma function, if $\hat{x}_h$ is taken to be constant as a function of $z_t$ for fixed $p_{T_t}$, $p_{T_a}$:
\begin{equation}
{d^2\sigma_\pi \over dp_{T_t} dp_{T_a}} = 
{B^2\over\hat{x}_h} {A\over p_{T_t}^{n}} {1 \over {b'^{\,n}}}  \left[ \Gamma({n},b' x_{T_t}) - \Gamma (n, b'\hat{x}_h {p_{T_t}\over p_{T_a}}) \right] 
\label{eq:ans1_int_mikes_dsig} \qquad ,
\end{equation}
where $b'$ is given by:
 \begin{equation}
  b'=b(1+ {p_{T_a} \over{\hat{x}_h p_{T_t}}}) \qquad.
  \label{eq:bprime}
  \end{equation}

The conditional probability of the $p_{T_a}$ distribution for a given $p_{T_t}$ is the ratio of the joint probability Eq.~\ref{eq:ans1_int_mikes_dsig} to the inclusive probability Eq.~\ref{eq:ans1_int_mjt_sig_inclus}, or
   \begin{equation}
\left.{dP_{\pi} \over d p_{T_a}}\right|_{p_{T_t}} = {B\over{b p_{T_t}\hat{x}_h}} {1\over {(1+ {p_{T_a} \over{\hat{x}_h p_{T_t}}})^{n}}} 
\, { {\left[ \Gamma({n},b' x_{T_t}) - \Gamma (n, b'\hat{x}_h {p_{T_t}\over p_{T_a}}) \right] } \over { \left[ \Gamma({n-1},b x_{T_t}) - \Gamma (n-1, b) \right]}}\qquad , 
\label{eq:ans1_condpta}
\end{equation}
and this answer is exact for the case of constant $\hat{x}_h$, with no assumptions other than a power law for the parton $\hat{p}_{T_t}$ distribution and an exponential fragmentation function. 
In the collinear limit, where, $p_{T_a}=x_E p_{T_t}$:
   \begin{equation}
\left.{dP_{\pi} \over dx_E}\right|_{p_{T_t}}= {1\over\hat{x}_h} {B\over b} {1\over
{(1+ {x_E \over{\hat{x}_h}})^{n}}}  
\, { {\left[ \Gamma({n},b' x_{T_t}) - \Gamma (n, b'{\hat{x}_h\over x_E}) \right] } \over { \left[ \Gamma({n-1},b x_{T_t}) - \Gamma (n-1, b) \right]}} 
\qquad . 
\label{eq:ans1_condxe}
\end{equation}

	With the same approximation for the incomplete gamma functions used previously (Eq.~\ref{eq:result2_int_mjt_sig_inclus}), namely taking the upper limit of the integral (Eq.~\ref{eq:int_mikes_dsig}) to infinity and the lower limit to zero, the ratio of incomplete gamma functions in Eq.~\ref{eq:ans1_condxe} becomes equal to $n-1$ and the $x_E$ distribution takes on a very simple and very interesting form:
	     \begin{equation}
\left.{dP_{\pi} \over dx_E}\right|_{p_{T_t}}\approx {\mean{m}(n-1)}{1\over\hat{x}_h} {1\over
{(1+ {x_E \over{\hat{x}_h}})^{n}}} \, \qquad , 
\label{eq:condxe2}
\end{equation}
where the only dependence on the fragmentation function is in the mean multiplicity of charged particles in the jet $\mean{m}\approx B/b\approx b$. The dominant term in Eq.~\ref{eq:condxe2} is the Hagedorn function $1/(1+x_E/\hat{x}_h)^n$ so that Eq.~\ref{eq:condxe2} exhibits $x_E$-scaling in the variable $x_E/\hat{x}_h$. The shape of the $x_E$ distribution is given by the power $n$ of the partonic and inclusive single particle transverse momentum spectra and does not depend on the exponential slope of the fragmentation function. However, the integral of the $x_E$ distribution (from zero to infinity) is equal to 
$\mean{m}$, the mean multiplicity of the unbiased away-jet. 

	The reason that the $x_E$ distribution is not very sensitive to the fragmentation function is that the integral over $z_t$ for fixed $p_{T_t}$ and $p_{T_a}$ (Eqs.~\ref{eq:mikes_dsig}, \ref{eq:int_mikes_dsig}) is actually an integral over the jet transverse momentum $\hat{p}_{T_t}$. However since both the trigger and away jets are always roughly equal and opposite in transverse momentum, integrating over $\hat{p}_{T_t}$ simultaneously integrates over $\hat{p}_{T_a}$, and thus also integrates over the away jet fragmentation function. This can be seen directly by the presence of $z_t$ in both the same and away fragmentation functions in Eq.~\ref{eq:mikes_dsig}, so that the integral over $z_t$ integrates over both fragmentation functions simultaneously.

\subsubsection{Why did we believe that the $x_E$ distribution measured the fragmentation function?}
\begin{figure}[thb]
\begin{center}
\includegraphics[width=0.75\linewidth]{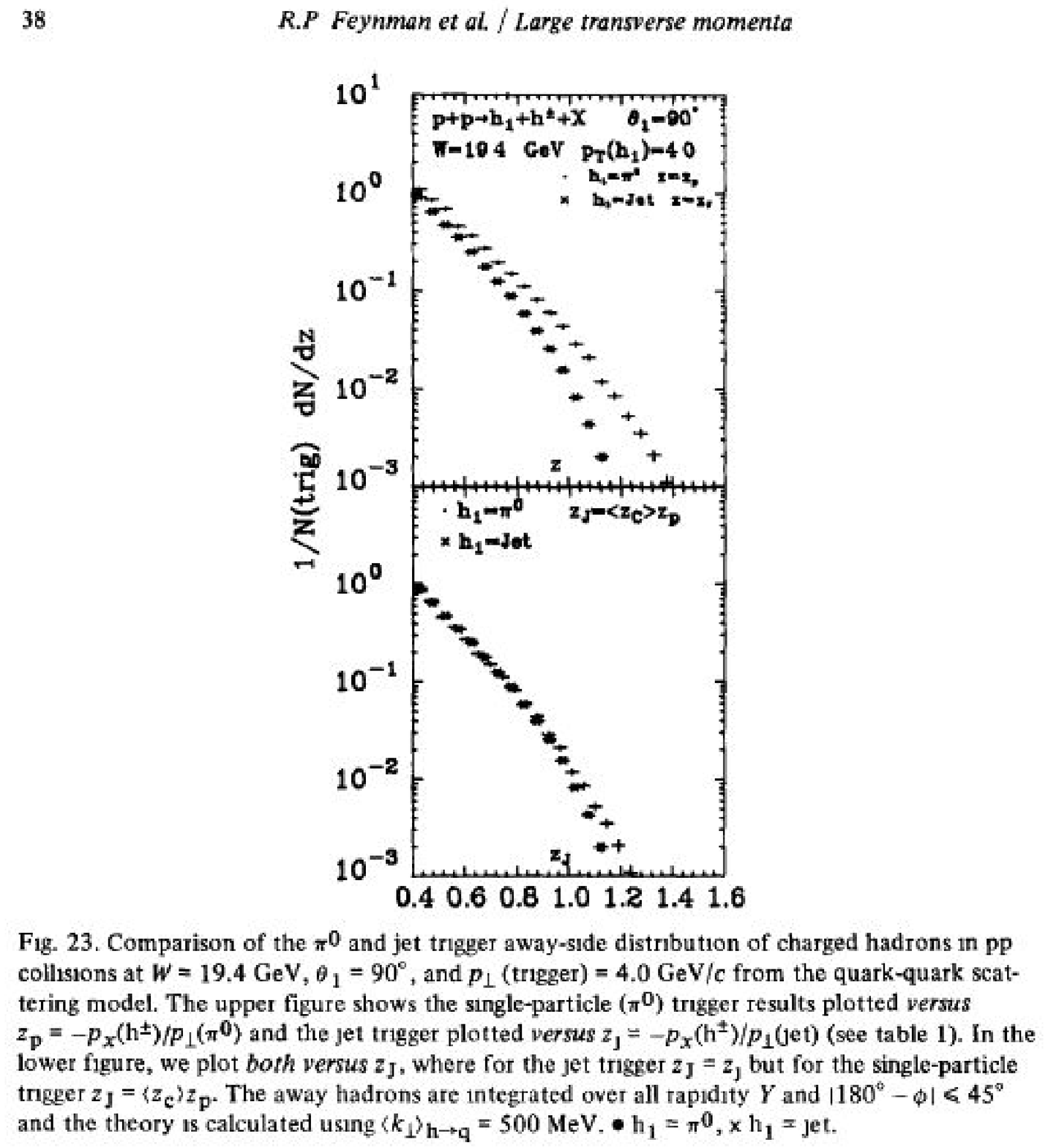}
\end{center}\vspace*{-0.25in}
\caption[]{Figure 23 from FFF~\cite{FFF} \label{fig:FFFfig23}}
\end{figure}

	The seminal paper of Feynman, Field and Fox (FFF)~\cite{FFF} was especially influential in forming the belief that the $x_E$ distribution measured the fragmentation function. To cite directly from Ref.~\cite{FFF}, p 25, ``There is a simple relationship between experiments done with single-particle triggers and those performed with jet triggers.
The only difference in the opposite side correlation is due to the fact that the
`quark', from which a single-particle trigger came, always has a higher $p_{\perp}$ than the trigger (by factor $1/z_t$). The away-side correlations for a single-particle trigger at $p_{\perp}$ should be roughly the same as the away side correlations for a jet trigger at $p_{\perp}$(jet)=$p_{\perp}$(single particle)$/\mean{z_t}$''. This point is reinforced in  the conclusions (p 59), ``2. The distribution of away-side hadrons from a jet trigger should be the same as that from a single particle trigger except for a correction due to $\mean{z_t}$ (see Fig.23)'' [which is shown as Fig.~\ref{fig:FFFfig23} below].  
Another interesting point is, ``8. Because the quarks scatter elastically (no quantum number exchange - except perhaps color), the away-side distribution of hadrons in pp collisions should be essentially independent of the quantum numbers of the trigger hadron.''---i.e. the jets fragment independently. Note that in FFF the notation is $a+b\rightarrow c+d$ (as in Eq.~\ref{eq:QCDabscat})  where $a,b,c,d$ are called `quarks', so FFF call $z_t$, $z_c$. 

\begin{figure}[ht]
\begin{center}
\begin{tabular}{cc}
\includegraphics[width=0.40\linewidth]{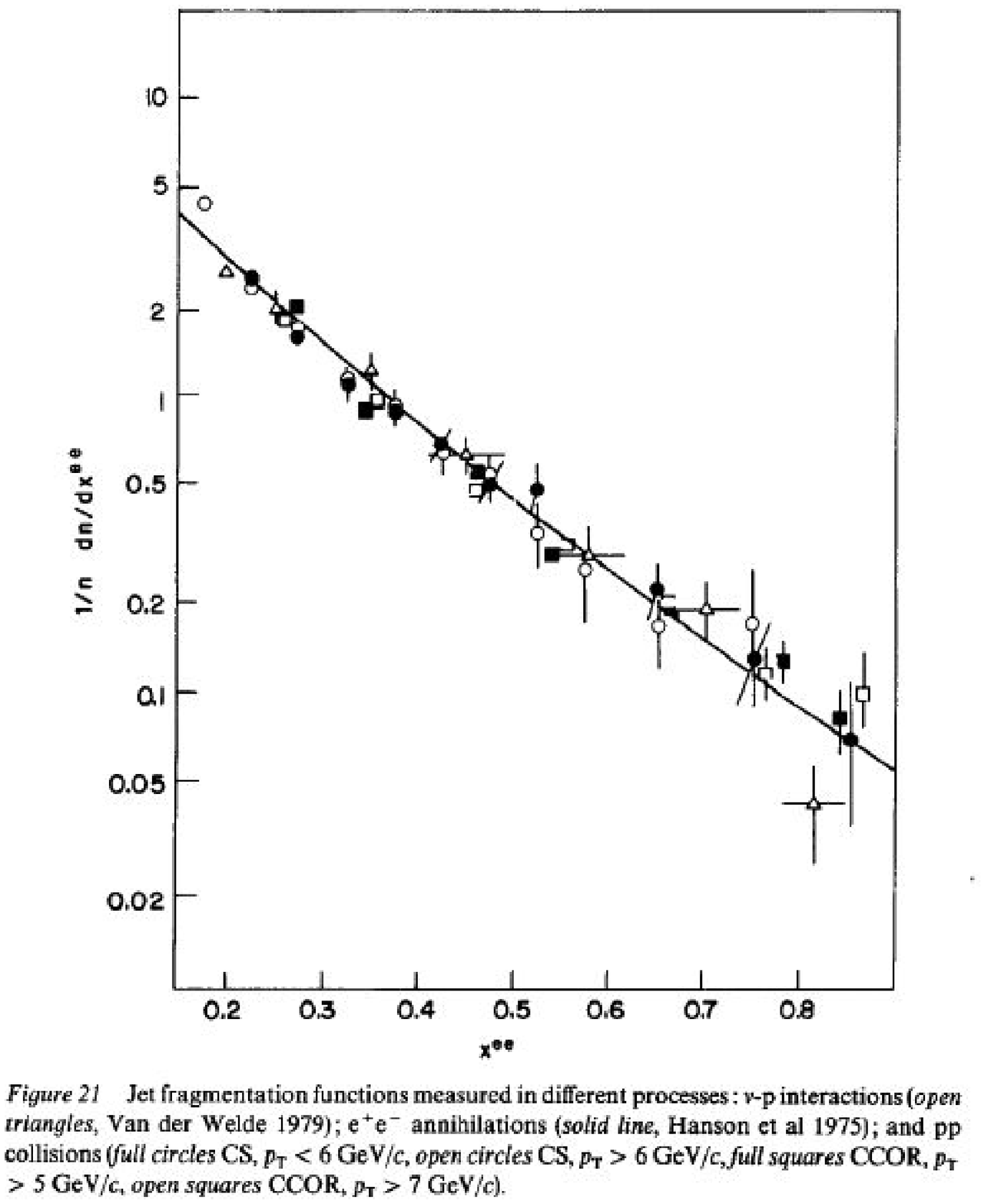} &
\includegraphics[width=0.55\linewidth]{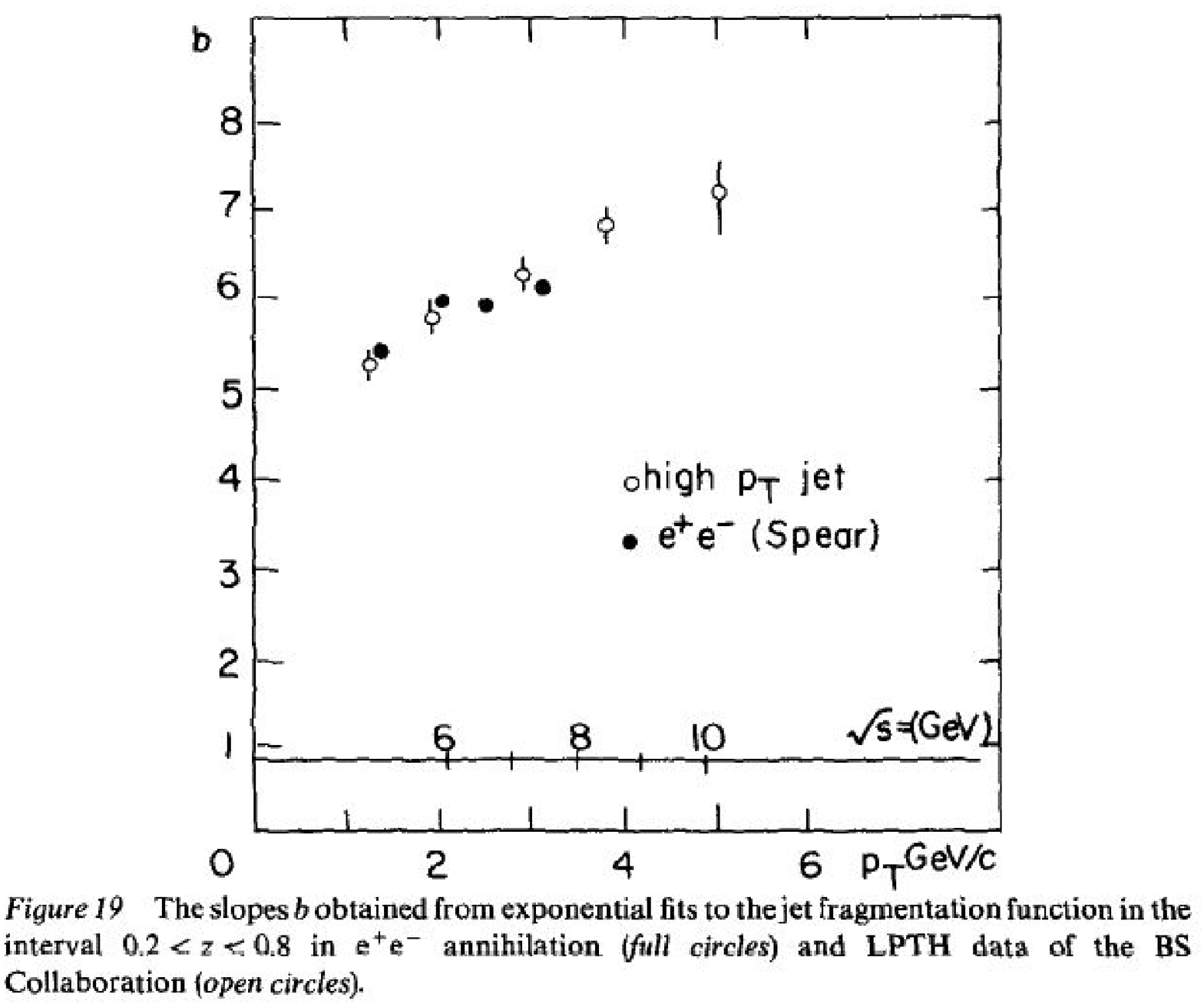}
\end{tabular}
\end{center}
\vspace*{-7mm}
\caption[]
{(left) Jet fragmentation functions from $\nu$-p, $e^+ e^-$ and p-p reactions. (right) $b$-slopes from this data, where `LPTH' is an acronym for Large $p_T$ Hadron production~\cite{Darriulat}.  
\label{fig:DARNPS} }
\end{figure}

	This belief was thought to have been verified by measurements at the CERN-ISR which showed (Fig.~\ref{fig:DARNPS}-(left)) that jet fragmentation functions in $\nu$-p, $e^+ e^-$ and p-p reactions (CCOR Fig.~\ref{fig:mjt-ccorazi}c~\cite{Angelis79,JacobEPS79b}) are the same, with the same dependence of the exponential slope $b$ on $\hat{s}$ (Fig.~\ref{fig:DARNPS}-(right)~\cite{Darriulat}. 

\subsubsection{A very interesting formula} \label{sec:interesting}
     Equation Eq.~\ref{eq:condxe2} (repeated below in a slightly different format) is very interesting.  
\begin{equation}
\left.{dP_{\pi} \over dy}\right|_{p_{T_t}}\approx {\mean{m}(n-1)} {1\over
{(1+ y)^{n}}} \qquad \mbox{ where }\qquad y={x_E \over{\hat{x}_h}} \qquad .\label{eq:condxe3}
\end{equation}
 It relates the ratio of the transverse momenta of the away and trigger particles, $p_{T_a}/p_{T_t}=x_h\approx x_E$, which is measured, to the ratio of the transverse momenta of the away to the trigger jet, $\hat{p}_{T_a}/\hat{p}_{T_t}$, which can thus be deduced. Although derived for p-p collisions, Eq.~\ref{eq:condxe2} (\ref{eq:condxe3}) should work just as well in A+A collisions since the only assumptions are independent fragmentation of the trigger and away-jets with the same exponential fragmentation function and a power-law parton $\hat{p}_{T_t}$ distribution. The only other (and weakest) assumption is that $\hat{x}_h$ is constant for fixed $p_{T_t}$ as a function of $x_E$. Thus in A+A collisions, Eq.~\ref{eq:condxe2} for the $x_E$ distribution provides a method of measuring the ratio $\hat{x}_h=\hat{p}_{T_a}/\hat{p}_{T_t}$ and hence the relative energy loss of the away to the same side jet assuming that both jets fragment outside the medium with the same fragmentation function as in p-p collisions. 
     \subsubsection{Test of determination of $\hat{x}_h$ from the $x_E$ distribution in p-p collisions}\label{sec:testpp}

  \begin{figure}[ht]
\begin{center}
\begin{tabular}{cc}
\includegraphics[width=0.45\linewidth]{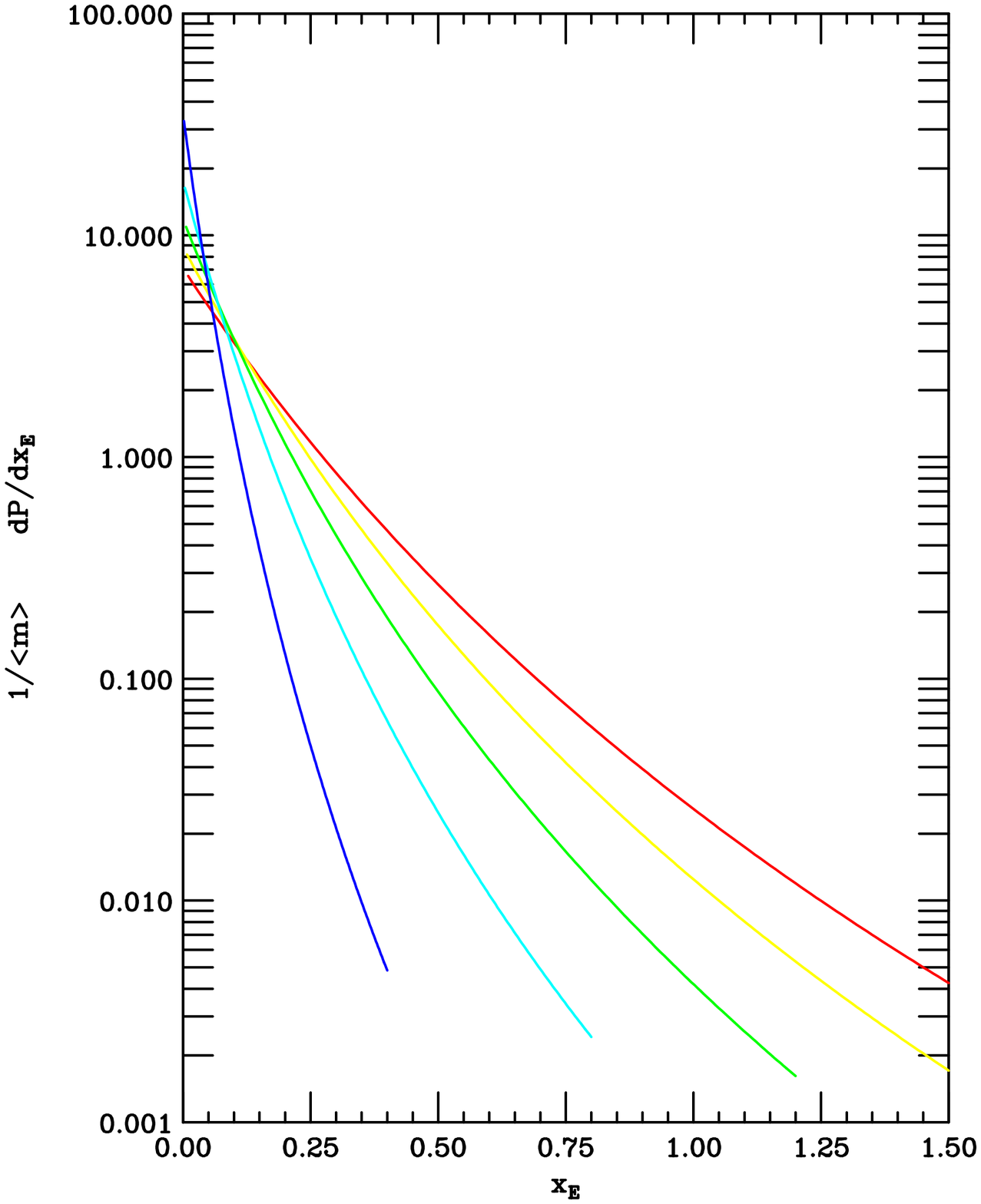} &
\includegraphics[width=0.45\linewidth]{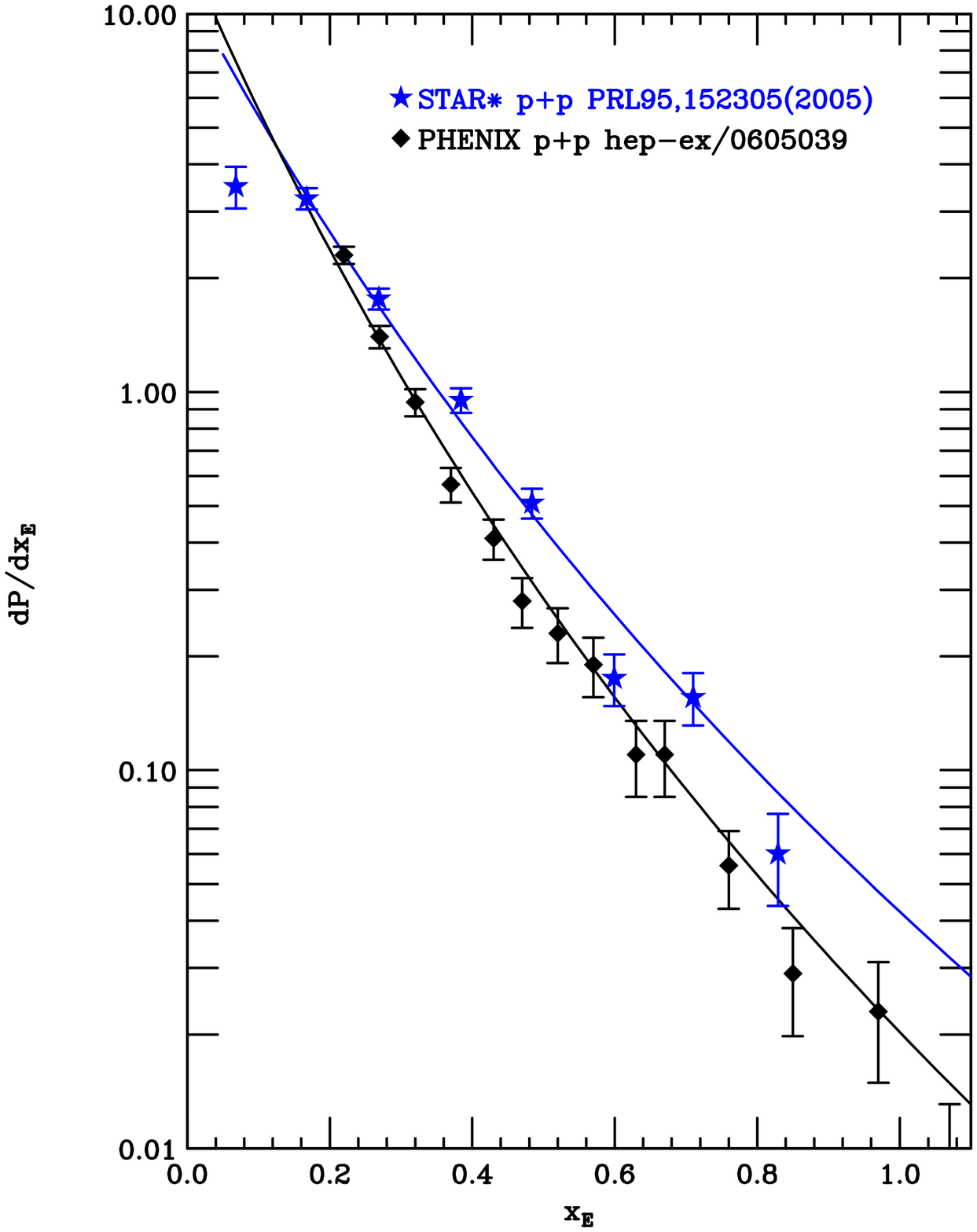} 
\end{tabular}
\end{center}
\caption[]{(left) Eq.~\ref{eq:condxe2} for n=8.1 divided by $\langle m\rangle$. The integral should be equal to 1. Curves are for $\hat{x}_h=$1.0 (red), 0.8, 0.6, 0.4, 0.2 (blue), with intercept=$7.1/\hat{x}_h$ at $x_E=0$. (right)-PHENIX $x_E$ distribution for $5<p_{T_t}<6$ GeV/c (Fig.~\ref{fig:wow}-(left)) with Eq.~\ref{eq:condxe2} for $\hat{x}_h=0.8$ (black); STAR~\cite{STARPRL95} $x_E$ distribution (Fig~\ref{fig:s42-44}) with  Eq.~\ref{eq:condxe2} for $\hat{x}_h=1.0$ (blue). Curves are from left panel multiplied by a factor of 1.63 to agree with PHENIX data. STAR data have  been multiplied by a factor of 0.6 to agree with this normalization. 
\label{fig:s31-43}}
\end{figure}
	A plot of Eq.~\ref{eq:condxe2} is shown in Fig.~\ref{fig:s31-43}-(left) for $n=8.1$ for various values of $\hat{x}_h$. Clearly, the smaller the value of $\hat{x}_h$, the steeper is the $x_E$ distribution. However, all the curves in Fig.~\ref{fig:s31-43}-(left) are related by a simple scale transformation of Eq.~\ref{eq:condxe3}: $y\rightarrow {x}_E=\hat{x}_h\, y$. In general, the values of $n$ and $\hat{x}_h$ should be able to be determined from a simultaneous fit of the inclusive $p_{T_t}$ spectrum (Eq.~\ref{eq:simpower} or \ref{eq:ans1_int_mjt_sig_inclus}) and the $x_E$ distribution (Eq.~\ref{eq:ans1_condxe}).  On the other hand, when the value of $n$ is well determined from the inclusive $p_{T_t}$ spectrum (e.g. see Fig.~\ref{fig:pizpp200}), Eq.~\ref{eq:condxe3} can just be scaled to fit the measured $x_E$ distribution. This was done for the PHENIX p-p data~\cite{ppg029} of Fig.~\ref{fig:wow}-(left) ($5.0< p_{T_t}<6.0$ GeV/c). The value of $\hat{x}_h=0.8$ which gave the best ``eyeball'' agreement with the data (Fig.~\ref{fig:s31-43}-(right)) agrees with the value of $\hat{x}_h$ determined independently from the $k_T$-smearing analysis (Fig.~\ref{fig:ppg029kT})~\cite{ppg029}. Similarly, for the STAR p-p data~\cite{STARPRL95} (Fig.~\ref{fig:s42-44}) excellent agreement is found with $\hat{x}_h=1.0$, for $x_E>0.2$. Thus, the method works for p-p collisions. Note that the evident deviation of the STAR data from Eq.~\ref{eq:condxe2} for $x_E<0.2$ may be a limitation due to the simple approximations or may be the result of the absence of corrections for decay-in-flight for the low $p_{T_a}$ particles (presumably $\pi^{\pm}$) in the measurement. Also, as indicated in the caption of Fig.~\ref{fig:s31-43}, typically, Eq.~\ref{eq:condxe2} must be normalized in order to agree with the data.  
          \subsection{2-particle correlation measurements in Au+Au collisions}
          One of the first and still most striking measurements of two-particle correlations in Au+Au collisions was presented by STAR at the Quark Matter 2002 conference~\cite{HardtkeQM02}. 
     		\begin{figure}[ht]
	\begin{center}
	\includegraphics[width=0.7\linewidth]{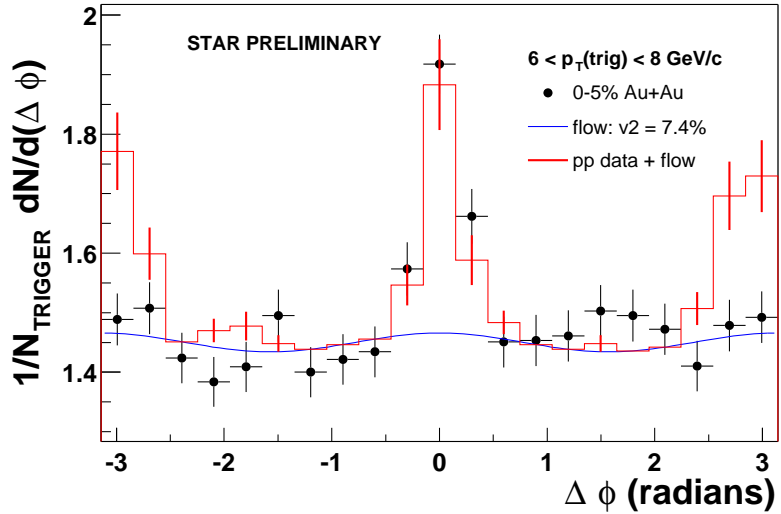}
	\end{center}
	\vspace*{-0.24in}
    \caption[] {STAR~\cite{HardtkeQM02} conditional probability ${1\over N_{trig}}{dN\over d\Delta\phi}$ of associated non-identified charged particles with $p_{T_a}$ in the range $2\,{\rm GeV/c}< p_{T_a} < p_{T_t}$ per trigger particle with $p_{T_t}$ between 6 and 8 GeV/c, all in the range $0<|\Delta\eta|<1.4$ for central Au+Au collisions at $\sqrt{s_{NN}}=200$ GeV as a function of the azimuthal angle difference $\Delta\phi$ between the trigger and associated particles (solid circles). Non-jet background modulated by elliptic flow $v_2$ is shown as the blue curve. The sum of the measured p-p correlation plus the flow is shown as the red histogram. }
	\label{fig:HardtkeCent}
	\end{figure}
In Fig.~\ref{fig:HardtkeCent}, the conditional probability---given a trigger particle with $p_{T_t}$ between 6 and 8 GeV/c---of detecting an associated particle with $p_{T_a}$ in the range $2\,{\rm GeV/c}< p_{T_a}<p_{T_t}$ is shown for central Au+Au collisions at $\sqrt{s_{NN}}=200$ GeV as a function of the azimuthal angle difference of the two particles, $\Delta\phi$. The 2-particle correlation function expressed as the conditional probability is the sum of the background of particles randomly associated to the trigger, which is modulated by the common hydrodynamic flow (represented by $v_2(p_T)$), plus the jet correlation function which was presumed to be the same as that measured in p-p collisions~\cite{jetnote}:
\begin{equation}
C_2^{AuAu}(\Delta\phi)=C_2^{pp}(\Delta\phi)+B(1+2 v_2(p_{T_t}) v_2(p_{T_a}) \cos(2\Delta\phi)) \qquad .
\label{eq:corrfn} 
\end{equation}
The trigger-side correlation peak in central Au+Au collisions appears to be the same as that measured in p-p collisions (corrected by the small flow effect) but the away side jet correlation in Au+Au appears to have vanished, the data seem to be saturated by the small flow effect. This observation appears consistent with a large energy loss in the medium, or a medium that is opaque to the propagation of high momentum partons, as originally indicated by the suppression observed~\cite{Adcox01} in single particle inclusive measurements for $p_{T_t}> 3$ GeV/c (recall Fig.~\ref{fig:RAA-pi-h}).

\begin{figure}[ht]
	\begin{center}
	\begin{tabular}{cc}
	\includegraphics[width=0.45\linewidth]{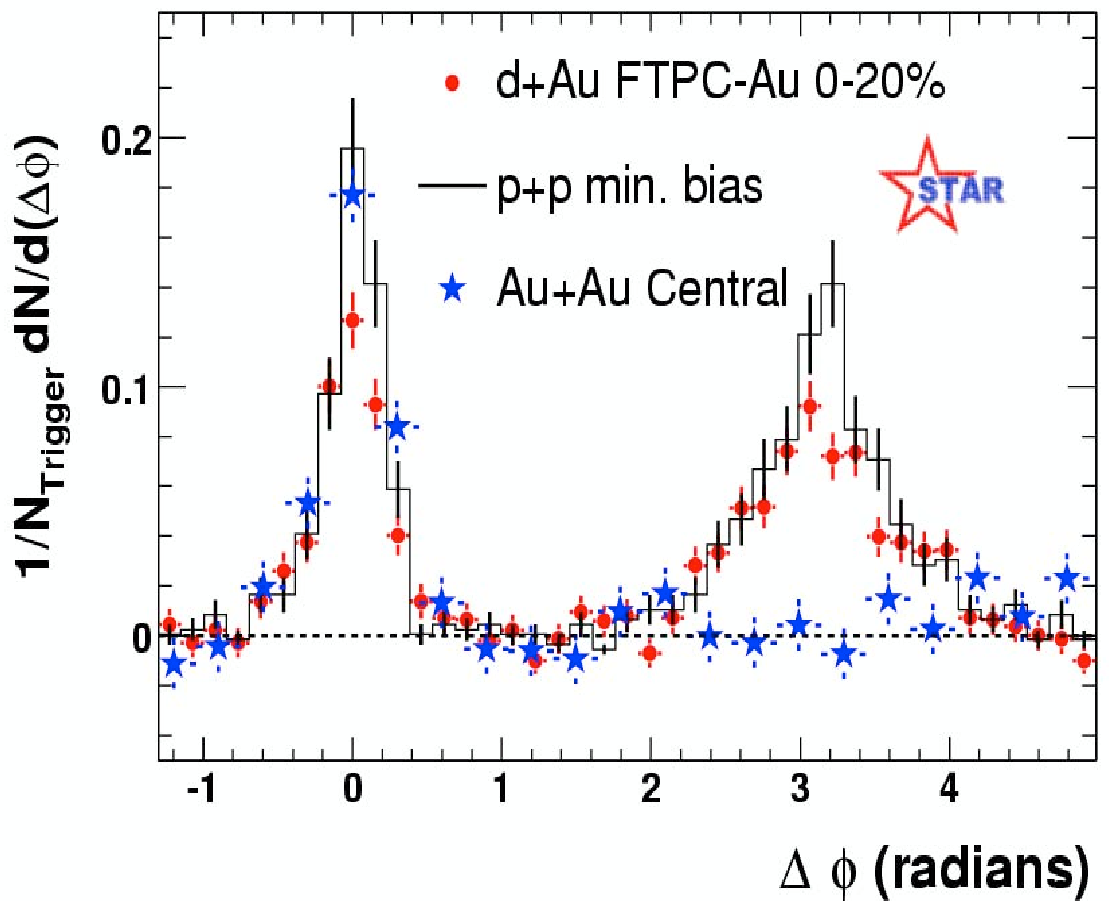}&
	\includegraphics[width=0.49\linewidth]{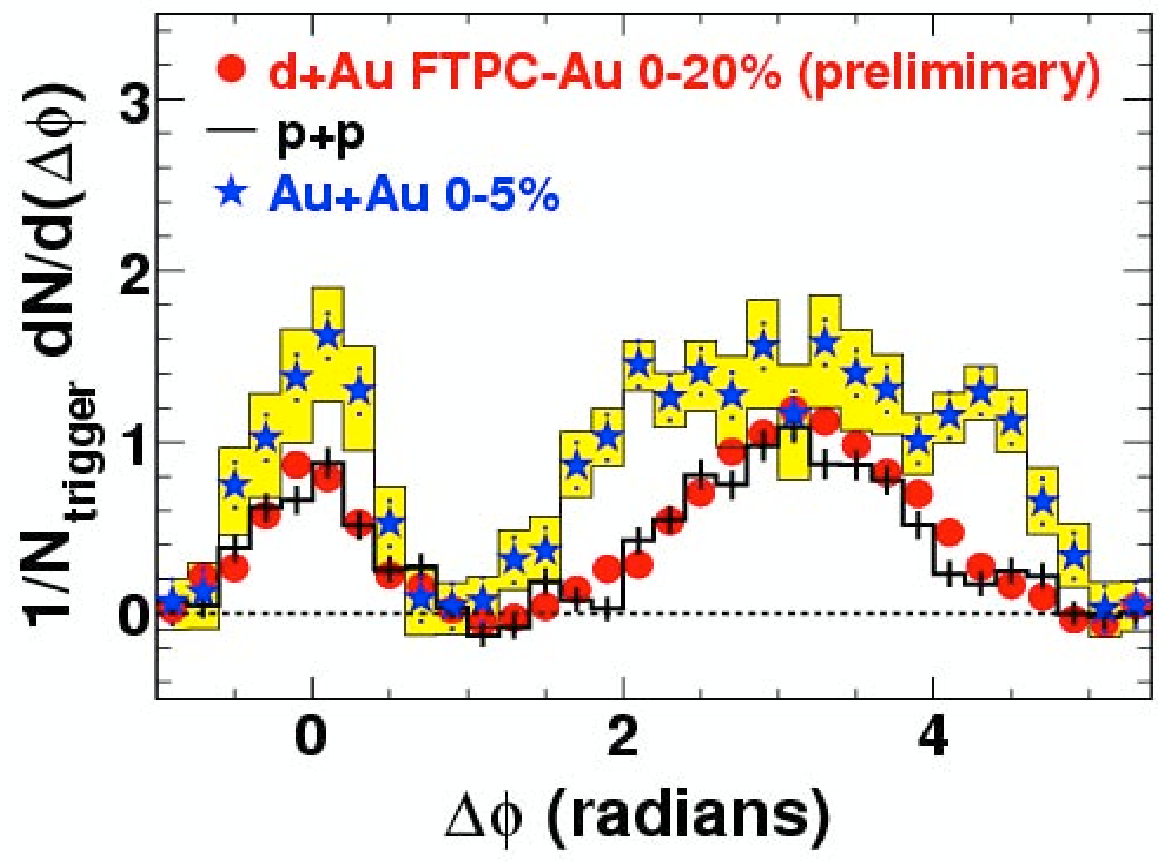}
	\end{tabular}
	\end{center}
	\vspace*{-0.24in}
     \caption[] {STAR conditional probability 2-particle correlation function with flow-modulated background subtracted: (left) Measurements in d+Au~\cite{STARdAu}, p-p and Au+Au central~\cite{STARPRL90} collisions at $\sqrt{s_{NN}}=200$ GeV with $4<p_{T_t}<6$ GeV/c and $2\,{\rm GeV/c}< p_{T_a} < p_{T_t}$; (right) STAR data with same trigger $p_{T_t}$ but with $0.15< p_{T_a} < 4$ GeV/c~\cite{FQWangQM04}. }
	\label{fig:nodisappear}
	\end{figure}

 Although the apparent vanishing of the away jet in central Au+Au collisions is fantastic from a public relations perspective, it is misleading from a scientific viewpoint as it suggests that the away-jet was totally absorbed by the opaque medium. Later work presented by STAR at Quark Matter 2004~\cite{FQWangQM04,STARPRL95} with $4<p_{T_t}<6$ GeV/c and $0.15< p_{T_a} < 4$ GeV/c  showed that the away jet didn't disappear, it just lost energy and the away-side correlation peak became much wider than in p-p collisions (Fig.~\ref{fig:nodisappear}).  		
	
\begin{figure}[ht]
	\begin{center}
		\includegraphics[width=0.55\linewidth]{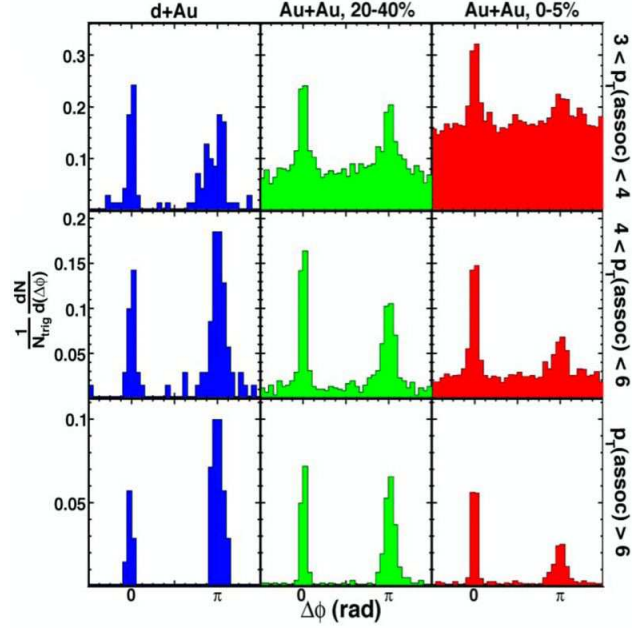}
	\end{center}
	\vspace*{-0.24in}
     \caption[] {STAR conditional probability 2-particle azimuthal correlation histograms for charged hadron triggers with $8 < p_{T_t}< 15$ GeV/c,
in minimum-bias d+Au, 20-40\% Au+Au and 0-5\% Au+Au collisions at $\sqrt{s_{NN}}$=200 GeV. $p_{T_a}$ increases from top to bottom as indicated~\cite{MagestroQM05,Magestro0604018}.}
	\label{fig:Magestro1}
	\end{figure}

	Still later work presented by STAR at Quark Matter 2005~\cite{MagestroQM05,Magestro0604018}, this past year, showed that an away-side jet correlation peak with the same width as in p-p collisions re-appeared when $p_{T_t}$ was raised to the range $8\leq p_{T_t}< 15$ GeV/c, with $p_{T_a}\geq 3$ GeV/c (Fig.~\ref{fig:Magestro1}). 	
Clearly, the study of jet phenomena by two-particle correlations in Au+Au collisions is much more complicated than the same subject in p-p collisions and one can expect a long learning curve. 

\begin{figure}[ht]
	\begin{center}
	\begin{tabular}{cc}
	\hspace*{-0.04in}\includegraphics[width=0.5\linewidth]{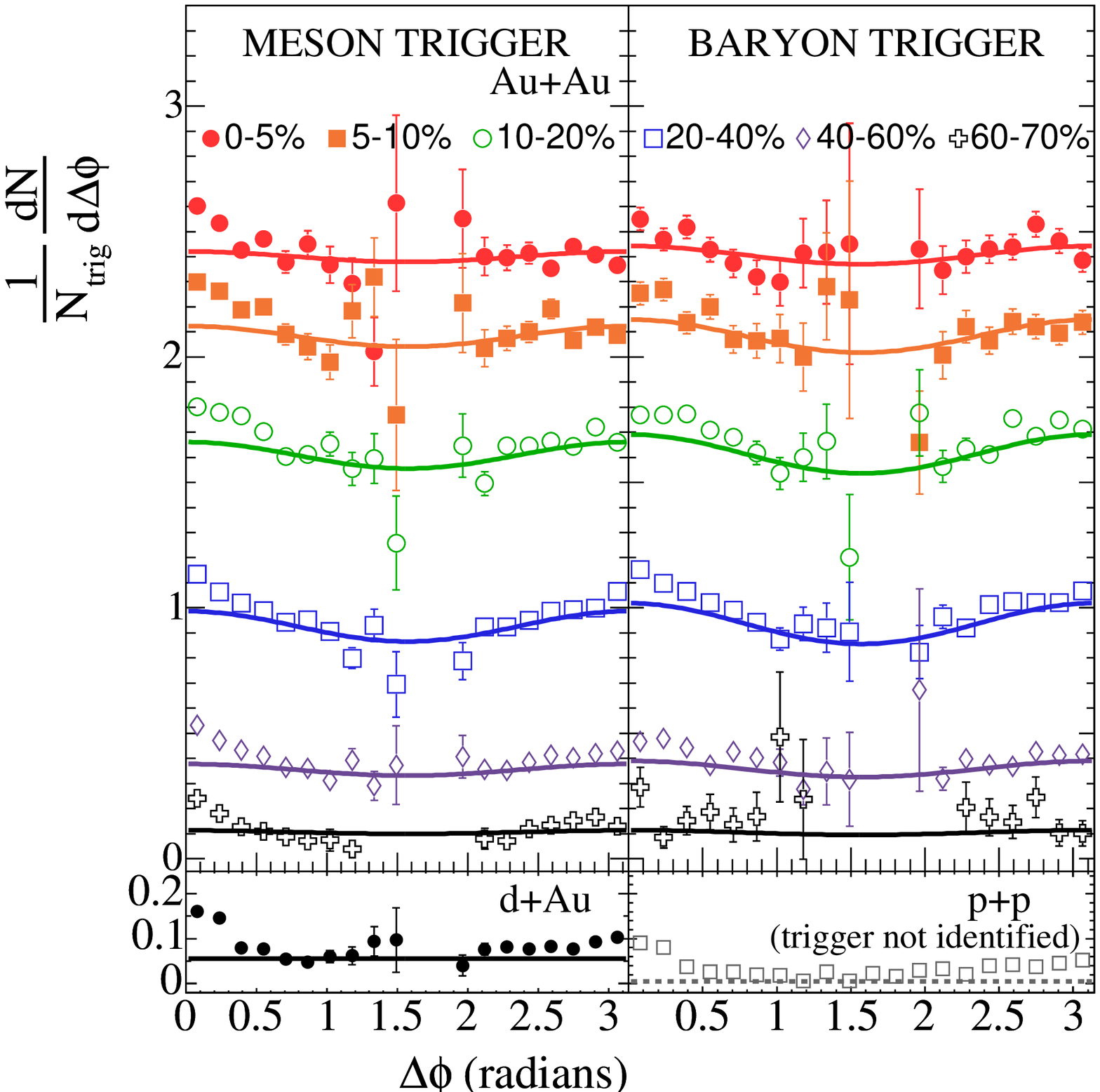}&
	\hspace*{-0.14in}\includegraphics[width=0.50\linewidth]{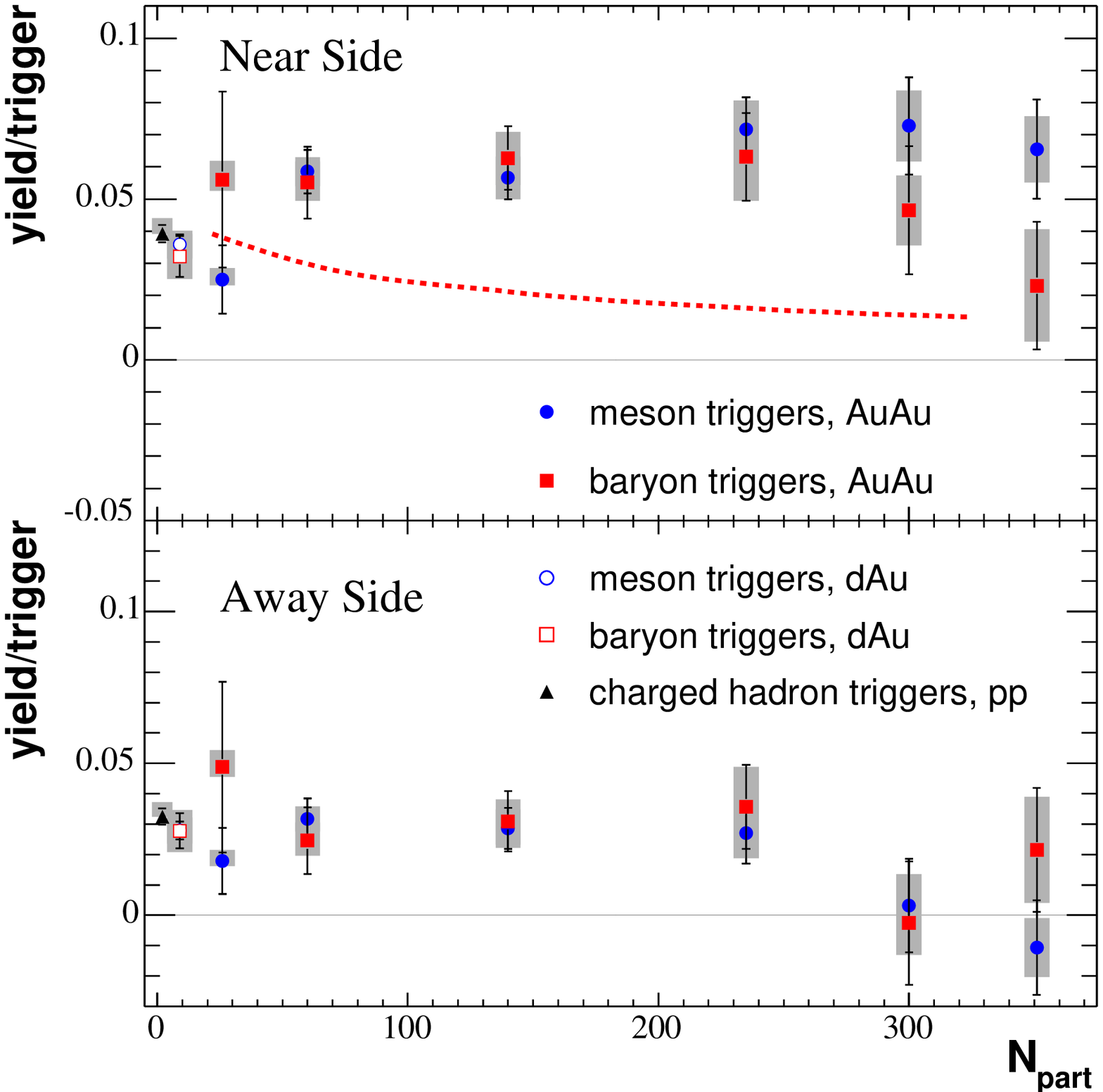}
	\end{tabular}
	\end{center}
	\vspace*{-0.24in}
     \caption[] {(left) PHENIX~\cite{PXPRC71} conditional probability 2-particle azimuthal correlation distributions ${1\over N_{trig}}{dN\over d\Delta\phi}$ for triggers by identified mesons and baryons with $2.5< p_{T_t} < 4$ GeV/c and   associated non-identified charged hadrons with $1.7 < p_{T_a} < 2.5$ GeV/c in collisions at $\sqrt{s_{NN}}=200$ GeV: Au+Au, as a function of centrality as indicated, d+Au minimum bias and p-p (non-identified triggers).      (right) Conditional yield per trigger meson (circles), baryon (squares) from this data,   integrated within $\Delta \phi=\pm 0.94$ radian of the trigger (Near Side) or the opposite azimuthal angle (Away Side), for Au+Au (full), d+Au (open) collisions at $\sqrt{s_{NN}}=200)$ GeV. Shaded boxes indicate centrality dependent systematic errors. An overall systematic error which moves all the points by 12\% is not shown. p-p data are shown for non-identified charged hadron triggers. }
	\label{fig:Sickles}
	\end{figure}

	However, even at this early stage, there is one definitive result from 2-particle jet correlations (Fig.~\ref{fig:Sickles})~\cite{PXPRC71}, in the sense that it casts serious doubt on the explanation of the `baryon anomaly' (recall Fig.~\ref{fig:banomaly}) by coalescence models~\cite{Greco,Fries,Hwa}. 
Fig.~\ref{fig:Sickles}-(left) shows the conditional probability 2-particle azimuthal correlation functions, with the integrated associated particle yields/trigger shown in Fig.~\ref{fig:Sickles}-(right), for p-p, d+Au and AuAu collisions in which the trigger is either an identified meson or baryon in the range $2.5\leq p_{T_t}\leq 4.0$ GeV/c and the associated particles, in the range $1.7\leq p_{T_a}\leq 2.5$ GeV/c, are not identified. The yield of associated particles/per trigger on the near side, from the same jet as the trigger hadron, is the same for meson and baryon triggers as a function of centrality, except perhaps in the most central bin; and the same effect is seen for the away-side yields. The red-dashed curve indicates the expected trigger-side conditional yield if all the anomalous protons in Au+Au collisions were produced by coalescence. This shows that meson and baryons at intermediate $p_T$ are produced by hard-processes with the same di-jet structure, and not by soft coalescence. 

\subsubsection{Jet energy loss or jet absorption?}
   In section \ref{sec:interesting}, I asserted that the ratio of the transverse momenta of the away-jet to the trigger-jet, $\hat{x}_h=\hat{p}_{T_a}/\hat{p}_{T_t}$, and hence the relative energy loss of the away to the same side jets in both p-p and A+A collisions could be determined from measurements of the $x_E$ distribution using Eq.~\ref{eq:condxe2}. A test of this method, which worked for p-p collisions, was presented in section~\ref{sec:testpp}. Now I apply the method to Au+Au collisions. 
   
\begin{figure}[ht]
\begin{center}
\begin{tabular}{cc}
\includegraphics[width=0.5\linewidth]{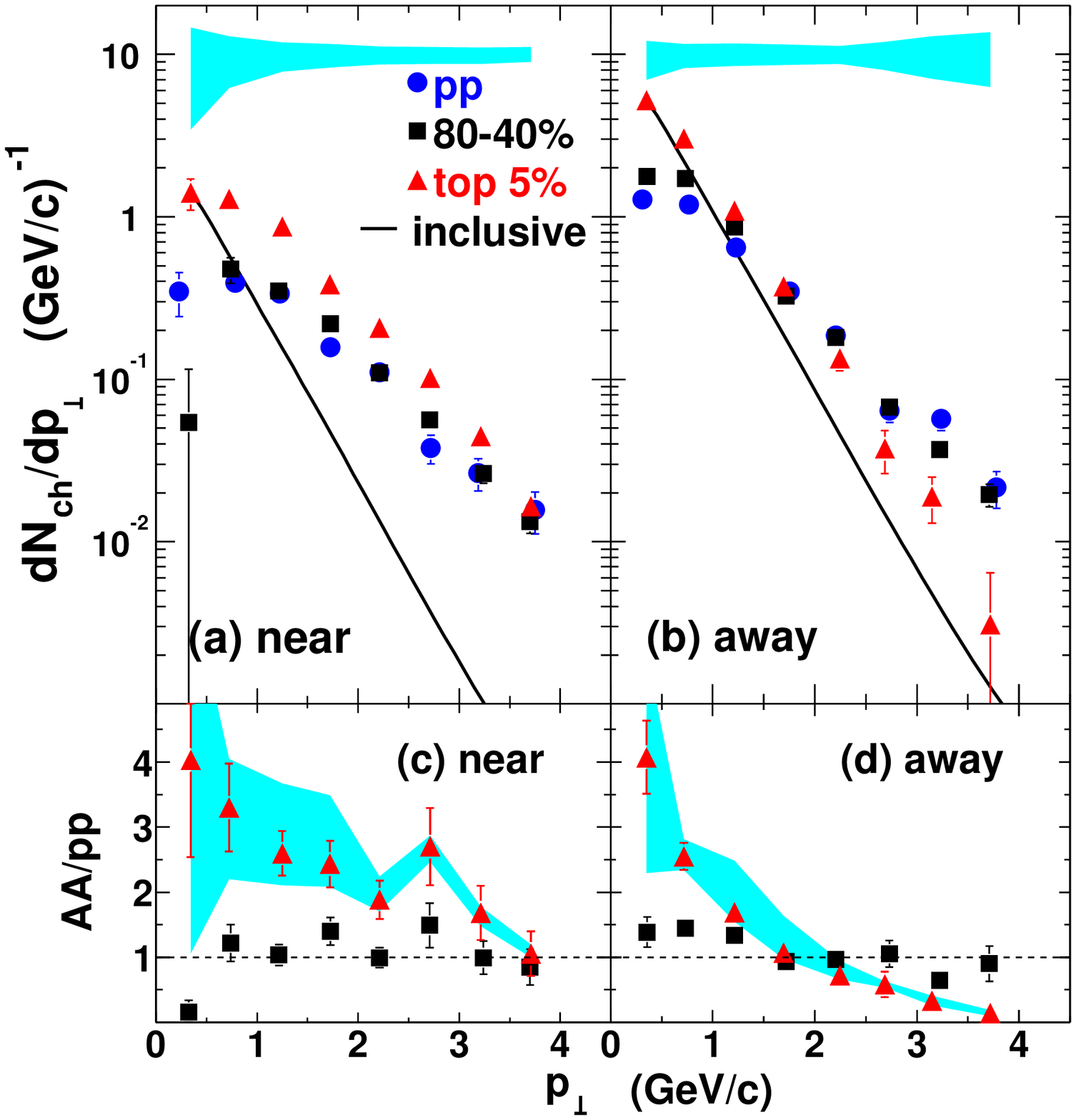} &
\includegraphics[width=0.4\linewidth]{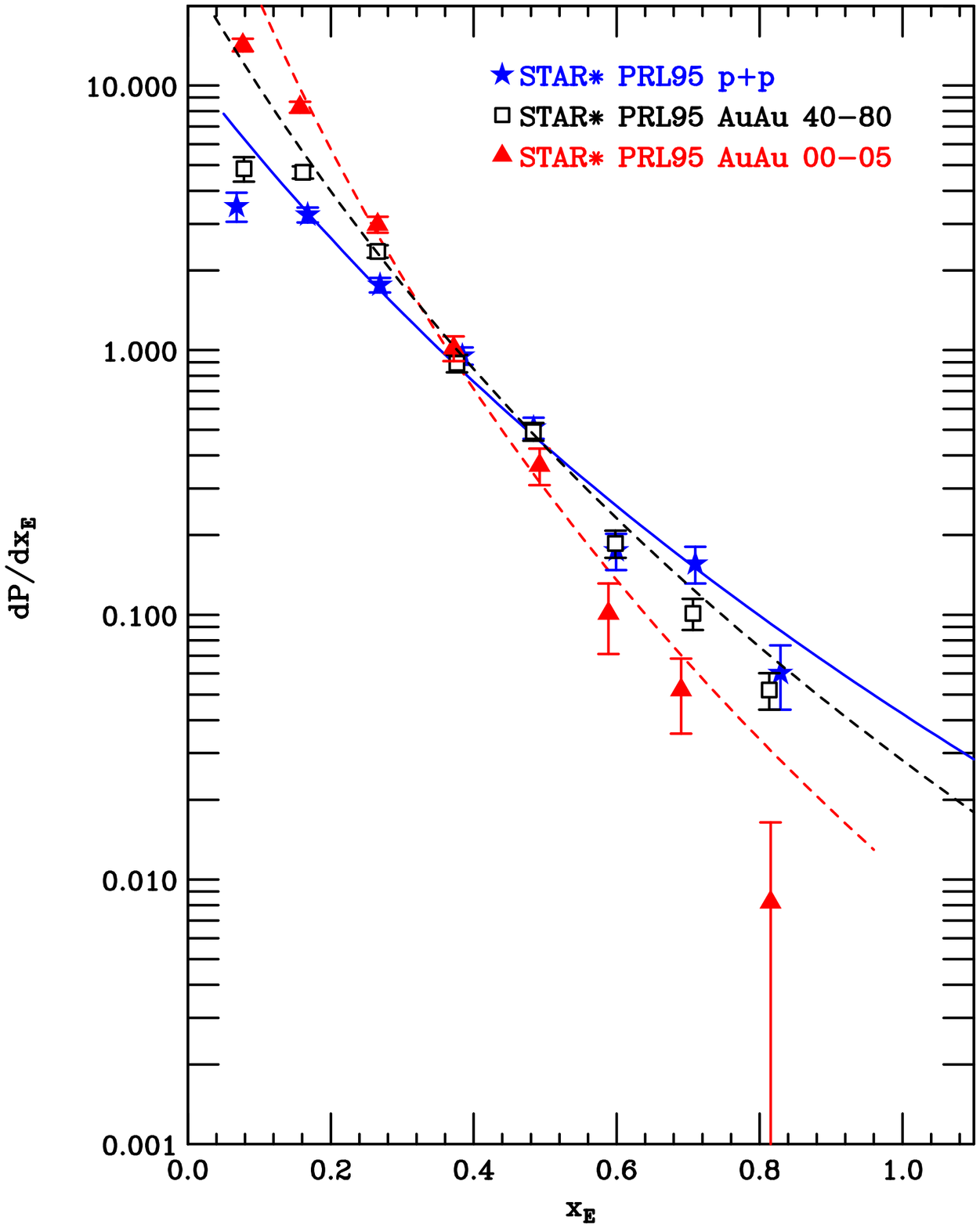} 
\end{tabular}
\end{center}
	\vspace*{-0.24in}
\caption[]{(left) STAR measurement~\cite{STARPRL95} of transverse momentum ($p_{\perp}$) distribution of associated charged hadrons for a trigger charged hadron  with $4 <p_{T_t}< 6$ GeV/c for pp, Au+Au peripheral(80-40\%), Au+Au central (top 5\%) collisions at $\sqrt{s_{NN}}=200$ GeV. a) near-side, b) away-side c,d) $I_{AA}$=ratio of AA to pp $p_{\perp}$ distributions for c) near side, d) away side. (right) data from (b) plotted as $dP/dx_E$ compared to Eq.~\ref{eq:condxe2} with $\hat{x}_h=1$ for p-p, $\hat{x}_h=0.75$ for Au+Au peripheral, and $\hat{x}_h=0.48$ for Au+Au central. The normalization is the same as in Fig.~\ref{fig:s31-43}-(left) for p-p collisions for both the data and the curve. The Au+Au data have been multiplied by the same factor of 0.6 to maintain the relative normalization as published, but the curves are normalized and $\hat{x}_h$-scaled by eye to agree with the measurements in the range $0.2\leq x_E\leq 0.8$. }
\label{fig:s42-44}
\end{figure}

   Fig~\ref{fig:s42-44}-(left) shows the STAR measurement~\cite{STARPRL95} of the $p_{T_a}$ distribution, given $p_{T_t}$, from the data shown in Fig.~\ref{fig:nodisappear}-(right).
In Fig.~\ref{fig:s42-44}-(right), these measurements are plotted as an $x_E\approx p_{T_a}/p_{T_t}$ distribution and shown together with Eqs.~\ref{eq:condxe2},~\ref{eq:condxe3}, with $n=8.1$, scaled to match the data, which are beautifully consistent with no relative energy loss of the two jets in p-p collisions as noted above (recall Fig.~\ref{fig:s31-43}-(right)). By contrast, in Au+Au collisions, agreement with the data is obtained with a ratio of away/trigger jet momenta of  0.75 in peripheral (40-80\%) and 0.48 in central  (0-5\%) collisions. This indicates a clear relative energy loss of the away jet compared to the trigger jet, which increases with increasing centrality. However, the trigger jets in Au+Au are surface biased by the falling power-law $p_T$ spectrum, an effect analogous to `trigger bias'---the jets which give trigger particles of a given $p_{T_t}$ are more likely to be produced near the surface and lose little energy, with $\hat{p}_{T_t}$ close to $p_{T_t}$, than to have been produced deeper in the medium with a larger $\hat{p}_{T_t}\gg p_{T_t}$ and then have lost significant energy in getting to the surface. Because of the trigger-jet surface bias, the away-jets must traverse the entire medium in order to be observed (except for the unlikely cases when the jet-pair is tangential to the medium). Hence, the decrease in $\hat{x}_h$ from 1.0 in p-p collisions to 0.75 in Au+Au peripheral (40-80\% ) collisions to 0.48 in Au+Au central (0-5\%) collisions indicates that the energy loss of the away-jet increases with distance traversed in the medium. 

\begin{figure}[ht]
	\begin{center}
	\begin{tabular}{ccc}
	\includegraphics[width=0.30\linewidth]{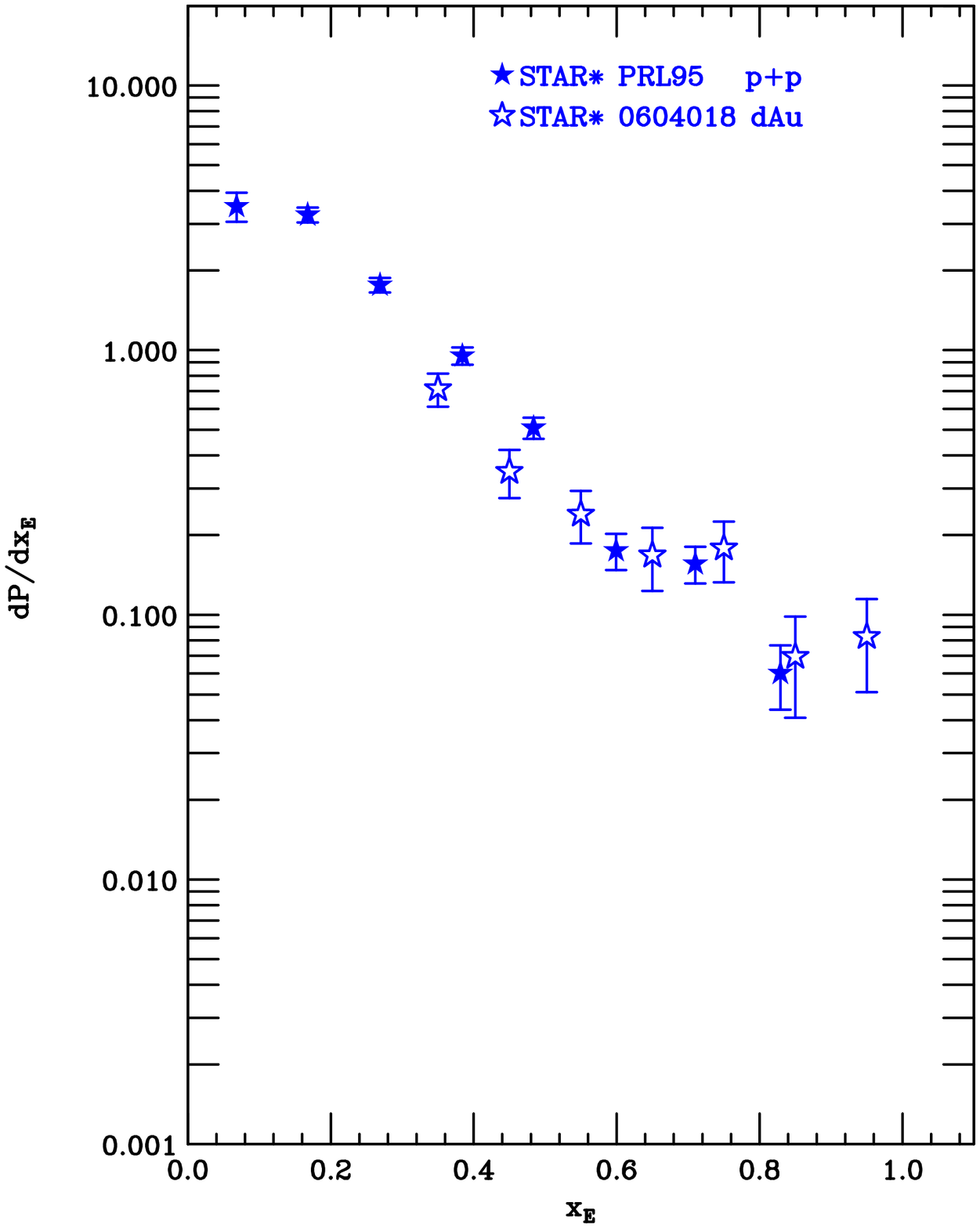}&
	\includegraphics[width=0.30\linewidth]{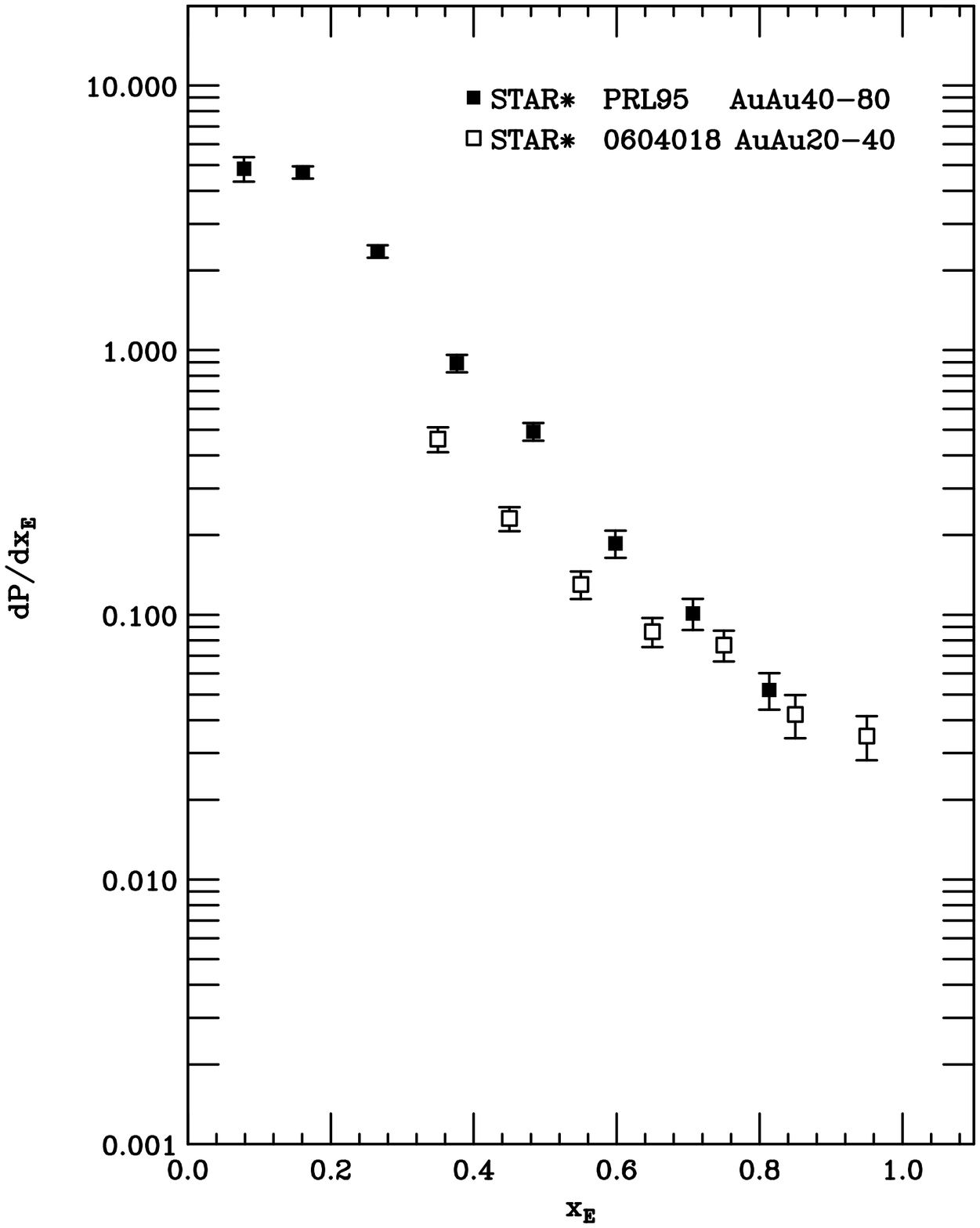}&
	\includegraphics[width=0.30\linewidth]{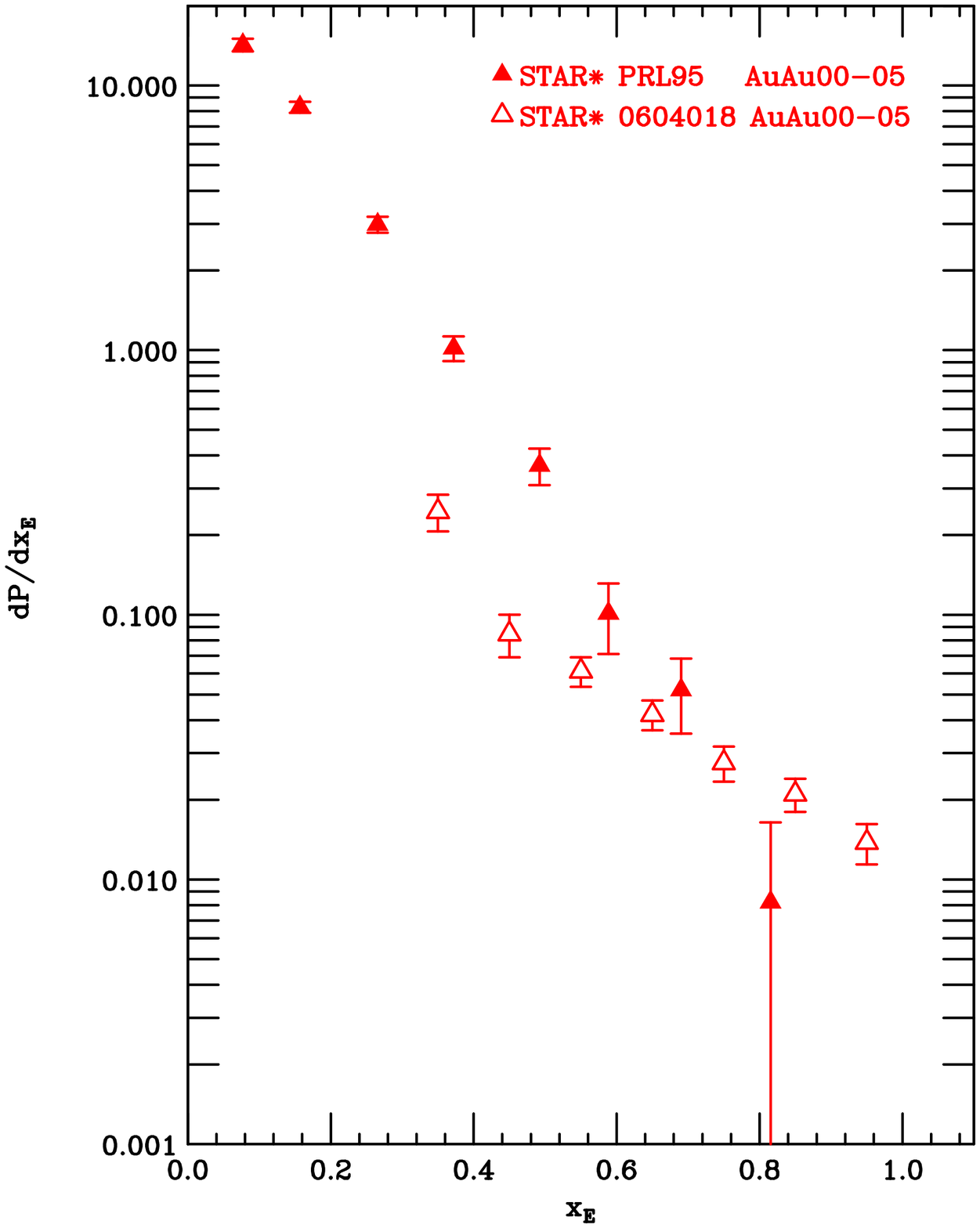}
	\end{tabular}
	\end{center}
	\vspace*{-0.24in}
     \caption[] {STAR data~\cite{STARPRL95} from Fig.~\ref{fig:s42-44} for d+Au, Au+Au (peripheral) and Au+Au central (0-5\%) collisions at $\sqrt{s_{NN}}=200$ GeV (filled points) compared to higher $p_{T_t}$ STAR data~\cite{Magestro0604018,thanksDan} (open points). }
	\label{fig:s46}
	\end{figure}

	I then tried to analyze the higher $p_{T_t}$ STAR away-jet measurement~\cite{MagestroQM05,Magestro0604018} by the same method~\cite{thanksDan}. First, I plotted the data from both STAR measurements as $x_E$ distributions on the same scale (Fig.~\ref{fig:s46}). The measurements appear to disagree, both in 
normalization and shape, so I tried normalizing the higher $p_{T_t}$  measurement~\cite{Magestro0604018} to agree with the lower $p_{T_t}$ measurement~\cite{STARPRL95},   
\begin{figure}[ht]
	\begin{center}
	\begin{tabular}{ccc}
	\includegraphics[width=0.30\linewidth]{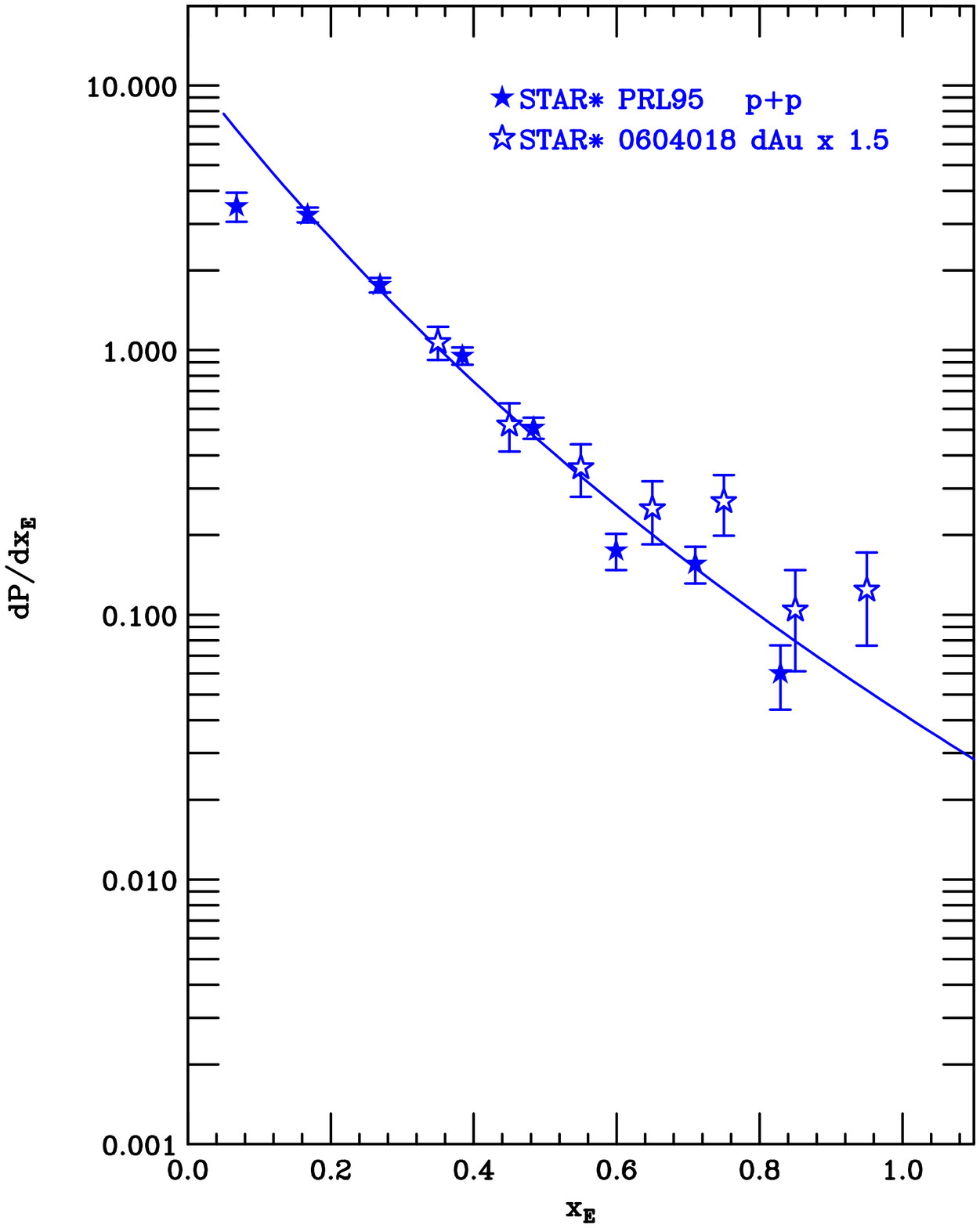}&
	\includegraphics[width=0.30\linewidth]{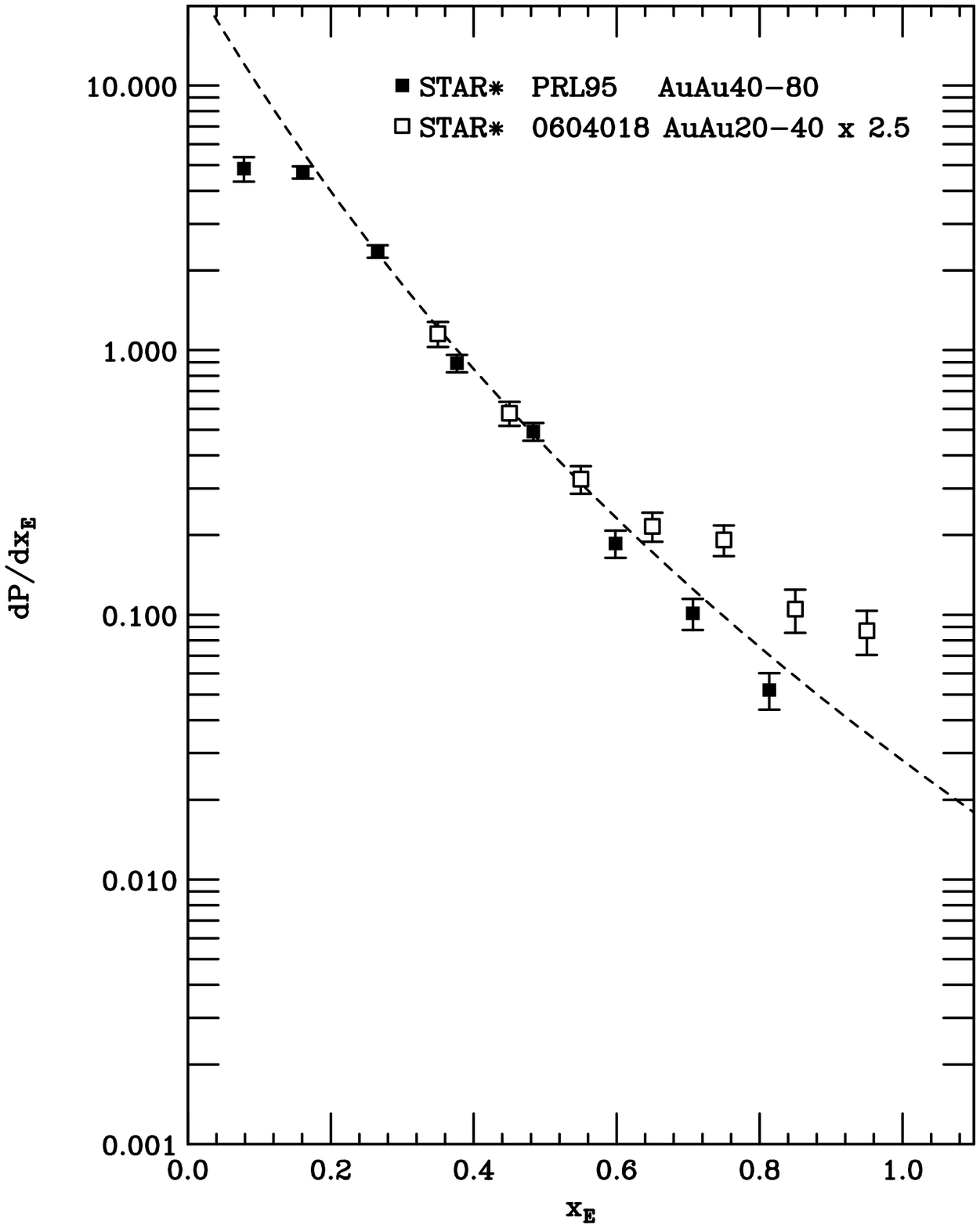}&
	\includegraphics[width=0.30\linewidth]{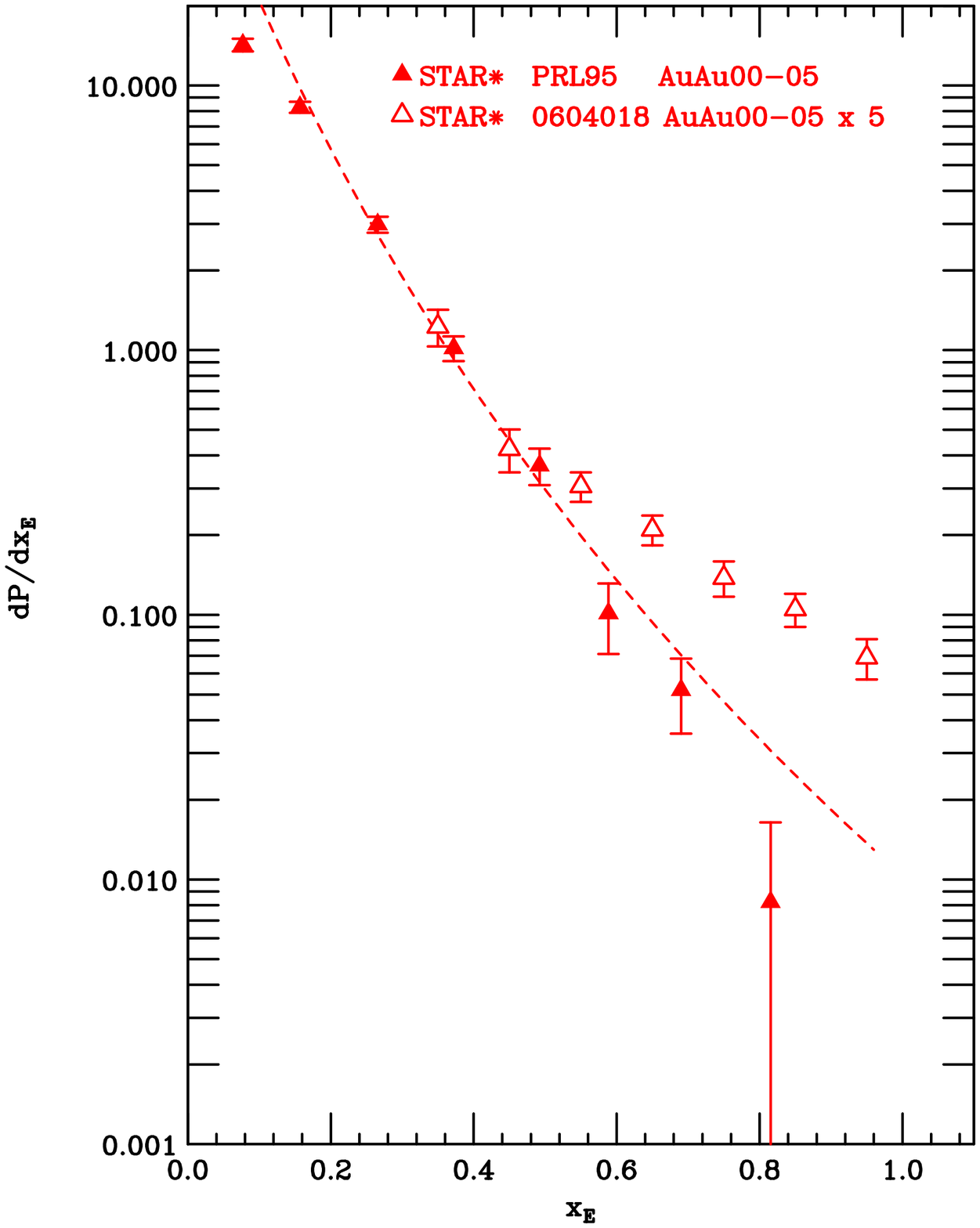}
	\end{tabular}
	\end{center}
	\vspace*{-0.24in}
     \caption[] {Data from Fig.~\ref{fig:s46}, with the higher $p_{T_t}$ data normalized to agree with the lower $p_{T_t}$ data for $x_E<0.45$, together with the curves from Fig.~\ref{fig:s42-44} which fit the lower $p_{T_t}$ data.    }
	\label{fig:s48}
	\end{figure}
which would be correct if $x_E$ scaling were valid in Au+Au collisions (see Fig.~\ref{fig:s48}). The results are quite interesting. It appears that the points at lower $x_E$ for the higher $p_{T_t}$ measurement are consistent with the shape of the lower $p_{T_t}$ data for $x_E < 0.45$, with a dramatic break and a flattening of the slope for $x_E\geq 0.5$. This could be suggestive of a two-component distribution where some jets, which pass through the medium, lose energy, while other jets, such as those emitted tangentially, punch through without any energy loss. However it is difficult to understand why the punch-through of tangential jets would depend on the trigger $p_{T_t}$. The comparison of the two STAR measurements and the possibility of a dramatic break in the $x_E$ distribution  would be greatly clarified if a few lower $x_E$ points could be obtained for the higher $p_{T_t}$ data. 

\begin{figure}[ht]
	\begin{center}
	\begin{tabular}{cc}
	\includegraphics[width=0.33\linewidth]{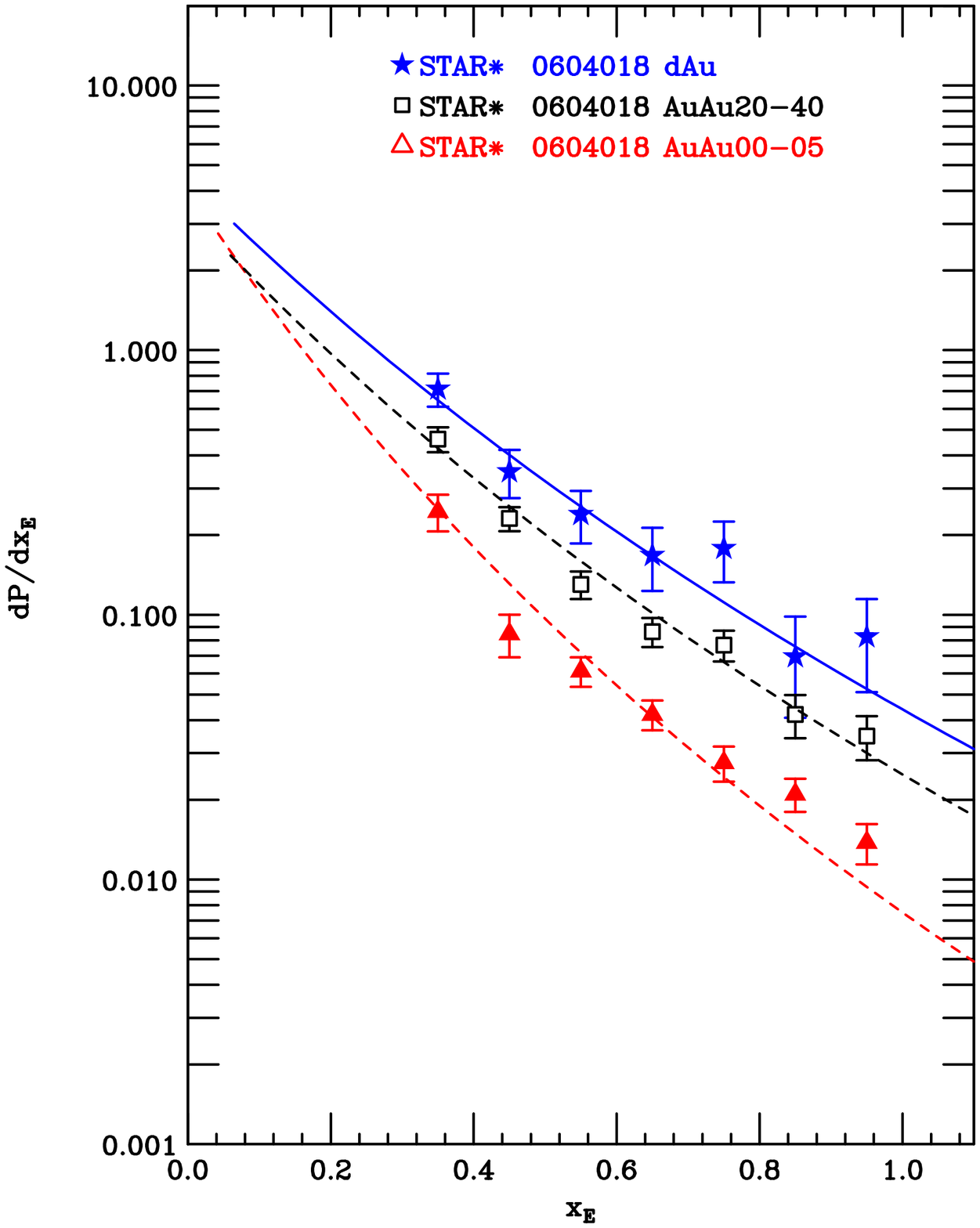}&
	\includegraphics[width=0.33\linewidth]{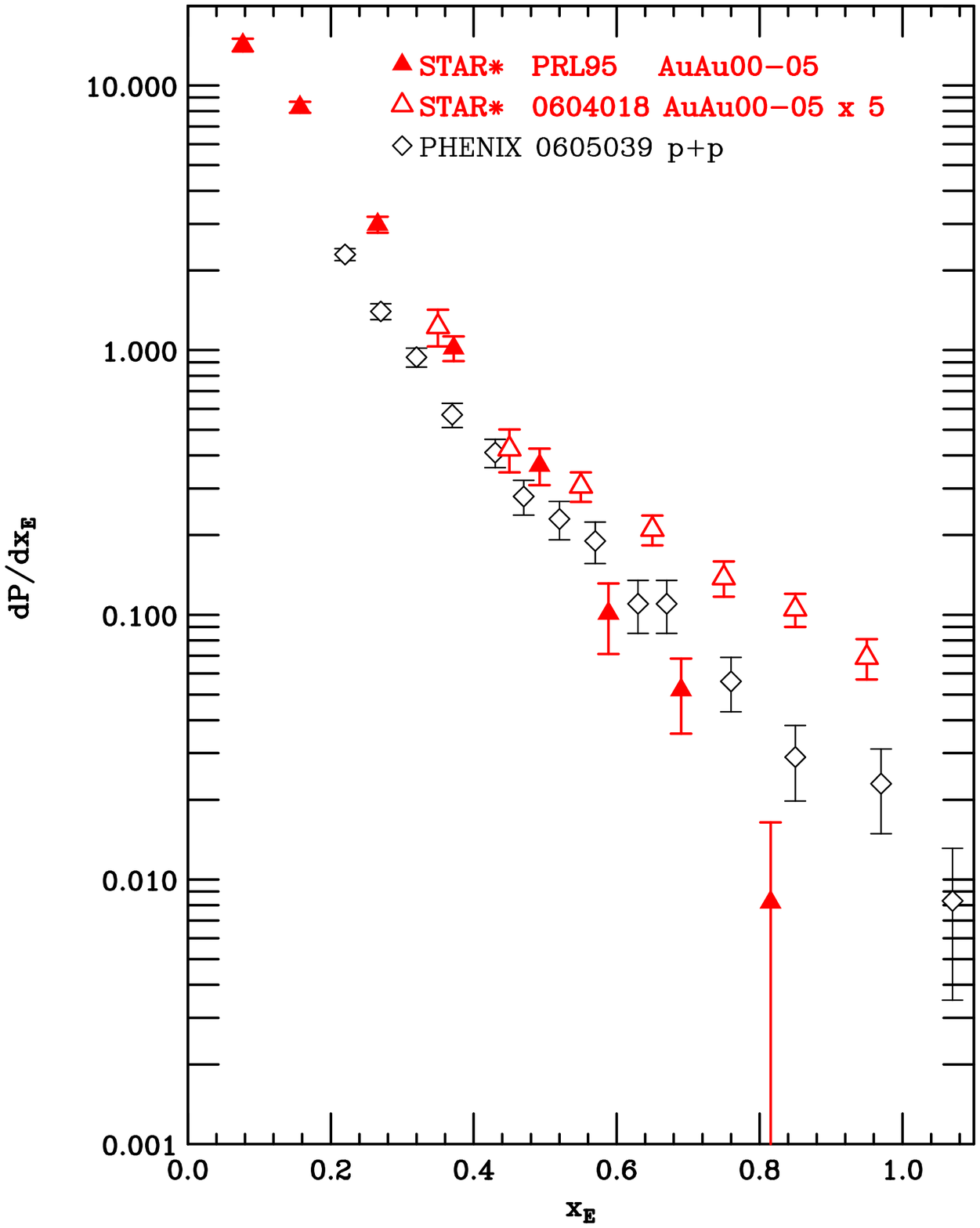}	\end{tabular}
	\end{center}
	\vspace*{-0.24in}
     \caption[] {(left) STAR higher $p_{T_t}$ data~\cite{Magestro0604018,thanksDan} as plotted in Fig.~\ref{fig:s46} together with curves of Eq.~\ref{eq:condxe2} adjusted by eye to best represent the data. (right) STAR higher and lower $p_{T_t}$ data from Fig.~\ref{fig:s48} for Au+Au central collisions together with PHENIX p-p data~\cite{ppg029} from Fig.~\ref{fig:s31-43}-(right).   }
	\label{fig:s49}
	\end{figure}
	
\begin{figure}[ht]
	\begin{center}
		\includegraphics[width=0.66\linewidth]{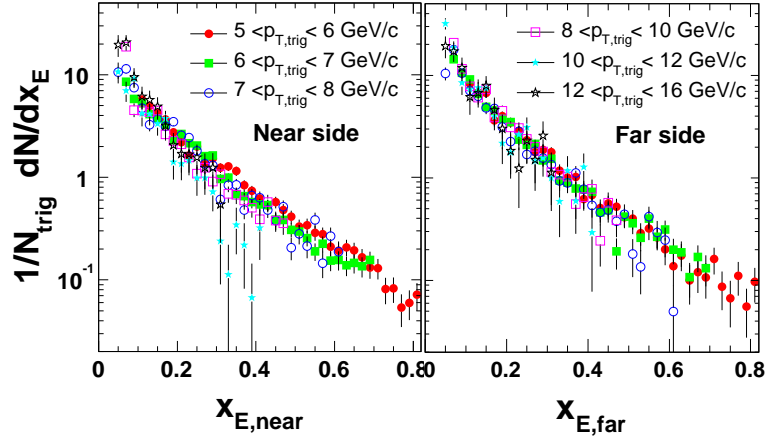}
	\end{center}
	\vspace*{-0.24in}
     \caption[] {PHENIX measurement~\cite{ppg039} of conditional yield as a function of $x_E$ for near- and far-side $\pi^{\pm}- h^{\pm}$ correlations for several values of trigger $p_{T_t}$ (indicated) in minimum bias d+Au collisions at $\sqrt{s_{NN}}=200$ GeV.  }
	\label{fig:ppg039}
	\end{figure}

	It is also possible to compare Eq.~\ref{eq:condxe2} to the STAR higher $p_{T_t}$ data without reference to the lower $p_{T_t}$ data (see Fig.~\ref{fig:s49}-(left)). Here, another troubling effect is revealed. The best values  of $\hat{x}_h$ are 1.30 for the d+Au data, 1.20 for the Au+Au peripheral (20-40\%) data and 0.85 for the Au+Au central (0-5\%) data. Thus, to within the error of the simplistic ``eyeball'' scaling, the away-jet in Au+Au central collisions  with higher $p_{T_t}$, $p_{T_a}$ also seems to lose about half it's energy relative to d+Au, consistent with the lower $p_{T_t}$ measurement. However the $x_E$ slope for the higher $p_{T_t}$ data is much flatter than other measurements in p-p and d+Au collisions in the same $p_{T_t}$ range (see Fig.~\ref{fig:s49}-(right) and Fig~\ref{fig:ppg039}) as reflected in the anomalous  
value of $\hat{x}_h=1.30$. In any model of jets with $k_T$ smearing, $\hat{x}_h$ must be $\leq 1$ as indicated by the other STAR and PHENIX data at RHIC (Figs.~\ref{fig:s31-43}-(right),~\ref{fig:s49}-(right),~\ref{fig:ppg039}-(right)). This clearly warrants further investigation.
\subsection{Possible new effects revealed by correlation measurements in Au+Au collisions at RHIC}
\subsubsection{The ridge}
\begin{figure}[ht]
	\begin{center}
	\includegraphics[width=0.90\linewidth]{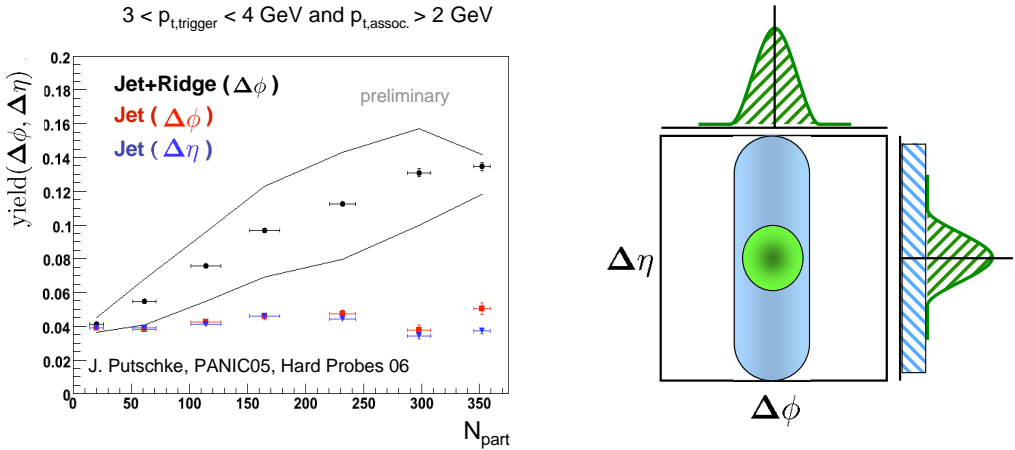}	
	\end{center}
	\vspace*{-0.24in}
     \caption[] {(left) STAR~\cite{PutschkePanic05} extraction of the near-side jet yields for $3<p_{T_t}< 4$ GeV/c and $p_{T_a}>2$ GeV/c: Jet+Ridge, when the whole $\Delta\eta$ acceptance is included; Jet($\Delta\phi$), Jet($\Delta\eta$), when account is taken of the ridge by two different methods. (right) Schematic drawing of the extent of the same-side ridge and jet correlation in the $\Delta\eta$ and $\Delta\phi$ directions.}
	\label{fig:ridge}
	\end{figure}

    Due to the large acceptance of the STAR detector, a near-side correlation in pseudo-rapidity ($\eta$) covering the full STAR $\eta$ acceptance was detected in addition to the flow modulated background and the near-side jet correlation~\cite{PutschkePanic05}. As indicated schematically in Fig.~\ref{fig:ridge}, the width of the ridge in the $\Delta\phi$ direction is comparable to the near-side jet correlation and must be taken into account in extracting the near-side jet yields. See Ref.~\cite{PutschkePanic05} for details. 
	
\subsubsection{Wide jets and/or Mach Cones--2 particle correlations} 	 
\begin{figure}[ht]
	\begin{center}
	\begin{tabular}{cc}
	\includegraphics[height=9.8cm,width=0.49\linewidth]{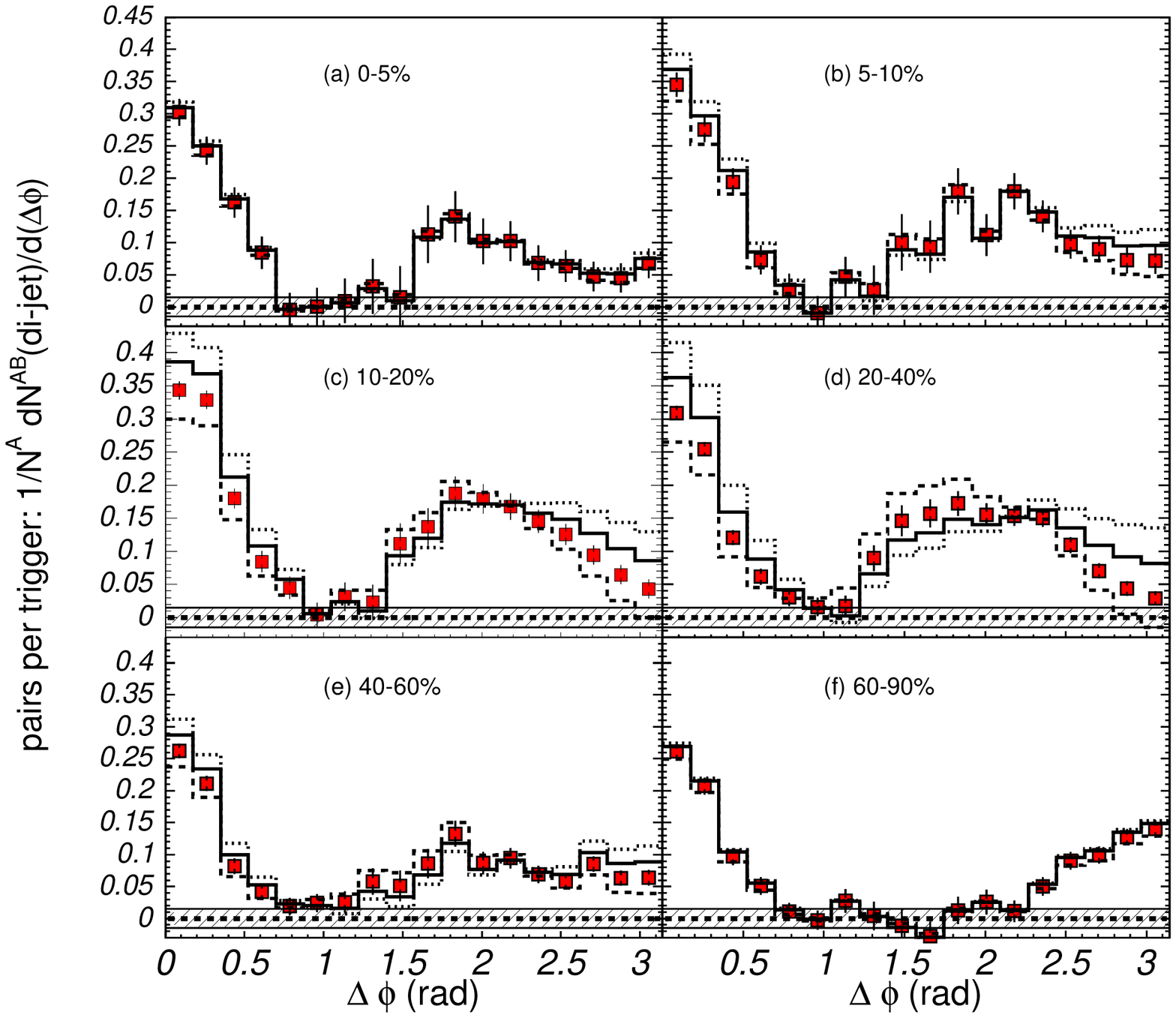}&
		\hspace*{-0.1in}\includegraphics[width=0.49\linewidth]{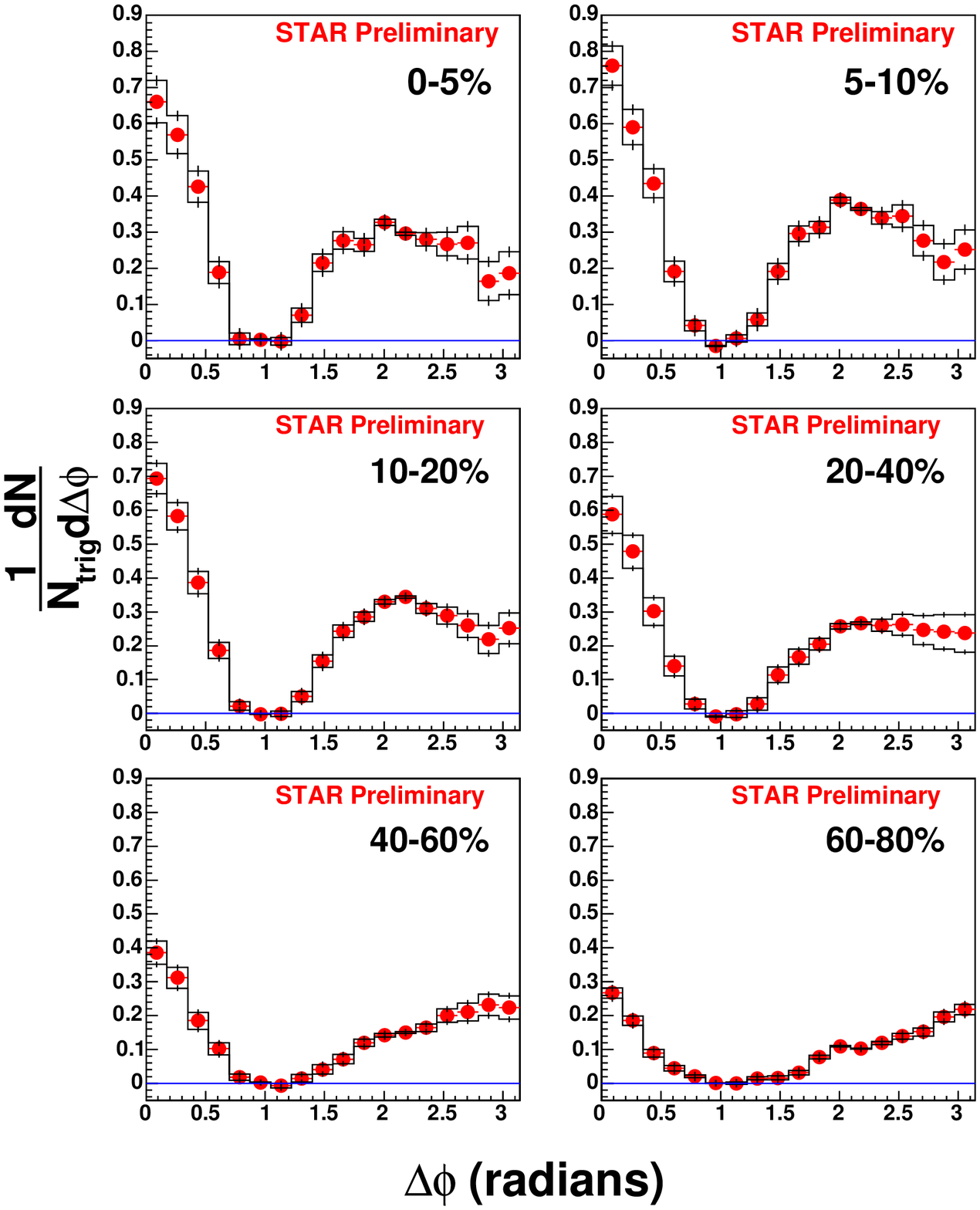}
	\end{tabular}
	\end{center}
	\vspace*{-0.24in}
     \caption[] { Conditional yields of associated particles with $1.0< p_{T_a}< 2.5$ GeV/c per trigger particle with $2.5 < p_{T_t}<4.0$ GeV/c as a function of $\Delta\phi$ for various centralities in Au+Au collisions at $\sqrt{s_{NN}}=$200 GeV after subtraction of the flow-modulated non-jet background: (left) PHENIX~\cite{ppg032} data where both 1 and 2 unit bands are shown for the (rms) systematic uncertainty in $v_2$; (right) STAR~\cite{UleryQM05} data with systematic uncertainties indicated as histograms. }
	\label{fig:widejet}
	\end{figure}

Measurements of non-identified charged hadron correlations by both PHENIX~\cite{ppg032} and STAR~\cite{UleryQM05} in the ``intermediate $p_T$'' region (where the `baryon anomaly' is found) are shown in Fig.~\ref{fig:widejet}.  
For both PHENIX and STAR the trigger and associated particles have $2.5 < p_{T_t}<4.0$ GeV/c and $1.0< p_{T_a}< 2.5$ GeV/c but the PHENIX range in pseudorapidity  is $|\eta|<0.35$ for both particles while for STAR $|\eta|<0.7$ for the trigger particle and $|\eta|<1$ for the associated particles. Although disagreeing in absolute value, presumably due to the different $\eta$ acceptances, the PHENIX and STAR measurements both exhibit a striking widening of the away-side correlation in going from peripheral to central collisions, with a strong hint of a local minimum (dip) developing at $\Delta\phi=\pi$ for centralities less than $\sim 60$\%.  The existence of these local minima per se is not significant once the systematic errors on $v_2$ are taken into account but it is clear that all the away-side distributions in the more central samples for both PHENIX and STAR have a very different shape than in the most peripheral sample and all seem to exhibit a dip at $\Delta\phi=\pi$. 
	
\subsubsection{Deflected jets and/or Mach Cones--3 particle correlations} 
\begin{figure}[ht]
	\begin{center}
	
	\includegraphics[width=0.90\linewidth]{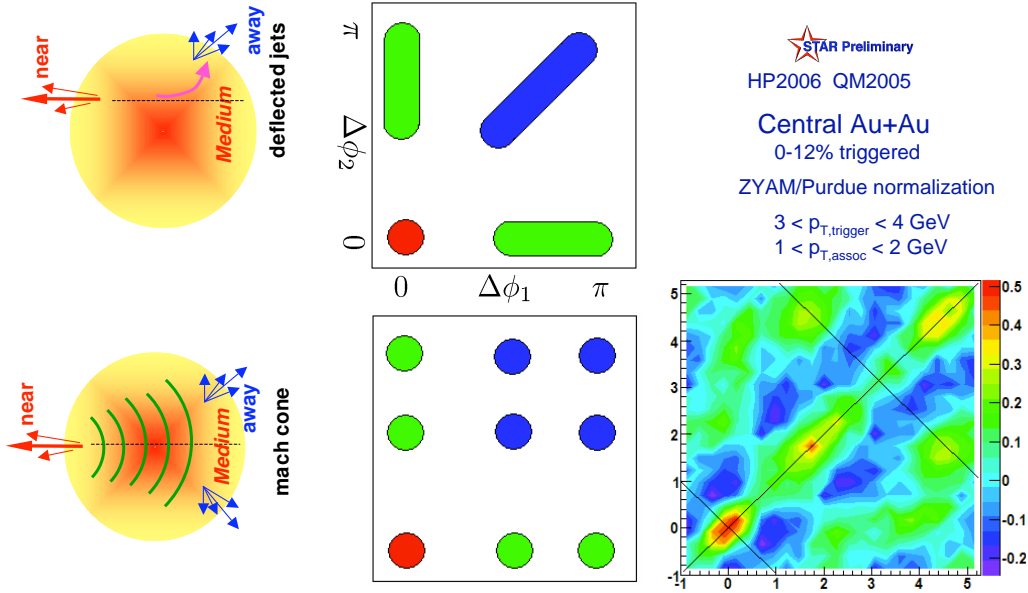}	
		\end{center}
	\vspace*{-0.24in}
     \caption[] {STAR method of 3 particle-correlations~\cite{UleryHP06}: (left) Schematic drawing of deflected jets (top) and Mach cone (bottom) opposite to a trigger; (center) A schematic plot of the azimuthal angle differences $\Delta\phi_1$ versus $\Delta\phi_2$ of each of the two associated particles with respect to the trigger for deflected jets (top) and a Mach cone (bottom); (right) STAR measurement of $\Delta\phi_1$ versus $\Delta\phi_2$ in central Au+Au collisions at $\sqrt{s_{NN}}=200$ GeV.}
	\label{fig:STAR3jet}
	\end{figure}

   There are numerous explanations for the possibly two-peaked structure, roughly 1 radian away from $\pi$, in the away-side distributions shown in Fig.~\ref{fig:widejet}, of which two are commonly discussed: a `Mach cone'~\cite{Mach}  due to the away parton exceeding the speed of sound in the medium and causing the QCD equivalent of a sonic-boom; or deflected jets, due to the strong interaction with the medium which, e.g. for mid peripheral collisions where the overlap region has a large eccentricity, might prevent directly back-to-back jets from penetrating through the medium (see Fig.~\ref{fig:STAR3jet}-(left)). 

	Both STAR and PHENIX try to distinguish a Mach cone from deflected jets using 3-particle correlations. In Fig.~\ref{fig:STAR3jet} STAR~\cite{UleryHP06}  studies the correlation of a trigger particle with $3< p_{T_t}<4$ GeV/c to  2 associated particles with $1< p_{T_a} < 2$ GeV/c by making a plot of  $\Delta\phi_1$ versus $\Delta\phi_2$, the difference in azimuth of each associated particle with the trigger particle. In Fig.~\ref{fig:STAR3jet}-(center) a schematic of the expected results are shown on top for the case of deflected jets for which $\Delta\phi_1=\Delta\phi_2$ when both associated particles are on the away-side and where $\Delta\phi\approx 0$ when one or both of the associated particles are on the trigger-side. The diagonal elongation near ($\pi,\,\pi$) is consistent with $k_T$ smearing, since the typical fragmentation transverse momentum $\sqrt{\mean {j_T^2}}\approx 0.6$ GeV/c is much less than $\sqrt{\mean{k_T^2}}\approx 2.7$ GeV/c~\cite{ppg029}, so that $\Delta\phi_1\approx\Delta\phi_2\neq \pi$.  For away-side particles which form a cone roughly around the direction opposite to the trigger, there are off-diagonal as well as on-diagonal $\Delta\phi_1=\Delta\phi_2$ correlations (bottom). This is not obviously the best projection to understand this problem as illustrated by the measurement in Fig.~\ref{fig:STAR3jet}-(right) which is difficult to understand but does appear to show off-diagonal activity.

\begin{figure}[ht]
	\begin{center}
	\begin{tabular}{c}
	\includegraphics[width=1.0\linewidth]{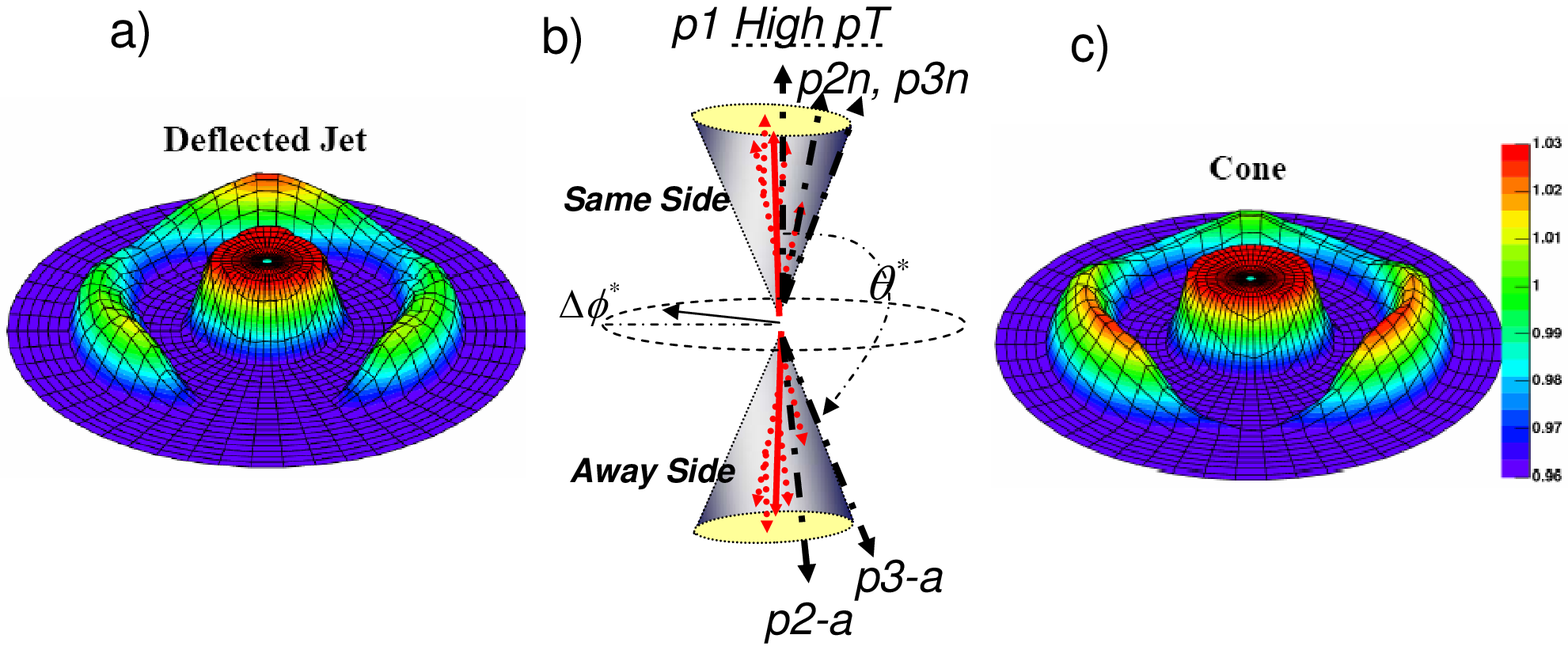}\cr
		d) \includegraphics[width=0.5\linewidth]{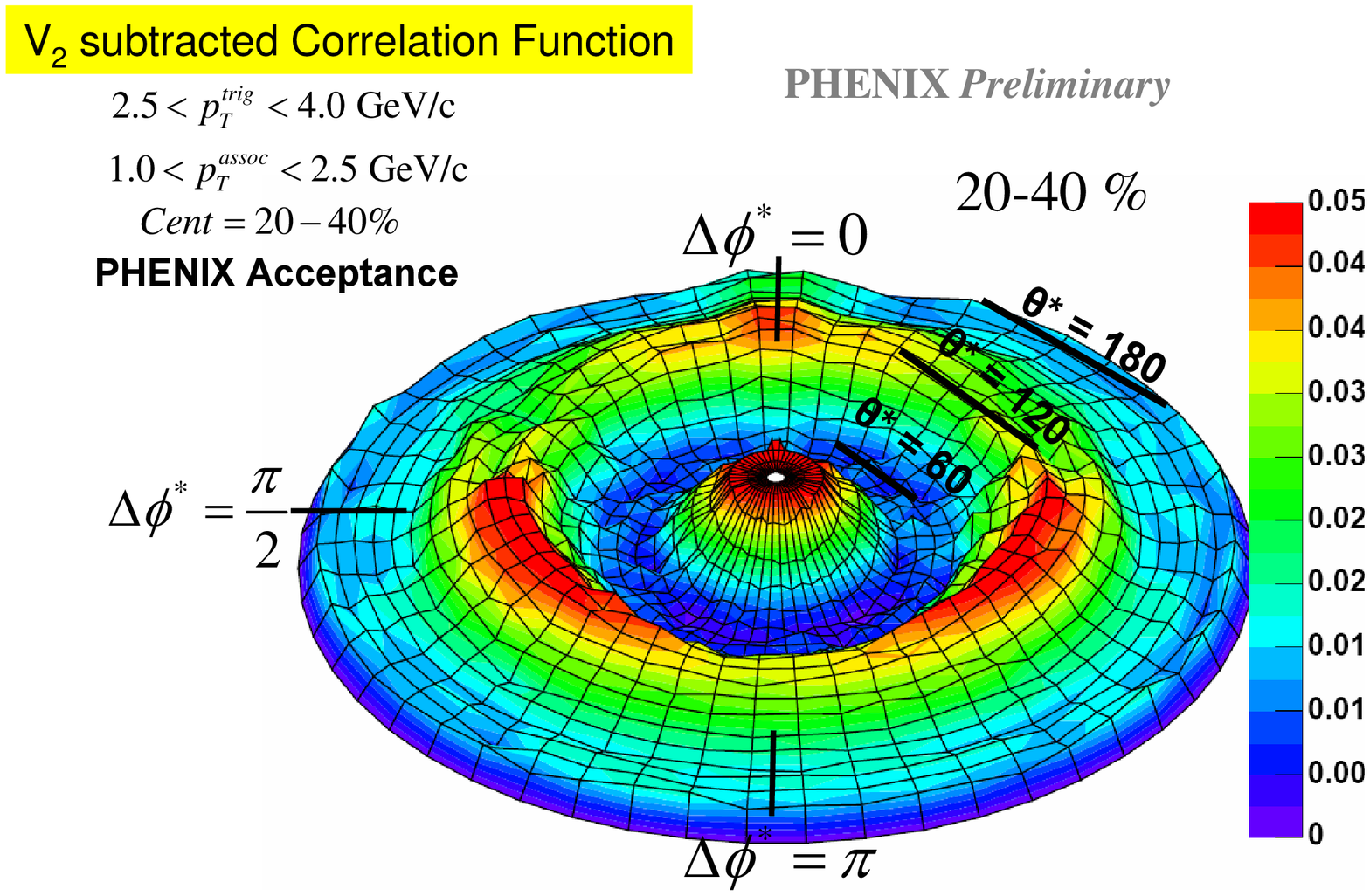}
	\end{tabular}
	\end{center}
	\vspace*{-0.24in}
     \caption[] {PHENIX method of 3-particle correlations~\cite{AjitHP06}. b) Schematic illustration of the coordinate system used for 3-particle correlations. The high $p_{T_t}$ trigger particle serves as the near-side jet axis and $\theta^*$ represents $\pi-\theta'$, where $\theta'$ would be the half-angle of a cone centered opposite in azimuth to the trigger $\vec{p}_{T_t}$;   the variable $\Delta\phi^*$ represents the `azimuthal' angle around the cone. a) Simulated 3-particle correlation for a deflected away-side jet in the PHENIX acceptance. c) Same as (a) but for a conical away-side jet. d) Measurement of $\theta^*$, $\Delta\phi^*$ 3-particle correlation surface for charged hadrons in central (0-5\%) Au+Au collisions within the PHENIX acceptance.  }
	\label{fig:PX3jet}
	\end{figure}

   PHENIX~\cite{AjitHP06} defines a coordinate system for the correlation of two associated particles ($1.0<p_{T_a}<2.5$ GeV/c) to a trigger particle ($2.5<p_{T_t}<4.0$ GeV/c) in which a conical correlation would be directly visible (Fig~\ref{fig:PX3jet}b). 
The angle $\theta^*$ represents $\pi-\theta'$, where $\theta'$ would be the half-angle of a cone centered opposite in azimuth to the trigger $\vec{p}_{T_t}$; and the variable $\Delta\phi^*$ for this analysis represents the `azimuthal' angle around the cone. The data are displayed as a polar plot of $\Delta\phi^*$ as a function of $\theta^*$ (Fig.~\ref{fig:PX3jet}a,c,d).  
For a Mach cone, there should be a `$\delta$-function' at a fixed half-angle $\theta^*$ smeared by $k_T$, and a uniform distribution in $\Delta\phi^*$ (Fig.~\ref{fig:PX3jet}c), while a deflected jet would show correlations that are close in both $\theta^*$ and $\Delta\phi^*$ and would favor orientations of $\Delta\phi^*$ in the $\eta$ direction, since the same-side and away-side jets are relatively uncorrelated in pseudorapidity (Fig.~\ref{fig:PX3jet}a).  The measurement shown in Fig.~\ref{fig:PX3jet}d seems to exhibit both types of activity. 
\section{Conclusion}
Much has been learned, both in the 1970's and recently at RHIC, by the study of jets and hard-scattering via single particle, two-particle and 3-particle measurements in p-p and A+A collisions. Clearly, for measurements in A+A collisions at RHIC, we are still at the early stages of a long and interesting learning process.


\begin{thebibliography}{99}
\bibitem{BDPS} R.~Baier, Yu.~Dokshitzer, S.~Peign\'e and D.~Schiff, \Journal{\PLB}{345}{277--286}{1995} [{\tt hep-ph/9411409}].
  \bibitem{BaierQCD98} R. Baier {\em QCD, Proc. IV Workshop-1998 (Paris)}  Eds, H. M. Fried, B. M\"uller (World Scientific, Singapore, 1999) pp 272--279.   
\bibitem{MJTQCD98}M.~J.~Tannenbaum {\it ibid.}, 
pp 280--285, pp 312--319.
\bibitem{Strasbourg} e.g. see {\em Proc. Int'l Wks. Quark Gluon Plsama 
Signatures (Strasbourg)} Eds. V.~Bernard, {\it et al.}, (Editions Frontieres, Gif-sur-Yvette, 
France, 1999).   
\bibitem{MJTEPS04} M.~J.~Tannenbaum, \Journal{\NPA}{749}{219c--228c}{2005}.
\bibitem{MJTHP04} M.~J.~Tannenbaum, \Journal{\EPJC}{43}{329--332}{2005}.
\bibitem{MJTCF05} M.~J.~Tannenbaum, \Journal{\JPCS}{27}{1--10}{2005}.
\bibitem{MJTRHIC97} M.~J.~Tannenbaum, {\it ``How to discover the QGP by single particle semi-inclusive measurements,''} Proc. RHIC '97 Summer Study, Upton, NY, July 7--16, 1997, Eds. D. Kahana and S. Kahana, see {\tt http://www.phenix.bnl.gov/{$\sim$}sapin/rhic97.ps.gz}.
\bibitem{MJTDPF79} M.~J.~Tannenbaum, {\em Particles and Fields-1979}, AIP Conference Proceedings Number 59, Eds. B. Margolis, D.~G.~Stairs (American Institute of Physics, New York, 1980) pp. 263--309. 
\bibitem{ppg029} S.~S.~Adler, {\it et al.}, PHENIX Collaboration, \Journal{\PRD}{74}{072002}{2006} [{\tt hep-ex/0605039}].
\bibitem{FFF} R.~P.~Feynman, R.~D.~Field and G.~C.~Fox, \Journal{\NPB}{128}{1--65}{1977}.
\bibitem{egsee1} e.g. see Refs.~\cite{MJTQCD98,MJTEPS04,MJTCF05} for a review.
\bibitem{Bj} J.~D.~Bjorken, \Journal{\PRD}{179}{1547}{1969}.  
\bibitem{BBK} S.~M.~Berman, J.~D.~Bjorken and J.~B.~Kogut, 
\Journal{\PRD}{4}{3388}{1971}.
\bibitem{CIM} R.~Blankenbecler, S.~J.~Brodsky, J.~F.~Gunion, 
\Journal{\PLB}{42}{461}{1972}. 
\bibitem{CGKS} R.~F.~Cahalan, K.~A.~Geer, J.~Kogut and Leonard Susskind, 
\Journal{\PRD}{11}{1199}{1975}.
\bibitem{CCHK} M.~Della Negra, {\it et al.}, 
\Journal{\NPB}{127}{1}{1977}. 
\bibitem{MJT79} For a contemporary view of the excitement of this period, and some more details, see Ref.~\cite{MJTDPF79}.
\bibitem{Owens78} J.~F.~Owens, E.~Reya, M.~Gl\"uck, 
\Journal{\PRD}{18}{1501}{1978}; J.~F.~Owens and J.~D.~Kimel, 
\Journal{\PRD}{18}{3313}{1978}.  
\bibitem{Owens} J.~F.~Owens, \Journal{\RMP}{59}{465}{1987}.
\bibitem{CutlerSivers} R.~Cutler and D.~Sivers, \Journal{\PRD}{17}{196}{1978}; \Journal{\PRD}{16}{679}{1977}.
\bibitem{Combridge:1977dm} B.~L.~Compridge, J.~Kripfganz and J.~Ranft, \Journal{\PLB}{70}{234}{1077}.
\bibitem{Paris82} { Proc. 21st Int'l Conf. HEP}, Paris, 1982, Eds.  
P.~Petiau, M.~Porneuf, {\it J. Phys.} C{\bf 3}\ (1982): see J.~P.~Repellin, p.  
C3-571; also see M.~J.~Tannenbaum, p. C3-134, G.~Wolf, p. C3-525.  
\bibitem{CCOR82NPB} A.~L.~S.~Angelis, {\it et al.}, 
\Journal{\NPB}{209}{284}{1982}.
\bibitem{Gordon} T.~{\AA}kesson, {\it et al.}, 
\Journal{\PLB}{128}{354}{1983}. 
\bibitem{MJTIJMPA} e.g. for a review, see M.~J.~Tannenbaum, 
\Journal{\IJMPA}{4}{3377}{1989}. 
\bibitem{Darriulat} P.~Darriulat,\Journal{\ARNPS}{30}{159}{1980}.
\bibitem{DiLella} L.~DiLella, \Journal{\ARNPS}{35}{107}{1985}.
\bibitem{CDF}F.~Abe {\it et al.}, CDF Collaboration \Journal{\PRL}{61}{1819}{1988}.
\bibitem{Adler:2003pb} S.~S. Adler, {\it et al.}, PHENIX Collaboration, \Journal{\PRL}{91}{241803}{2003}.
\bibitem{Aversa:1988vb}
F.~Aversa, P.~Chiappetta, M.~Greco, J.~P. Guillet, \Journal{\NPB}{327}{105}{1989}. 
\bibitem{Jager:2002xm}
B.~Jager, A.~Schafer, M.~Stratmann, W.~Vogelsang, \Journal{\PRD}{67}{054005}{2003}. 
\bibitem{ppg054} S.~S.~Adler, {\it et al.}, PHENIX Collaboration, {\it A detailed study of high-$p_T$ neutral pion supppression and azimuthal anisotropy in Au+Au collisions at $\sqrt{s_{NN}}=200$ GeV}, submitted to \PRC. 
\bibitem{BjPRD8} J.~D.~Bjorken, \Journal{\PRD}{8}{4098--4106}{1973}.
\bibitem{JacobLandshoff} M.~Jacob and P.~V.~Landshoff, \Journal{\it\PLC}{48}{285--350}{1978}.
\bibitem{prec-note} Eqs.~\ref{eq:gampi} and \ref{eq:eoverpi} are exact to the extent that the probability of a member of the $\gamma$-pair or $e^+ e^-$-pair to have any energy up to energy of the parent is constant. This is exact for $\pi^0$ decay, Eq.~\ref{eq:gampi}, but is only approximate for conversions, Eq.~\ref{eq:eoverpi}, where asymmetric energies of the pair are somewhat favored~\cite{CCRSNPB113}.
\bibitem{CCRSNPB113} F.~W.~B\"usser, {\it et al.}, CCRS Collaboration, \Journal{\NPB}{113}{189--245}{1976}.
\bibitem{MMay} M.~May {\it et al.} \Journal{\PRL}{35}{407--410}{1975}. Note that this article measures the ratio of $\mu-$p to $\mu-$A in DIS, but precisely the same factor of $A$ for scaling the point-like cross section applies. 
\bibitem{Vogt99} R.~Vogt \Journal{Heavy Ion Physics}{9}{399}{1999}  [{\tt nucl-th/9903051}].
\bibitem{CCRSPLB53} F.~W.~B\"usser, {\it et al.}, CCRS Collaboration, \Journal{\PLB}{53}{212--216}{1974}.
\bibitem{MJTqcd2003} M.~J.~Tannenbaum, {\it``Lepton and Photon Physics at RHIC''}, Proc. 7th Workshop on Quantum Chromodynamics, La Citadelle, Villefranche-sur-Mer, France, January 6--10, 2003, Eds. H.~M.~Fried, B. Muller, Y. Gabellini (World Scientific, Singapore, 2003) pp 25--38 [{\tt nucl-ex/0406023}]. 
\bibitem{MayaQM05} M.~Shimomura, {\it et al.}, PHENIX Collaboration, Proc. 18th Int'l Conf. on Ultra-Relativistic Nucleus-Nucleus Collisions--Quark Matter 2005 (QM'05)  Budapest, Hungary, Aug. 4--9, 2005,  \Journal{\NPA}{774}{457--460}{2006} [{\tt nucl-ex/0510023}].
\bibitem{explain1} It is important to note that the effective fractional energy loss estimated from the shift in the $p_T$ spectrum is less than the real average fractional energy loss of a parton at a given $p_T$. The effect is similar to that of trigger bias and for the same reason--the steeply falling $p_T$ spectrum. For a given observed $p_T$, the events at larger $p{'}_T$ with larger energy loss tend to be lost under the events with
smaller $p{'}_T$ with smaller energy loss. 
\bibitem{Adler:2003au}
S.~S. Adler, {\it et~al.}, PHENIX Collaboration, \Journal{\PRC}{69}{034910}{2004} [{\tt nucl-ex/0308006}]. 
\bibitem{PXscalingPRL91} S.~S.~Adler {\it et al.} PHENIX Collaboration \Journal{\PRL}{91}{172301}{2003}. 
\bibitem{Greco} V.~Greco, C.~M.~Ko and P.~Levai \Journal{\PRL}{90}{202302}{2003}.
\bibitem{Fries} R.~J.~Fries, B.~M\"uller and C.~Nonaka \Journal{\PRL}{90}{202303}{2003}.
\bibitem{Hwa} R.~C.~Hwa \Journal{\EPJC}{43}{233--237}{2005} and references therein.
\bibitem{PXPRC71} S.~S.~Adler {\it et al.} PHENIX Collaboration \Journal{\PRC}{71}{051902(R)}{2005}.
\bibitem{QCDcompton} H.~Fritzsch and P.~Minkowski, \Journal{\PLB}{69}{316}{1977}.  
\bibitem{PSU} {\em Proceedings of the Polarized Collider Workshop}, University 
Park, PA (1990), Eds. J.~Collins, S.~Heppelmann and R.~W.~Robinett, 
AIP conf. proc. No. 223, (AIP, New York, 1991). 
\bibitem{ppg060} S.~S.~Adler {\it et al.}, PHENIX Collaboration, {\it ``Measurement of direct photon production in $p+p$ collisions at $\sqrt{s}=200$ GeV''}, Submitted to \PRL, {\tt hep-ex/0609031}.
\bibitem{Werlen} M.~Werlen, {\it ``Perturbative photons in pp collisions at RHIC energies''}, seminar at BNL, Upton, NY, June 21, 2005. {\tt http://spin.riken.bnl.gov/rsc/write-up/Riken-BNL-werlen.pdf}
\bibitem{Aurenche99} P.~Aurenche, {\it et al.}, \Journal{\EPJC}{9}{107-119}{1999}. 
\bibitem{Blazey} G.~C.~Blazey and B.~L.~Flaugher, \Journal{\ARNPS}{49}{633--685}{1999}. 
\bibitem{STARPLB637} J.~Adams, {\it et al.}, STAR Collaboration, \Journal{\PLB}{637}{161--169}{2006} [{\tt nucl-ex/0601033}]. 
\bibitem{BPR06} S.~J.~Brodsky, H.~J.~Pirner and J.~Raufeisen, \Journal{\PLB}{637}{58--63}{2006}. 
\bibitem{AkibaQM05} Y.~Akiba, {\it et al.}, PHENIX Collaboration, Proc. 18th Int'l Conf. on Ultra-Relativistic Nucleus-Nucleus Collisions--Quark Matter 2005 (QM'05)  Budapest, Hungary, Aug. 4--9, 2005,  \Journal{\NPA}{774}{403--408}{2006} [{\tt nucl-ex/0510008}]. 
\bibitem{VG} I.~Vitev and M.~Gyulassy, \Journal{\PRL}{89}{252301}{2002}. 
\bibitem{egVG} e.g. see Ref.~\cite{VG} for detailed citations. 
\bibitem{egseeHP06} e.g. see {\em Discussion Sessions on Parton Energy Loss} in Proc. 2nd Int'l Conf. on Hard and Electromagnetic Probes of High Energy Nuclear Collisions (Hard Probes 2006), Asilomar, Pacific Grove, CA, June 9--16, 2006, to appear in {\it \NPA}.
\bibitem{coleQM05} B.~Cole, Proc. Quark Matter 2005, \Journal{\NPA}{774}{225--236}{2006}. 
\bibitem{EMC} J.~J.~Aubert, {\it et al.}, European Muon Collaboration (EMC), \Journal{\PLB}{123}{275--278}{1983}.
\bibitem{NMC} M.~Arneodo, {\it et al.}, New Muon Collaboration (NMC), \Journal{\NPB}{481}{23--39}{1996}.
\bibitem{PXdA} D.~Peressounko, {\it et al.} PHENIX collaboration, {\it ``Direct Photon Production in p+p and d+Au Collisions Measured with the PHENIX Experiment''}, Proc. 2nd Int'l Conf. on Hard and Electromagnetic Probes of High Energy Nuclear Collisions (Hard Probes 2006), Asilomar, Pacific Grove, CA, June 9--16, 2006, to appear in {\it \NPA}, {\tt hep-ex/0609037}.
\bibitem{EKS} K.~J.~Eskola, V.~J.~Kohlinen and C.~A.~Salgado, \Journal{\EPJC}{9}{61--68}{1999}. 
\bibitem{EKV} K.~J.~Eskola, V.~J.~Kohlinen and R.~Vogt, \Journal{\NPA}{696}{729--746}{2001}.
\bibitem{noteonn} Note that $n$ in the equations in sections \ref{sec:leading} and \ref{sec:frag-single} is the partonic power whereas in practice the power is measured from the $\pi^0$ spectrum, $n_{\pi}$.  To the extent that $x_T$ is small, the approximation of Eq.~\ref{eq:ans1_int_mjt_sig_inclus} to Eq.~\ref{eq:result2_int_mjt_sig_inclus} is sufficiently accurate that $n_{\pi}\approx n$. Furthermore, the neat results described in these sections plus section~\ref{sec:power} depend on a partonic power law spectrum for all $\hat{p}_{T_t}$ up to the endpoint $\sqrt{s}/2$, which may not be strictly true. Nevertheless, the simplifications based on the power-law work very well. 
\bibitem{Moriond79} e.g. see {\em Proc. XIV Rencontre de Moriond---``Quarks, Gluons and Jets" (Les Arcs)} (Editions Fronti\`eres, Dreux, France, 1979)  H.~Boggild, p. 321, M.~J.~Tannenbaum, p. 351, and references therein.  
\bibitem{Angelis79} A.~L.~S.~Angelis, {\it et al.}, CCOR Collaboration, \Journal{Physica Scripta}{19}{116--123}{1979}. 
\bibitem{JacobEPS79} M.~Jacob, {\em Proc. EPS Int'l Conf. HEP }, Geneva, Switzerland, 27 June-4 July, 1979 (CERN, Geneva, 1979) Volume 2, pp. 473-522. 
\bibitem{STARPRL95} J.~Adams, {\it et al.}, STAR Collaboration, \Journal{\PRL}{95}{152301}{2005}. 
\bibitem{DarriulatNPB107} P.~Darriulat, {\it et al.}, \Journal{\NPB}{107}{429--456}{1976}.  
\bibitem{JacobEPS79b} See p. 512 in reference \cite{JacobEPS79}.  
\bibitem{Sterman} e.g. see A.~Kulesza, G.~Sterman and W.~Vogelsang, \Journal{\PRD}{66}{014011}{2002}.
\bibitem{Levin} See also, E.~M.~Levin and M.~G.~Ryskin, \Journal{Sov. Phys. JETP}{42}{783}{1975}.
\bibitem{CCOR80} A.~L.~Angelis, {\it et al.}, CCOR Collaboration, \Journal{\PLB}{97}{163}{1980}.
\bibitem{OPAL} G.~Alexander, {\it et al.}, OPAL Collaboration, \Journal{\ZPC}{69}{543}{1996}.
\bibitem{DELPHI} P.~Abreu, {\it et al.}, DELPHI Collaboration, \Journal{\EPJC}{13}{573}{2000}.  
\bibitem{HardtkeQM02} D.~Hardtke, {\it et al.}, STAR Collaboration, \Journal{\NPA}{715}{272c--279c}{2003}. 
\bibitem{jetnote} It is not necessary to assume that the jet correlation is the same as in p-p collisions. In general, any jet correlation function with peaks can be added to the flow-modulated background and fit to the measurements. 
\bibitem{Adcox01} K.~Adcox, {\it et al.}, PHENIX Collaboration, \Journal{\PRL}{88}{022301}{2001}. 
\bibitem{FQWangQM04} F.~Wang, {\it et al.}, STAR Collaboration \Journal{\JPG}{30}{S1299--S1303}{2004}. 
\bibitem{STARdAu} J.~Adams, {\it et al.}, STAR Collaboration \Journal{\PRL}{91}{072304}{2003}. 
\bibitem{STARPRL90} C.~Adler, {\it et al.}, STAR Collaboration \Journal{\PRL}{90}{082302}{2003}.  
\bibitem{MagestroQM05} D.~Magestro, {\it et al.}, STAR Collaboration \Journal{\NPA}{774}{573--576}{2006}. 
\bibitem{Magestro0604018} J.~Adams, {\it et al.}, STAR Collaboration \Journal{\PRL}{97}{162301}{2006} [{\tt nucl-ex/0604018}]. 
\bibitem{thanksDan} Thanks to Dan Magestro for supplying me with a table of data points.  
\bibitem{ppg039} S.~S.~Adler, {\it et al.}, PHENIX Collaboration, \Journal {\PRC}{74}{054903}{2006}. 
\bibitem{PutschkePanic05} J.~Putschke, {\it et al.}, STAR Collaboration, {\em Particles and Nuclei}, AIP Conference Proceedings Number 842, Eds. P.~D.~Barnes, {\it et al.} (American Institute of Physics, Melville, NY, 2006) pp 119--121. 
\bibitem{ppg032} S.~S.~Adler, {\it et al.}, PHENIX Collaboration, \Journal{\PRL}{97}{052301}{2006}.
\bibitem{UleryQM05} J.~G.~Ulery, {\it et al.}, STAR Collaboration \Journal{\NPA}{774}{581--584}{2006}. 
\bibitem{Mach} J. Casalderrey-Solana, E.V. Shuryak, and D. Teaney, \Journal{\JPCS}{27}{22}{2005}. 
\bibitem{UleryHP06} J.~G.~Ulery, {\it et al.}, STAR Collaboration, {\tt nucl-ex/0609047}. 
\bibitem{AjitHP06} N.~N.~Ajitnand, {\it et al.}, PHENIX Collaboration, {\tt nucl-ex/0609038}.
\end{thebibliography}
\end{document}